\documentclass[aps,prapplied,superscriptaddress,twocolumn,floatfix,10pt]{revtex4-2}

\usepackage{placeins,etoolbox}
\preto\subsection{\FloatBarrier}
\usepackage[utf8]{inputenc}
\usepackage{amssymb,amsmath}
\usepackage{graphicx}
\usepackage[dvipsnames]{xcolor}

\usepackage{mathptmx}

\usepackage{times}

\usepackage{soul}
\usepackage{rotating}

\usepackage[normalem]{ulem}

\begin{document}

\title{Highly Efficient Functionalization of hBN with Lithium Oxalate: A Multifunctional Platform for Composites, Ion Transport, and Spin Labeling}

\author{Bence~G.~M\'arkus}\thanks{These three authors contributed equally.}
\affiliation{Stavropoulos Center for Complex Quantum Matter, Department of Physics and Astronomy, University of Notre Dame, Notre Dame, Indiana 46556, USA}

\author{Anna~Ny\'ary}\thanks{These three authors contributed equally.}
\affiliation{Stavropoulos Center for Complex Quantum Matter, Department of Physics and Astronomy, University of Notre Dame, Notre Dame, Indiana 46556, USA}

\author{D\'avid~Beke}\thanks{These three authors contributed equally.}
\affiliation{Stavropoulos Center for Complex Quantum Matter, Department of Physics and Astronomy, University of Notre Dame, Notre Dame, Indiana 46556, USA}
\affiliation{Institute for Solid State Physics and Optics, HUN-REN Wigner Research Centre for Physics, PO. Box 49, H-1525, Hungary}

\author{Lili~Vajtai}
\affiliation{Stavropoulos Center for Complex Quantum Matter, Department of Physics and Astronomy, University of Notre Dame, Notre Dame, Indiana 46556, USA}

\author{Sivaviswa~Radhakrishnan}
\affiliation{Mechanical, Materials and Aerospace Engineering Department, Illinois Institute of Technology, 10 W. 32nd St., Chicago, IL 60616, USA}

\author{Vignyatha~R.~Tatagari}
\affiliation{Mechanical, Materials and Aerospace Engineering Department, Illinois Institute of Technology, 10 W. 32nd St., Chicago, IL 60616, USA}

\author{Bradlee~J.~McIntosh}
\affiliation{Stavropoulos Center for Complex Quantum Matter, Department of Physics and Astronomy, University of Notre Dame, Notre Dame, Indiana 46556, USA}

\author{Adil~Saleem}
\affiliation{Mechanical, Materials and Aerospace Engineering Department, Illinois Institute of Technology, 10 W. 32nd St., Chicago, IL 60616, USA}

\author{Changlong~Chen}
\affiliation{Mechanical, Materials and Aerospace Engineering Department, Illinois Institute of Technology, 10 W. 32nd St., Chicago, IL 60616, USA}

\author{Bal\'azs~Zsirka}
\affiliation{University of Pannonia, Center for Natural Sciences, 8200 Veszprém, Egyetem u. 10., Hungary}

\author{Mandefro~Y.~Teferi}
\affiliation{Chemical Sciences and Engineering Division, Argonne National Laboratory, Lemont, Illinois 60439, USA}

\author{Jens~Niklas}
\affiliation{Chemical Sciences and Engineering Division, Argonne National Laboratory, Lemont, Illinois 60439, USA}

\author{Oleg~G.~Poluektov}
\affiliation{Chemical Sciences and Engineering Division, Argonne National Laboratory, Lemont, Illinois 60439, USA}

\author{Ira~D.~Bloom}
\affiliation{Chemical Sciences and Engineering Division, Argonne National Laboratory, Lemont, Illinois 60439, USA}

\author{Fulya~Dogan}
\affiliation{Chemical Sciences and Engineering Division, Argonne National Laboratory, Lemont, Illinois 60439, USA}

\author{Margit~Kov\'acs}
\affiliation{University of Pannonia, Center for Natural Sciences, 8200 Veszprém, Egyetem u. 10., Hungary}

\author{Ferenc~Simon}
\affiliation{Stavropoulos Center for Complex Quantum Matter, Department of Physics and Astronomy, University of Notre Dame, Notre Dame, Indiana 46556, USA}
\affiliation{Institute for Solid State Physics and Optics, HUN-REN Wigner Research Centre for Physics, PO. Box 49, H-1525, Hungary}
\affiliation{Department of Physics, Institute of Physics, Budapest University of Technology and Economics, M\H{u}egyetem rkp. 3., H-1111 Budapest, Hungary}

\author{G\'abor~Szalontai}
\affiliation{University of Pannonia, Center for Natural Sciences, 8200 Veszprém, Egyetem u. 10., Hungary}

\author{Leon~Shaw}
\affiliation{Mechanical, Materials and Aerospace Engineering Department, Illinois Institute of Technology, 10 W. 32nd St., Chicago, IL 60616, USA}

\author{L\'aszl\'o~Forr\'o}
\email{Corresponding author: lforro@nd.edu}
\affiliation{Stavropoulos Center for Complex Quantum Matter, Department of Physics and Astronomy, University of Notre Dame, Notre Dame, Indiana 46556, USA}

\author{K\'aroly~N\'emeth}
\email{Corresponding author: knemeth@illinoistech.edu}
\affiliation{Physics Department, Illinois Institute of Technology, 3101 South Dearborn St., Chicago, IL 60616, USA}

\begin{abstract}
    Multifunctional solid-state materials are crucial for enabling safer and simpler lithium battery architectures. Here we report a scalable, solvent-free mechanochemical strategy to overcome the chemical inertness of hexagonal boron nitride and produce Li$_2$(BN)$_6$C$_2$O$_4$ (LBNCO), a lithium-rich boron nitride nanostructure. High-energy milling simultaneously exfoliates hBN and induces robust covalent oxalate functionalization, yielding a nanocrystalline lamellar compound with intrinsically incorporated lithium. \mbox{LBNCO} exhibits Li$^+$ ionic conductivity, electronic insulation, and thermal stability exceeding $350~^{\circ}$C. Additional mechanochemical processing enables further lithium enrichment without compromising chemical stability, allowing tunable ionic conductivity. By integrating ion transport, mechanical robustness, and thermal stability within a single material, LBNCO establishes a new platform for solid-state lithium batteries and related technologies.
\end{abstract}

\maketitle

\section{Introduction}

Advancing contemporary lithium-ion battery technology will require more than incremental modifications to existing materials. Meaningful improvements in energy density and fast-charging capability must be accompanied by substantial enhancements in safety, thermal stability, and device-level simplicity. A particularly promising strategy is the development of multifunctional solid-state material platforms that integrate ionic transport, mechanical reinforcement, and electrode-electrolyte interfacial stabilization within a single material system.

Covalent functionalization of two-dimensional (2D) materials provides a robust and versatile framework for modulating both chemical reactivity and physical properties. Graphene oxide (GO) is the canonical example \cite{brodie1859xiii, staudenmaier1898verfahren, hummers1958preparation, szabo2006evolution, geim2007rise, dreyer2010chemistry, Eigler2013Chem, Eigler2014Angewandte, Eigler2017PSR}: its basal planes and edges are decorated with oxygen-containing functional groups, affording extensive chemical tunability and enabling facile processing for applications spanning energy storage, catalysis, and nanocomposite fabrication. Closely related sp$^3$-hybridized carbon derivatives, offering the highest specific energy among battery chemistries, including graphite fluoride (CF$_x$) \cite{ruff1934reaktionsprodukte, rudorff1947konstitution, watanabe1970primary, watanabe1972high, fukuda1981active, jang2011graphene, kim2014all, liu2014lithium, kim2014novel, krishnan2012energetic, kim2010self, huo2023high, Liu2026JCIS} and hydrogenated graphite \cite{Pekker2001JPhysChemB, Pumera2013RSC, Fei2020CEJ}, further demonstrate that targeted lattice modification can be leveraged to introduce and optimize electrochemical functionality.

\begin{figure*}[htp]
    \includegraphics[width=\textwidth]{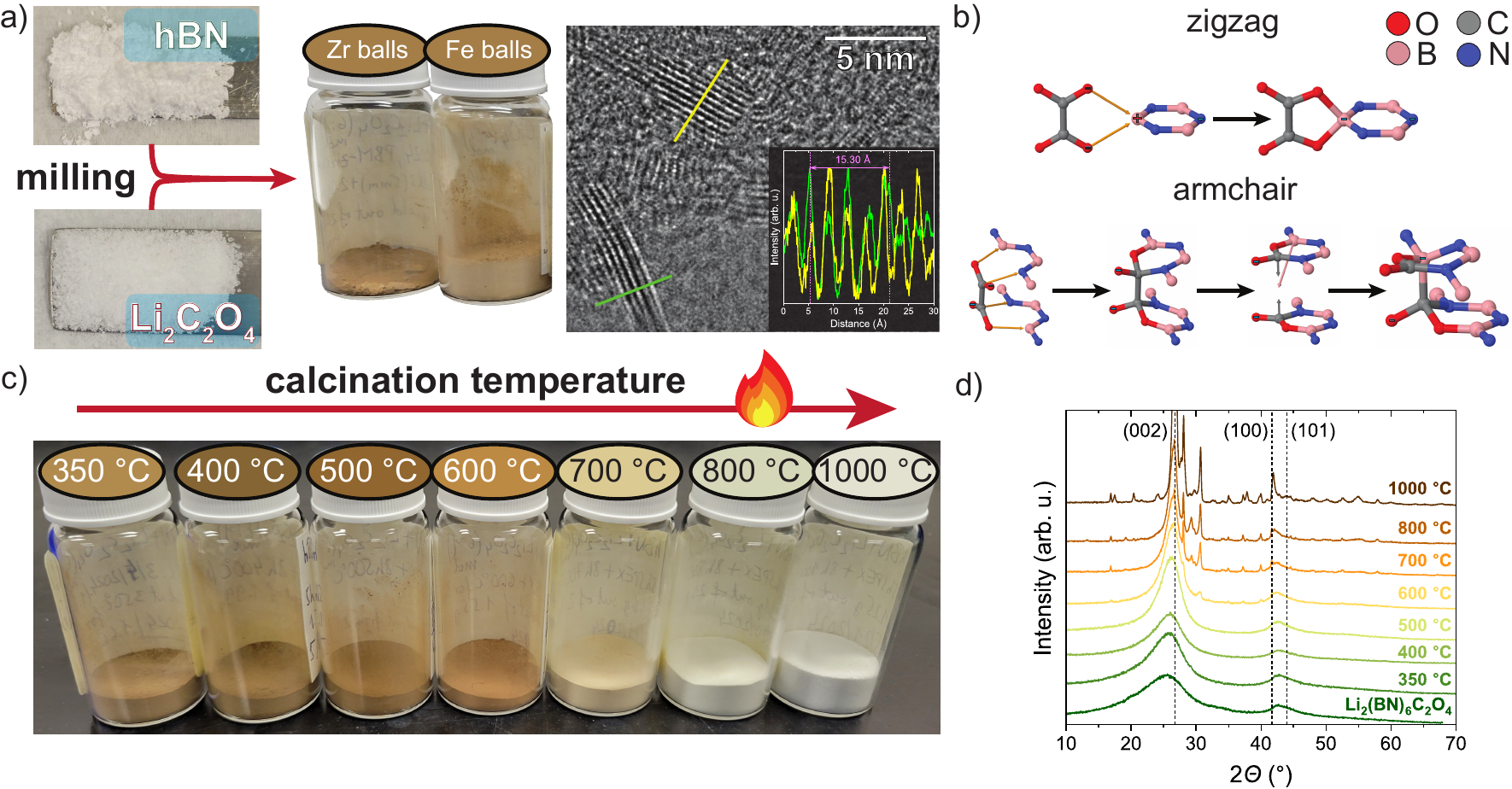}
    \caption{
        a) The synthesis of Li$_{2}$(BN)$_{6}$C$_{2}$O$_{4}$ (LBNCO) from its constituents using a high-energy ball mill (Fe balls) or a planetary ball mill (Zr balls), and an HRTEM image of the resulting material showing the lamellar nanocrystalline nature of it and the lattice periodicity in the $c$-axis direction. We note that, for clarity, the line profiles shown here are restricted to the central $30$ \AA\ section. The complete line profiles are provided in the Supporting Information.} b) Proposed reaction steps for both types of edges (zigzag and armchair) supported by DFT-based modeling and spectroscopic evidence. c) Calcination of the LBNCO material results in a color change of the material: while up to $400~^{\circ}$C it darkens, above $600~^{\circ}$C it immensely lightens, reaching a pastel white color above $800~^{\circ}$C. d) Powder X-ray diffraction patterns of LBNCO and its calcined derivatives. The vertical lines refer to the reflections of crystalline hBN; milling results in a turbostratic material.
    \label{Fig1:synthesis}
\end{figure*}

Building on these developments, hexagonal boron nitride (hBN) has emerged as a compelling two-dimensional building block, owing to its thermal stability, mechanical strength, and chemical inertness \cite{hbnbook2024}. However, these same attributes render hBN highly resistant to covalent functionalization and intentional doping. Consequently, reported functionalization protocols often rely on harsh reaction conditions and deliver low yields with limited control over functional group density and composition. Though early radical intercalation studies revealed unusual electronic behavior (e.g., enhanced conductivity and paramagnetism), these were limited by poor chemical stability and significant handling constraints \cite{bartlett1978novel, shen1999intercalation}. Lithium in particular resists incorporation: although Li-hBN chemistry is well established under extreme conditions, e.g., Li$_3$N/LiH catalysts in high-pressure cBN synthesis \cite{Du2007}, direct Li intercalation into hBN is thermodynamically forbidden due to its strongly endothermic nature and the inability of the insulating lattice to accommodate charge transfer \cite{Shirodkar2016, altintas2011intercalation}, and experimentally requires temperatures above $1500$~K, yielding metastable compounds that revert to insulating hBN upon Li removal \cite{Sumiyoshi2010}, even though stable lithiated hBN phases are predicted to host remarkable functionalities such as strong two-dimensional plasmons \cite{Despoja2018}. More recently, radical-mediated approaches have expanded the functionalization space accessible for hBN \cite{ikuno2007amine, nazarov2012functionalization, pakdel2014plasma, cui2014large, li2014strong, nemeth2017OzoneBF3, sainsbury2012oxygen, sainsbury2014dibromocarbene}. Owing to the electron-deficient character of boron, B sites in hBN can act as Lewis acids \cite{nemeth2018simultaneous, nemeth2021radical}, enabling covalent functionalization by strong Lewis bases~\cite{pal2007functionalization, cai2017facile, sheng2019polymer}. Reactivity is highest at edges and in exfoliated monolayers, where the energetic barrier to change sp$^2$ to sp$^3$ hybridization is lower than in bulk multilayers. These methods have led to the development of various functionalized boron nitride (FBN) platforms with enhanced dispersibility, tunable surface chemistry, and ion-transport functionality~\cite{nemeth2018simultaneous, nemeth2021radical, nemeth2020-US10693137B2, nemeth2022radicalAnion-US11453596B2}. In parallel, the renewed focus on Li-metal anodes, driven by their nearly tenfold higher theoretical capacity compared to graphite \cite{lee2023toward, asenbauer2020success}, has increased the demand for multifunctional separators and protective interfaces. Consequently, FBN-polymer composites have been explored for dendrite suppression and improved mechanical and thermal stability \cite{wang2024interfacial, ho2024boron, kim2024horizontal, shen2019chemically, zuo2023fast, heng2021stabilized, cheng2019stabilizing}, as well as in Li-S batteries to mitigate polysulfide migration \cite{fan2017functionalized, deng2018enhanced, mussa2021hexagonal}. 

However, scalable methods that enable high loading, precise chemical control, and the incorporation of intrinsic lithium functionality remain rare. Commonly employed synthesis routes suffer from low yields, lack of precise chemical control, extensive post-processing, and lithium-free FBN products that deplete active Li from the electrolyte or cathode, ultimately compromising cell capacity \cite{fan2017functionalized, wang2024interfacial}; a quantitative benchmarking of prior hBN functionalization and lithiation strategies, and of the resulting transport properties against established solid-state ion conductors, is provided in Tables~S5 and~S6 of the Supporting Information.

Here, we demonstrate a scalable, solvent- and byproduct-free mechanochemical route to lithium-oxalate-functionalized hBN via direct reaction of the reactants, achieving near-quantitative precursor conversion under optimized milling conditions. The resulting material uniquely combines redox inertness, room-temperature Li-ion conductivity, electronic insulation, and high thermal stability, while also hosting stable free radicals. This combination of properties, integrated within a single covalently bonded phase, points toward multifunctional solid-state platforms to integrate ionic transport, structural robustness, and interfacial functionality. Crucially, this approach produces an FBN with intrinsic Li content, overcoming the central limitation of conventional hBN functionalization strategies, with prospective uses in ion-conductive polymer composites, potentially dendrite-suppressive separators, and protective Li-metal interfaces in emerging battery architectures. Beyond energy storage, the presence of stable free radicals may render this material of interest for potential applications such as polymerization initiation, catalysis, combustion tracing, spin labeling in biochemical studies, and anti-counterfeiting technologies. Here, we focus on the solvent-free synthesis and comprehensive characterization of this intrinsically Li-containing FBN, thereby establishing a foundation for a potentially new class of functionalized boron nitride materials.

\section{Synthesis of Li$_2$(BN)$_6$C$_2$O$_4$}\label{sec:synth}

In a typical reaction, a $6{:}1$ molar ratio of hexagonal boron nitride to lithium oxalate was loaded into a stainless-steel high-energy ball mill, and the mixture was milled for $5$ hours, with $30$-minute cooling pauses after each hour to prevent overheating. For comparison, the same mixture was milled in a planetary ball mill using zirconia balls in a zirconia jar sealed under argon for a total of $92$ hours. These two milling setups differ substantially in milling geometry, impact energy, and the hardness of the milling media, resulting in different milling times as discussed in detail in the Methods section. The completion of the reaction was confirmed by the disappearance of the Li$_2$C$_2$O$_4$ peaks in the X-ray diffraction (XRD) pattern. In both cases, the resulting product was a brown powder (Fig. \ref{Fig1:synthesis}a), denoted as Li$_2$(BN)$_6$C$_2$O$_4$, or LBNCO, indicating that the coloration was intrinsic rather than caused by contamination from the milling media. Pure hBN was milled under otherwise identical conditions to provide a reference material, with milling times of both $5$ and $10$ h used for comparison (Fig.~S4). A more detailed description of the material synthesis can be found in the Materials and Methods section.

The progress of the reaction and its completion in $5$ hours was monitored using XRD as a function of milling time (see Figs.~S2, S3, and Table~S1 of the SI). XRD patterns of $11$ independently prepared batches showed no discernible variation in peak positions or relative intensities, demonstrating the reproducibility of the synthesis (Fig.~S1). Furthermore, pristine hBN milled under identical conditions showed sharp reflections of well-ordered hBN after $5$~h and disappeared only after $10$~h (Fig.~S4), indicating that the presence of the oxalate phase promotes the mechanochemical breakdown of the stacking order. The HRTEM images of the samples (Figs. \ref{Fig1:synthesis}a, S5, and S16) showed typical hBN grains smaller than $5$~nm and a lattice periodicity of $3.83$~{\AA} along the $c$-axis. All these interlayer distances exceed the $3.60$~{\AA} typically observed in exfoliated and restacked hBN \cite{dorn2020identifying}, or the $3.33$~{\AA} measured on the milled hBN, and fall within the predicted range for lithium-intercalated hBN $(3.67-3.99$~{\AA}) \cite{altintas2011intercalation}. For comparison, the interlayer spacing in LiC$_6$ graphite intercalation compounds is $3.70$~{\AA}~\cite{dolotko2014understanding}.

\begin{figure*}[!htp]
    \includegraphics[width=0.93\textwidth]{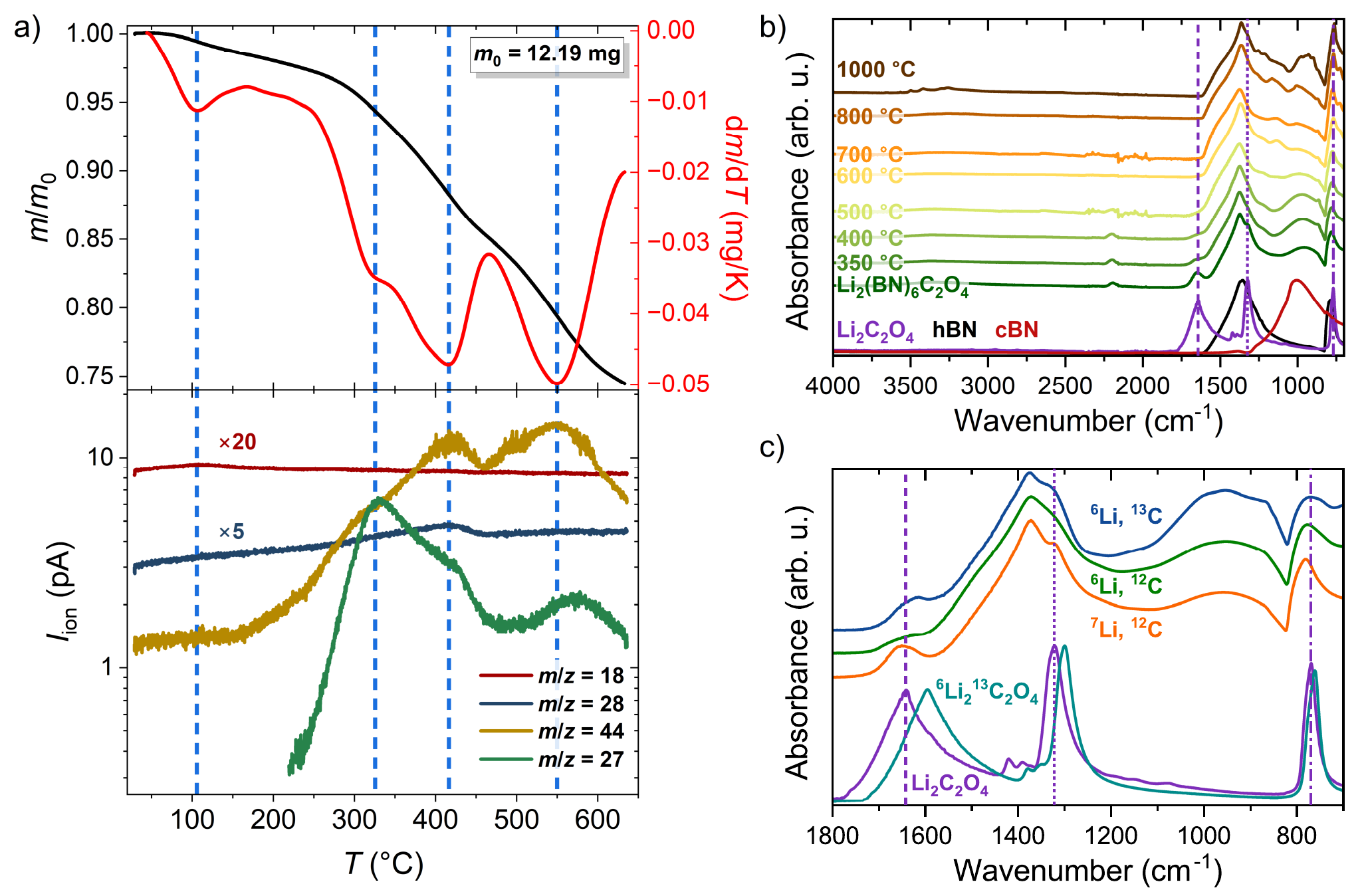}
    \caption{
        a) Thermal decomposition of LBNCO as monitored by TGA (upper panel) and MS (lower panel) with Ar as carrier gas. The mass is normalized to the initial sample mass, $m/m_0$. The right axis of the upper panel shows a numerical derivative, $\mathrm{d}m/\mathrm{d}T$, where negative peaks indicate mass-loss events, emphasized by vertical dashed lines for direct comparison with the MS ion current signals. The MS traces are identified as H$_{2}$O ($m/z = 18)$, N$_{2}$ ($m/z = 28$), HCN ($m/z = 27$) and CO$_{2}$ ($m/z = 44$). b) FTIR absorbance spectra of LBNCO and its calcined derivatives, together with hBN, cBN, and Li$_{2}$C$_{2}$O$_{4}$. The vertical lines refer to the peaks of the main bands of Li$_{2}$C$_{2}$O$_{4}$. c) Isotope effects in the FTIR spectra of LBNCO (upper panel) and Li$_{2}$C$_{2}$O$_{4}$ (lower panel). The vertical lines refer to the peaks of the main bands of Li$_{2}$C$_{2}$O$_{4}$ with natural abundance isotopes.}
    \label{Fig2:TGA-MS-IR}
\end{figure*}

Figure~\ref{Fig1:synthesis}b depicts possible attachments of lithium oxalate to the zigzag and armchair edges of the hBN crystallite, a mechanism further supported by spectroscopic evidence and DFT-based modeling discussed in the following sections.

\section{Thermal Stability and Electronic Structure}\label{sec:xps-tgams}

The thermal stability of the synthesized LBNCO is of particular importance for its potential application as an ionic conductor and as a separator material in Li batteries, where operation at elevated temperatures is common. Moreover, investigating its thermal stability provides valuable insight into the underlying reaction mechanisms and transformations that govern its performance. In the following, the reaction between hBN and Li$_2$C$_2$O$_4$ is studied by a broad range of techniques, and at the same time, the signatures of the degradation of the product by increasing temperature are also followed.

The thermal stability of LBNCO was examined by calcination under flowing nitrogen at $350-1000~^{\circ}$C for $8$ hours (Fig.~\ref{Fig1:synthesis}c). The as-prepared LBNCO darkened slightly up to $400~^{\circ}$C and turned pastel white above $800~^{\circ}$C. Corresponding XRD patterns (Fig.~\ref{Fig1:synthesis}d) showed no significant changes up to $400~^{\circ}$C. The material remained largely amorphous up to $600~^{\circ}$C, displaying broad features of turbostratic BN \cite{gross2019unravelling}. From $600~^{\circ}$C and higher temperatures, crystalline decomposition products including LiBO$_2$, Li$_3$BO$_3$, Li$_6$B$_4$O$_9$, and crystalline hBN emerged. No peaks of unreacted Li$_2$C$_2$O$_4$ or its typical decomposition product, Li$_2$CO$_3$, were detected, even though Li$_2$CO$_3$ decomposes at $410~^{\circ}$C \cite{dollimore1971thermal}. Together with the absence of measurable mass loss during milling, these results confirm that the mechanochemical reaction between hBN and Li$_2$C$_2$O$_4$ was complete under the applied conditions, and LBNCO appears stable up to $400~^{\circ}$C.

The thermal stability of LBNCO was followed by thermogravimetric analysis (TGA), mass spectrometry (MS), differential scanning calorimetry (DSC) (Figs.~\ref{Fig2:TGA-MS-IR}a, S14, and S15), and FTIR spectroscopy as well. Samples showed gradual mass loss starting just below $100~^{\circ}$C, which accelerated above $300~^{\circ}$C. MS detected products at $m/z = 18$ (H$_2$O), $28$ (N$_2$ and CO), and $44$ (CO$_2$). The $m/z = 18$ signal appeared near $100~^{\circ}$C and persisted, reflecting continuous desorption of physisorbed or weakly chemisorbed water rather than decomposition of organics. The $m/z = 44$ signal (CO$_2$) emerged at ${\sim}300~^{\circ}$C and peaked above $400~^{\circ}$C, consistent with decomposition of the organic backbone. The $m/z = 28$ signal, accompanied by $m/z = 14$, indicates primarily desorbed N$_2$, with minor CO formation supported by a weak $m/z = 16$ signal. A transient HCN peak ($m/z = 27$) near $300~^{\circ}$C suggests intermediate cyanide formation during decomposition. These gas-evolution processes correlate with the mass loss observed in the TGA curve (Fig.~\ref{Fig2:TGA-MS-IR}a upper panel). It should be noted that the gas profiles excluded the signatures of oxalic acid, Li$_2$C$_2$O$_4$, and Li$_2$CO$_3$, which decompose at higher temperatures or generate CO in amounts not observed here. Likewise, decomposition pathways of LiBOB and LiBOB$\cdot$H$_2$O are inconsistent with the data, as both produce substantial CO.

Overall, TGA-MS analysis confirmed structural stability up to ${\sim}350~^{\circ}$C and minor degradation between $350$ and $420~^{\circ}$C. In this range, the most pronounced MS signal corresponds to increased desorption of adsorbed water, accompanied by partial degradation of the organic structure, consistent with XRD data. LBNCO undergoes further decomposition in two distinct steps, at approximately $420~^{\circ}$C and $550~^{\circ}$C, each associated with significant mass loss and CO$_2$ evolution.

The Raman spectra of LBNCO and its annealed derivatives (Fig.~S9) were dominated by a strong luminescent background (Fig.~S18), which partially obscured the vibrational features. The characteristic G-band of hBN at $1367$~cm$^{-1}$, corresponding to the $E_{2\text{g}}$ in-plane B$-$N stretching mode, was significantly broadened after the reaction and exhibited only marginal changes upon annealing at the selected temperatures.

Attenuated total reflection Fourier-transformed infrared (ATR-FTIR) absorption spectra of LBNCO, its calcined derivatives, and reference compounds (hBN, cBN, and Li$_2$C$_2$O$_4$) are shown in Fig.~\ref{Fig2:TGA-MS-IR}b. A comparison between the hBN+lithium oxalate physical mixture and LBNCO is presented in Fig.~S19. Pristine hBN displays characteristic vibrational bands at $1356$~cm$^{-1}$ (in-plane B$-$N stretching, $E_{1\text{u}}$(TO)) and $786$~cm$^{-1}$ (out-of-plane bending, $A_{2\text{u}}$(TO)) \cite{PhysRev.146.543, gil2020boron}. Li$_2$C$_2$O$_4$ exhibits three principal bands: $1643$ and $1328$~cm$^{-1}$ (asymmetric O$-$C$-$O stretching) and $772$~cm$^{-1}$ (O$-$C$-$O bending) \cite{shippey1980vibrational}. 

The FTIR spectrum of the as-prepared LBNCO showed weak residual oxalate signatures, undetectable by XRD, which diminished upon calcination at $350-400~^{\circ}$C, indicating near-complete consumption of lithium oxalate. Alternatively, the residual oxalate signature may also be due to oxalate ions bound to hBN only at one terminal CO$_{2}^{-}$ while the other CO$_{2}^{-}$ is still unbound. A broad band emerged between $1260$ and $830$~cm$^{-1}$, centered at $950$~cm$^{-1}$. This peak is often associated with OH-functionalized hBN \cite{sainsbury2012oxygen} and corresponds to the transitions between trigonal and tetragonal B$-$O/N stretching vibration. Trigonal (sp$^2$-like) modes appear at higher wavenumbers, whereas tetragonal (sp$^3$-like) B coordination gives rise to lower-frequency modes, very similar to the cBN vibration~\cite{morsy2025structural}. In lithium borate glasses, this region splits into sub-bands depending on the presence of non-bridging oxygens (NBOs), where higher-frequency bands ($>1200$~cm$^{-1}$) reflect the presence of NBOs~\cite{morsy2025structural}. The gradual attenuation of the broad band upon calcination suggests thermal decomposition of the oxalate ligands and loss of sp$^3$-hybridized boron, analogous to LiBOB decomposition \cite{melin2024elucidating}. At temperatures $\geq 600~^{\circ}$C, the FTIR signal stabilized, consistent with the crystallization of borate phases detected by XRD. A new band appeared at ${\sim}930$~cm$^{-1}$ only after calcination at $1000~^{\circ}$C, which is consistent with BO$_4$ vibrational modes in distorted trigonal BO$_3$ units of lithium borate glasses \cite{morsy2025structural}. Broad features in the $3000-3500$~cm$^{-1}$ region corresponded to N$-$H and O$-$H stretching modes, including hydrogen-bonded edge terminations and adsorbed water, respectively.

A minor band appeared at ${\sim}2200$~cm$^{-1}$ selectively in samples calcined at $\leq 400~^{\circ}$C. While it may have originated from atmospheric CO$_2$ or diamond ATR tip artifacts, it also overlaps with known signatures of C$\equiv$N or $-$NCO groups in BCN-type materials \cite{chen2018carbon, goyal2016single}. Alternatively, it may reflect the formation of a carbamate intermediate at armchair hBN edges \cite{sainsbury2012oxygen}, as illustrated in Fig.~\ref{Fig1:synthesis}b, and this assignment appears very likely in light of the XPS results (see below), which revealed oxidation and the emergence of multiple-bond species only at elevated temperatures. The subtle changes observed in the TGA-MS spectra may also be associated with the dissociation of the proposed intermediate structure.

The dominant FTIR band in the samples is centered at ${\sim}1400$~cm$^{-1}$, described as a convolution of the hBN $E_{1\text{u}}$ stretching mode and the asymmetric O$-$C$-$O stretching modes of Li$_2$C$_2$O$_4$. Isotopic substitution was applied to unambiguously trace the origin of the newly detected vibrational modes. Substitution with $^{13}$C and $^{6}$Li produces clear shifts in this region (Fig.~\ref{Fig2:TGA-MS-IR}c), confirming that the new modes arise from chemical incorporation of oxalate species. Notably, these bands are absent in pure Li$_2$C$_2$O$_4$, indicating covalent binding to the hBN lattice. The spectral features closely resemble those of oxalate-chelated boron in LiBOB \cite{shippey1980vibrational, zhuang2004study, melin2024elucidating, zor2021guide}, which confirms the formation of chelate complexes between oxalate and edge B sites in hBN.

Overall, the ATR-FTIR spectra signaled only minor changes below $600~^{\circ}$C. At $600~^{\circ}$C, the broad band near $950$~cm$^{-1}$ decreased markedly, and the shoulder of the ${\sim}1400$~cm$^{-1}$ peak appeared at higher wavenumbers, while the maximum started to shift back to the position of the pristine hBN $E_{1\text{u}}$ mode.

\begin{figure}[!htp]
    \includegraphics[width=0.9\columnwidth]{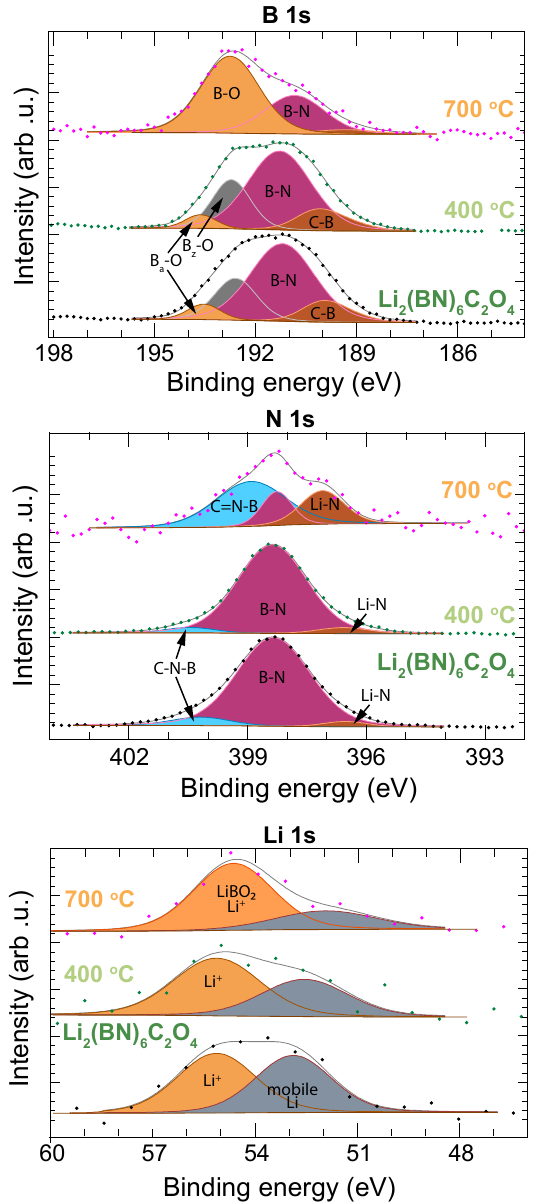}
    \caption{
        The B, N, and Li 1s XPS spectra and their deconvolution of LBNCO and its $400~^{\circ}$C and $700~^{\circ}$C calcined derivatives.}
    \label{Fig3:XPS}
\end{figure}

X-ray Photoelectron Spectroscopy (XPS) spectra of LBNCO and its calcined derivatives at $400~^{\circ}$C and $700~^{\circ}$C are presented in Fig. \ref{Fig3:XPS} for B 1s, N 1s, Li 1s, and in Fig.~S8 for C 1s and O 1s core levels. Peak assignments were guided by standard reference databases and literature comparisons.

\textbf{B 1s:} The B 1s spectrum featured a dominant peak at $191.3$ eV (B$-$N bonds) and two distinct B$-$O environments: one at $192.6$ eV (B$_z-$O, zigzag edge) and another at $193.5$ eV (B$_a-$O, armchair edge). The assignments are supported by the relative intensity ratio of the B$-$O peaks and correlated with $^{13}$C NMR results, which suggest more oxalate binding to zigzag edges (see below). A low-energy shoulder at $190.1$ eV was attributed to C$-$N$-$B units, which have previously been reported in carbon-doped hBN \cite{chen2018carbon}. Upon calcination to $700~^{\circ}$C, the B$-$N signal diminished while B$-$O became dominant, consistent with the formation of LiBO$_2$ and Li$_3$BO$_3$ (as confirmed by XRD). The C$-$N$-$B related peak also decreased in intensity and shifted by ${\sim}0.5$ eV to lower energy, consistent with electronic redistribution following CO$_2$ elimination.

\textbf{N 1s:} The primary peak in the N 1s spectrum appeared at $398.3$ eV (B$-$N bonding), with a secondary peak at $400.2$ eV attributed to C$-$N$-$B. Upon thermal treatment, this peak shifted to ${\sim}399.2$ eV, reflecting conversion to N$=$C$-$B species. The overall B$-$N intensity decreased with calcination. A third peak at $396.5$ eV was tentatively assigned to Li$-$N environments, which may arise from Li adsorption or intercalation in proximity to nitrogen sites. Although no Li$_3$N is present based on XRD, this feature becomes more prominent in the $700~^{\circ}$C sample, potentially due to lithium ion migration into the hBN lattice as CO$_2$ departs, leaving behind electron-rich sites.

\textbf{Li 1s:} The Li 1s spectrum displayed two peaks: one at $55.1$ eV, characteristic of typical lithium salts (e.g., Li$_2$CO$_3$, LiOH, LiBO$_2$), and another at $52.9$ eV, commonly associated with mobile species: metallic \cite{wood2018xps, hensley1994xps}, or hBN-intercalated Li atoms, as predicted by DFT studies, where Li exhibits a small partial positive charge $({\sim}+0.2)$ \cite{altintas2011intercalation}. Calcination at $700~^{\circ}$C resulted in the disappearance of this low-energy band, coinciding with the formation of LiBO$_2$.

\textbf{C 1s:} The recorded C 1s spectrum of LBNCO was dominated by C$-$O and C$=$O bonds in the $288-292$ eV region, consistent with oxalate functionalities. Peaks at $287.3$ eV and $285.7$ eV are assigned to N$=$C$-$O and N$-$C$-$O groups, respectively. A feature at $284.8$ eV corresponds to the C$-$C bond within the oxalate ion, while the $283.7$ eV peak is attributed to C$-$N$-$B linkages. Calcination at $400~^{\circ}$C induced only minor changes, whereas treatment at $700~^{\circ}$C led to a substantial increase in N$=$C$-$O intensity and a corresponding decrease in N$-$C$-$O, C$-$C, and C$-$N$-$B signals. These changes likely reflect thermal decomposition of oxalate groups, particularly the elimination of CO$_2$. The slight shift of the peaks to lower binding energies can be interpreted as a result of electron absorption in the conjugated $\pi$-system of the hBN lattice, following the loss of neutral CO$_2$ from negatively charged oxalate anions, where the oxalate may also be present in its split form of $\bullet$CO$_2^-$ radical anions (see the ESR analysis below). This charge redistribution facilitates planarization of N$=$C$-$O and C$-$N$-$B motifs, integrating them into the $\pi$-conjugated system, thus producing a reductive effect on the local electronic structure.

\textbf{O 1s:} The O 1s region exhibited overlapping contributions from C$=$O, C$-$O, and B$-$O bonds. A shoulder at $529.0$ eV is assigned to N$-$C$-$O groups, consistent with the carbamate structure proposed for oxalate binding at armchair edges (see Fig. \ref{Fig1:synthesis}b). This peak increased upon calcination but did not shift, suggesting that the electronic structure of the N$=$C$-$O unit remains stable. A new peak emerged at $531.9$ eV in the $700~^{\circ}$C product and is attributed to LiBO$_2$ \cite{hensley1994xps}.

High-resolution XPS spectra of various core levels also indicated reasonable stability up to $400~^{\circ}$C, followed by significant decomposition at higher temperatures.

\section{Solid-State NMR Characterization}\label{sec:NMR}

Solid-state NMR spectroscopy provides detailed insight into the local structure and bonding environments in LBNCO and its calcined derivatives, which contain NMR-active nuclei including $^{11}$B, $^{13}$C, $^{14}$N, and $^{7}$Li. High-power decoupled magic angle spinning NMR (HPDEC-MAS) spectra and $^{7}$Li $T_1$ values were acquired at room temperature. The technique has proven to provide crucial information on related, relevant materials \cite{insinna2023graphite, pecher2017materials, zhang2016experimental, dorn2020identifying, jeschke1998high, jeschke1998comprehensive, gervais200511b, jakobsen2007long, Ricco2009PRL, Gadermaier2025SSI}.

The dataset underlying this section is extensive, comprising four nuclei, several reference compounds, relaxation measurements, and the corresponding DFT-calculated parameters. Rather than reproducing the full analysis here, we present the conclusions in a condensed form and refer the reader to the Supporting Information, where the complete calculated and experimental NMR spectrum parameters are collected (Tables~S2 and S3); the bonding models derived from these data are summarized schematically in Fig.~\ref{Fig1:synthesis}b.

\begin{figure*}[htp]
    \includegraphics[width=\textwidth]{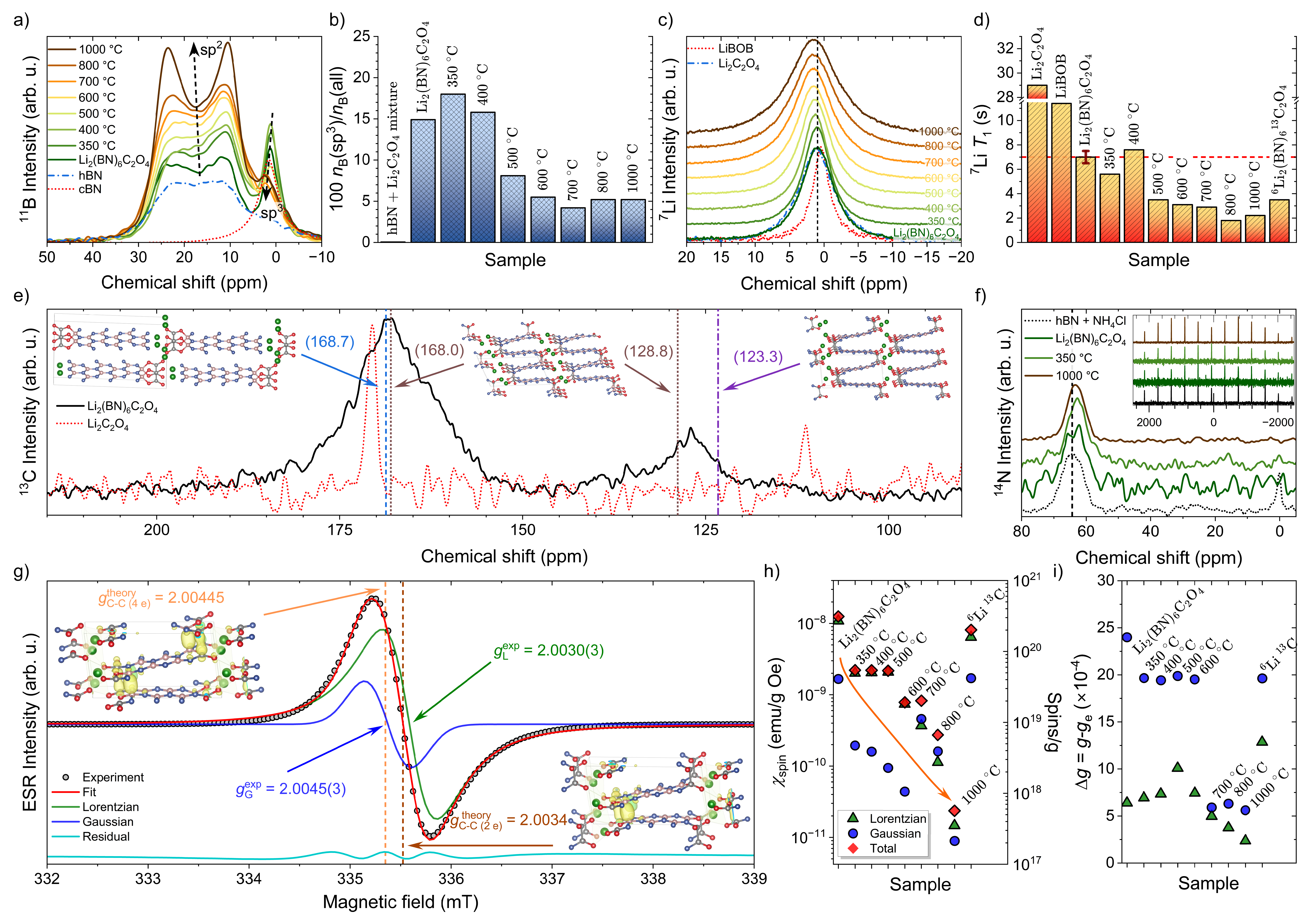}
    \caption{
        NMR (a-f) and ESR (g-i) spectra of LBNCO and its calcined derivatives. The intensities are scaled for a better visualization of relative changes. a) $^{11}$B together with hBN and cBN as reference materials. b) Ratio of sp$^3$ boron compared to the total number of B atoms. c) $^{7}$Li shown together with Li$_{2}$C$_{2}$O$_{4}$ and LiBOB for reference. d) $^{7}$Li spin-lattice, $T_1$, relaxation times obtained using the inversion-recovery pulse sequence. Please note the $4$-fold shortening of the relaxation time in the functionalized materials compared to lithium oxalate. Calcination further decreases the relaxation time. LiBOB is also given as a reference. e) $^{13}$C with $^{13}$C-enriched LBNCO and Li$_{2}$C$_{2}$O$_{4}$ for reference (natural abundance; Savitzky--Golay filtered). The structures, as well as the chemical shifts obtained from DFT calculations, are also depicted. f) $^{14}$N, a $95$ w\% hBN (sp$^2$) mixed with $5$ w\% NH$_{4}$Cl sp$^3$ used for reference. The inset shows a full rotational spectrum, noting that no further nitrogen configurations are present. g) Measured derivative X-band cw ESR signal and its deconvolution to a Lorentzian and a Gaussian component of the as-prepared LBNCO. Graphics show the corresponding structures and electron densities. h) Mass spin-susceptibility and spin-counts. i) $g$-factors (relative to the free electron $g$-factor, $g_{\text{e}}$) associated with the Lorentzian and Gaussian ESR signals.}
    \label{Fig4:NMR-ESR}
\end{figure*}

\textbf{$^{11}$B NMR:} The $^{11}$B MAS spectra (Fig.~\ref{Fig4:NMR-ESR}a) showed contributions from both sp$^2$- and sp$^3$-hybridized boron in the as-synthesized material, confirming the proposed edge functionalization. The sp$^3$ boron fraction (${\sim}15-18\%$) matched the initial molar ratio of Li$_2$C$_2$O$_4$ to hBN, consistent with the conversion of approximately every sixth boron atom from sp$^2$ to sp$^3$ hybridization. This conversion may occur via chelation of edge boron atoms by oxalate ligands, leading to a BOB-like binding motif. Alternatively, it may result from the binding of $\bullet$CO$_2^-$ radical anions through the formation of B$-$C bonds, in which case the radical anion is incorporated at the armchair edge of hBN. These mechanisms are illustrated in Fig.~\ref{Fig1:synthesis}b and are further supported by DFT-based modeling of the NMR and ESR spectra, as discussed later. Upon calcination, the sp$^3$ boron content decreased, with residual tetrahedral environments (${\sim}5\%$) most likely associated with disordered B$-$O/N species or lithium borate phases, as shown in Fig.~\ref{Fig4:NMR-ESR}b and discussed previously.

\textbf{$^{7}$Li NMR:} The $^{7}$Li spectra (Fig.~\ref{Fig4:NMR-ESR}c) revealed Li environments similar to those in pristine Li$_2$C$_2$O$_4$, indicating that Li$^+$ remained in oxalate-like coordination even after functionalization. It also indicates that there is only one Li-containing phase instead of separate Li$_2$C$_2$O$_4$ and other phases, which supports the completeness of the reaction. Minor broadening upon calcination suggests increased structural disorder and the emergence of secondary Li$-$B$-$O phases, corroborated by XPS. The shortening of the $T_1$ relaxation time upon calcination suggests that, although the local coordination environment around the $^{7}$Li nuclei did not change significantly during calcination, the electron density in their vicinity increased, thereby enhancing nuclear spin perturbation (Fig.~\ref{Fig4:NMR-ESR}d).

\textbf{$^{13}$C NMR:} The $^{13}$C spectrum of the $^{13}$C-enriched material showed two distinct resonances at $168.4$ and $127.0$~ppm (Fig.~\ref{Fig4:NMR-ESR}e), attributed to B-chelated oxalate (BOB-type) units at zigzag and cyclic carbamate groups at armchair edges, respectively. The ${\sim}4{:}1$ integrated area ratio indicates coexisting edge terminations with an approximate $3{:}2$ molar distribution of these configurations.

\textbf{$^{14}$N NMR:} The $^{14}$N spectra (Fig.~\ref{Fig4:NMR-ESR}f) were dominated by the characteristic signal of sp$^2$-hybridized nitrogen in hBN, with no evidence of sp$^3$ nitrogen or nitridic defects. The absence of any observable deviation from the hBN $^{14}$N signal is unexpected in view of the C$-$N bonds clearly identified by XPS and $^{13}$C NMR. 
However, according to DFT calculations (Tables S2 and S3), the quadrupole coupling is extremely large for the edge nitrogen atoms (compared to the in-plane ones), which could make these atoms challenging to detect in NMR, especially when the respective chemical shifts also vary considerably due to structural disorder in the edges.

\textbf{Density functional theory (DFT)} simulations were conducted to identify structural models of oxalate-functionalized hBN consistent with the experimental NMR data. Out of numerous configurations explored, only two models -- corresponding to functionalization at the zigzag and armchair edges -- reproduced the experimental $^{13}$C chemical shifts within $2$ ppm. The zigzag model predicts a single $^{13}$C resonance at $168.7$ ppm. Within the armchair model, two different structures are identified: armchair(B$-$C) with an interlayer B(sp$^3$)$-$C(sp$^3$) bond and armchair(C$-$C) with C(sp$^3$)$-$C(sp$^3$) (see Fig. \ref{Fig1:synthesis}b). The armchair(B$-$C) structure contains one C(sp$^3$) and one C(sp$^2$) per formula unit with resonances at $168.0$ and $128.8$ ppm. The armchair(C$-$C) structure has two C(sp$^3$)s with resonances at $123.3$ and $122.0$ ppm. The resonances of the zigzag and armchair(B$-$C) structures closely match the experimental values of $168.4$ and $127.0$ ppm, respectively, as shown in Fig. \ref{Fig4:NMR-ESR}e. The signal from the armchair(C$-$C) structure is a less good match but still fits in the broad peak at $127.0$ ppm. This structure does not contribute B(sp$^{3}$) atoms; therefore, it is less likely to occur.

Both zigzag and armchair(B$-$C) models predict a $1{:}1$ molar ratio of B(sp$^3$) centers to oxalate units, in excellent agreement with quantitative $^{11}$B NMR analysis and other spectroscopic data. Calculated NMR parameters are provided in Tables S2 and S3 of the Supporting Information, respectively.  

Energetically, the armchair(C$-$C) model is the most stable. The armchair(B$-$C) and zigzag structures are $0.24$ and $1.69$ eV higher in total energy per formula unit, respectively. The high energy of the zigzag structure likely stems from van der Waals repulsion between adjacent chelate rings in the zigzag geometry. The area ratio of the two experimental $^{13}$C peaks (${\sim}4{:}1$) further supports the \emph{coexistence of multiple edge configurations}, with the more intense resonance corresponding to the zigzag model.

In the zigzag configuration, each oxalate ion forms two B$-$O bonds with neighboring edge B atoms, despite their close proximity (${\sim}2.56$ \AA), highlighting the high chemical reactivity of hBN edge sites. The edge terminations alternate between B and N atoms, reflecting the intrinsic lattice symmetry of hBN.

In contrast, the armchair model undergoes a complex relaxation process involving the transformation of oxalate into embedded radical species. It is assumed that the oxalate ions participate in a cycloaddition with the armchair edge of hBN. The C$=$O unit of the oxalate ion provides the double bond, which reacts with the dangling bonds in the armchair edge of hBN, forming a new B$-$O and N$-$C bond and converting the C$=$O bond to C$-$O. This reaction is also supported by the Lewis acid-base reactivity of the components. The cycloaddition happens at both $\bullet$CO$_2^-$ ends of the oxalate ion and involves two neighboring hBN layers. As a result, the hybridization of both carbons changes from sp$^2$ to sp$^3$. Upon relaxation, the oxalate C$-$C bond may cleave to form two $\bullet$CO$_2^-$ radical anions embedded in the armchair edge. This intermediate further reacts with a neighboring hBN layer, establishing an interlayer B(sp$^3$)$-$C(sp$^3$) bond (Fig. \ref{Fig1:synthesis}b). It may also recombine to an interlayer C(sp$^3$)$-$C(sp$^3$) bond. In some cases, these reactions are not possible due to geometric mismatch, leading to the preservation of the embedded $\bullet$CO$_2^-$ radical anions.

Because the embedded $\bullet$CO$_2^-$ moiety contributes only one electron to the B(sp$^2$) atoms (while the sp$^2$ boron of hBN could take up two additional electrons), the formation of a complete two-electron B(sp$^3$)$-$C(sp$^3$) bond requires an additional electron, which is acquired from a nearby embedded $\bullet$CO$_2^-$ radical center. This internal redox reaction leads to disproportionation: one carbon is reduced to C(sp$^3$) and the other oxidized to C(sp$^2$), in case a B(sp$^3$)$-$C(sp$^3$) bond forms. The resulting C(sp$^3$)$-$O$^{-}$ bond ($1.35$ \AA) is significantly longer than the carbonyl C(sp$^2$)$=$O bond ($1.28$ \AA), reflecting this charge redistribution (the oxygens refer to the oxygens pointing outside the carbamate ring, the oxygen in C(sp$^3$)$-$O$^{-}$ carries a negative charge while the carbonyl C(sp$^2$)$=$O is not charged). The long B(sp$^3$)$-$C(sp$^3$) bond ($1.70$ \AA) suggests an energetic cost to interlayer crosslinking, but the calculated $^{13}$C chemical shifts confirm the structural plausibility.

Notably, the armchair(B$-$C) model contains both C(sp$^3$)$-$N(sp$^2$) and C(sp$^2$)$-$N(sp$^2$) bonds, with calculated lengths of $1.49$ and $1.36$ \AA, respectively. The armchair(C$-$C) model contains only C(sp$^3$)$-$N(sp$^2$) bonds. Upon calcination, XPS data show preferential loss of C(sp$^3$)$-$N(sp$^2$) bonding, consistent with CO$_2$ evolution and the retention of thermodynamically more stable C(sp$^2$)$-$N(sp$^2$) species. Concurrently, the number of C$-$B bonds decreases, further supporting this mechanism.

The above models also align with $^{14}$N NMR data, which show no evidence of N(sp$^3$) environments (Fig. \ref{Fig4:NMR-ESR}f and its inset). They localize Li$^+$ ions near negatively charged oxygen and nitrogen atoms at edge sites, consistent with the observed $^{7}$Li NMR shifts and XPS spectra (Figs. \ref{Fig4:NMR-ESR}c and \ref{Fig2:TGA-MS-IR}d, respectively). The zigzag model, in particular, replicates the N 1s feature associated with Li$-$N interactions. Analysis of both models reveals the possibility of fractional Li occupation at additional nearby interstitial sites, which could account for the Li 1s XPS component attributed to mobile Li.

Finally, although the embedded $\bullet$CO$_2^-$ radical is localized at the armchair edge, the B(sp$^2$) acceptor involved in interlayer bonding may originate from any site on the opposing layer. As a result, some B(sp$^3$)$-$C(sp$^3$) bonds may form away from the edge, placing Li$^+$ ions in environments resembling intercalated states. This provides a structural explanation for the Li 1s binding energy observed at $52.9$ eV, characteristic of low-valent, intercalated lithium.

\section{Dangling Bonds and Free Radicals -- Electron Spin Resonance characterization}

The electron paramagnetic resonance (ESR or EPR) spectra of LBNCO and its calcined derivatives were analyzed to characterize unpaired electron species, as presented in Figs. \ref{Fig4:NMR-ESR}g-i and S10. In the as-prepared sample milled in a stainless-steel jar, a broad low-field signal ($<200$ mT) was present, attributed to contamination from steel abrasion. This signal was absent in material synthesized using zirconia milling media, confirming its extrinsic origin. At higher fields (centered at $335.4$ mT), a sharp ESR signal was consistently observed and is the focus of subsequent analysis.

Spin quantification indicates a significant unpaired electron population of ${\sim}0.08$ spins per Li$_2$(BN)$_6$C$_2$O$_4$ formula unit (Fig. \ref{Fig4:NMR-ESR}h), stable over at least six months under dry storage conditions. The high number of spins, even after substantial shelf storage, attests to a stable free radical nature and opens the possibility for the material to initiate radical polymerization (Fig.~S13) and to be used as a spin label in composites (Fig.~S11).

The ESR signal can be deconvolved into two overlapping components with distinct lineshapes: a dominant Lorentzian and a weaker Gaussian (Fig. \ref{Fig4:NMR-ESR}a). Isotope enrichment with $^{13}$C and $^{6}$Li significantly broadened both components (Fig.~S10), which suggests that both types of unpaired electrons are associated with carbon atoms that are relatively mobile. This broadening is consistent with a mechanism involving fragmentation of the interlayer B(sp$^3$)$-$C(sp$^3$) and C(sp$^3$)$-$C(sp$^3$) bonds of the armchair edge (whose presence was indicated by XPS (Fig. \ref{Fig3:XPS}a) and $^{13}$C NMR (Fig. \ref{Fig4:NMR-ESR}e)). These bonds become much longer than normal due to geometric mismatch. The resulting dangling bonds act as radicals and carry spins as they cannot recombine in $8\%$ of the formula units. 

The established theoretical model also accounts for ESR observations: ${\sim}8\%$ of formula units contain unpaired spins, likely originating from edge radicals that remain unreacted due to geometric frustration, preventing bond formation with nearby B(sp$^2$) or C(sp$^2$) atoms. These rather localized spins are consistent with the Lorentzian-dominated ESR spectra. The geometric frustration relaxes upon annealing at $350~^{\circ}$C, resulting in a decrease of the spin-count by an order of magnitude.

DFT-based modeling of the ESR spectra associates the more intense Lorentzian component with the scenario when only one of the two interlayer B(sp$^3$)$-$C(sp$^3$) or C(sp$^3$)$-$C(sp$^3$) bonds per unit cell is fragmented. The fragmentation of both bonds is associated with the less intense Gaussian line. This is in line with the observed one-order-smaller intensity for this component, as breaking two bonds is less probable than breaking only one. The fewer interlayer bonds are present, the farther the structure departs from the accordion-fold and the more rigid it becomes. This is the reason why the Lorentzian linewidths are broader than the Gaussian ones.
 
Despite the $^{13}$C enrichment, no clear hyperfine splitting was detected, likely due to significant inhomogeneous broadening that masks hyperfine interactions from the $I=1/2$ $^{13}$C nuclei (Fig.~S10). Additionally, $^{13}$C MAS NMR spectra did not reveal signals characteristic of carbon-centered radicals, suggesting the spin density is not localized only on carbon itself but may be delocalized onto adjacent atoms to some extent. This is also consistent with DFT results.

This view is supported by the measured $g$-factors, which were consistently greater than the free-electron value ($g_{\text{e}} = 2.0023$), indicating the presence of hole-like states (Fig. \ref{Fig4:NMR-ESR}i). The larger $g$-values associated with the Gaussian component indicate a larger spatial separation of the two sides of the fragmented bonds and a stronger (less recombined) radical anion character.

The ESR calculations found model structures that recovered the experimental $g$-factors ($2.0030$ and $2.0045$) with good accuracy using the armchair models. The best match was achieved when the interlayer bond was C(sp$^3$)$-$C(sp$^3$) ($2.0034$ and $2.00445$) as visualized in Fig. \ref{Fig4:NMR-ESR}. A less accurate match was achieved for B(sp$^3$)$-$C(sp$^3$) bonds ($2.0036$ and $2.0051$, Fig. S20). This suggests that the spins may predominantly come from frustrated C(sp$^3$)$-$C(sp$^3$) bonds. A complete list of the calculated energies, interlayer bond lengths, and $g$-factors at the relaxed geometries can be found in Table S4 of the Supporting Information.

\section{Ionic Electrical Transport}

\begin{figure}[htp]
    \includegraphics[width=\columnwidth]{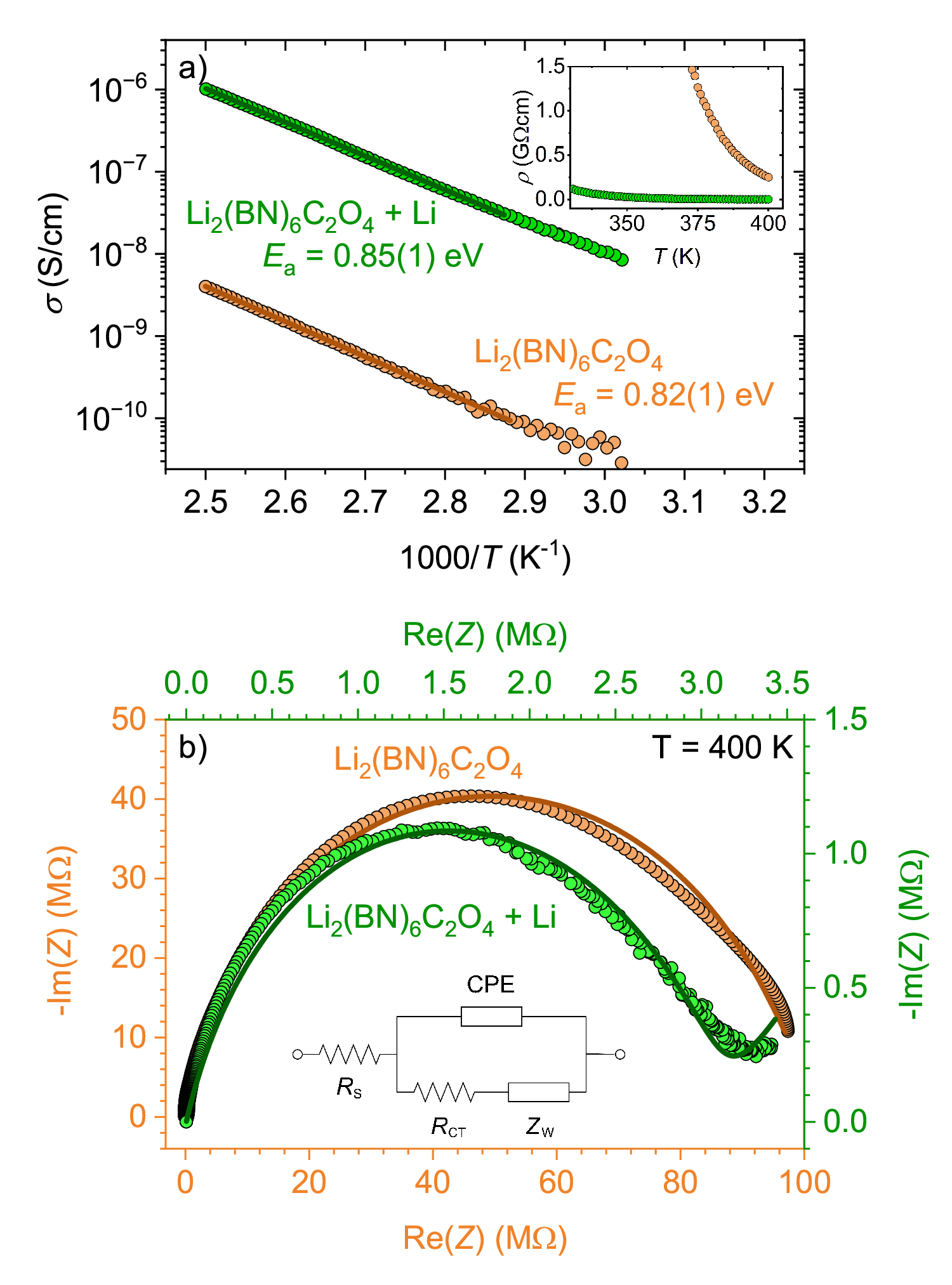}
    \caption{
     a) Arrhenius plot of the electrical transport measurement showing activated temperature dependence consistent with fixed-range hopping. Lithium doping increases the conductivity of the material by nearly three orders of magnitude. The inset presents the same data as resistivity versus temperature. b) Nyquist plots of the impedance spectra of the two materials shown in a, at the temperature of $400$ K. Note the different scales of the color-coded axes. The low-frequency range of the lithium-doped sample shows a clear diffusive character. The inset shows the equivalent Randles circuit used for fitting.}
    \label{Fig5:sigma} 
\end{figure}

The electrical transport measurements on LBNCO indicate predominantly ionic conduction, which we assign primarily to thermally activated Li$^+$ transport. Because the oxalate groups are covalently bound to hBN edges and defect sites, Li$^+$ is the only evident mobile ionic species in the present material. A direct Li$^+$ transference-number measurement, however, will be required to establish quantitatively whether LBNCO behaves as a true single-ion conductor. Such behavior would be attractive for lithium-metal batteries, where high Li$^+$ transference numbers can help reduce concentration polarization and dendrite formation \cite{deng2021flame}. In LBNCO-based polymer composite electrolytes, this type of transport should persist even after introducing common lithium salts (e.g., LiPF$_6$, Li-TFSI, LiBF$_4$, etc.), since sp$^2$-coordinated boron sites on exposed LBNCO surfaces can bind and immobilize their anions through Lewis acid–base interactions, analogous to exfoliated hBN \cite{ho2024boron}. Optical absorption measurements on LBNCO and its heat-treated derivatives revealed a wide band gap of $5.82-6.06$~eV, characteristic of an insulating material and consistent with the ${\sim}6$~eV indirect band gap of hBN. Indeed, in strongly disordered systems, electronic conduction typically proceeds via variable‐range hopping (VRH) \cite{Hill1976PSSA, Tsigankov2002PRL, Yu2004PRL}. However, for heavy ions in most solids, the hopping rate decreases rapidly with distance, since a long‐range ion jump requires coordinated motion of many lattice atoms and the crossing of multiple energy barriers. Consequently, the \emph{long‐distance/small‐energy‐mismatch} mechanism is not compatible -- the prefactor for such jumps is essentially negligible. As a result, the optimum hop length is almost always close to the nearest‐neighbor distance, corresponding to fixed‐range hopping \cite{Adkins1998JPCM}. In this regime, conductivity is governed by jumps over a single, well‐defined barrier, leading to  
\begin{equation}
    \sigma(T) \propto \exp{\left(-E_{\text{a}}/k_{\text{B}}T\right)},
\end{equation}
as shown in Fig.~\ref{Fig5:sigma}a.

To further probe lithium transport, $2.0$ g of the as-synthesized LBNCO was doped with $200$ mg Li using the same ball-milling process. As shown in Fig.~\ref{Fig5:sigma}, the doped samples exhibited the same temperature dependence but with higher conductivity. This is reinforced by Fig.~\ref{Fig5:sigma}b, which shows Nyquist plots of the impedance spectra for the \mbox{LBNCO} material and its lithium-doped counterpart. In the low-frequency regime, the doped sample shows clear signs of charge diffusion. The absence of this regime in the LBNCO data might be a result of frequency limitations. The straightforward mechanochemical doping of LBNCO with metallic lithium, together with the expanded interlayer spacing measured on HRTEM images, suggests that lithium incorporation is thermodynamically more favorable in oxalate-functionalized hBN than in pristine hBN, where intercalation is endothermic \cite{altintas2011intercalation}. This observation may have practical implications for the design of interfacial layers that wet Li-metal anodes effectively and promote uniform lithium deposition in lithium-metal batteries. Further details of the lithium doping experiments, along with a comprehensive transport study, will be reported separately. Our preliminary electrochemical results indicate that LBNCO can be used as a nanofiller in separators based on solid polymer electrolytes, as demonstrated by stable cycling of symmetric cells and lithium iron phosphate half cells, see discussion in Sec.~S9 and Fig.~S12.

\section{Conclusions}

This study introduces a scalable, solvent-free mechanochemical route for the functionalization of hexagonal boron nitride with lithium oxalate, yielding LBNCO (Li$_2$(BN)$_6$C$_2$O$_4$) as a nanocrystalline lamellar compound with distinctive, technologically relevant properties. LBNCO uniquely integrates ionic conductivity, chemical and thermal stability, and processability within a single-phase material. Comprehensive structural, chemical, and spectroscopic characterization (HR-TEM, XRD, XPS, FTIR, Raman, UV-Vis, TGA-DSC-MS, NMR, and ESR) confirms covalent oxalate functionalization at hBN edge sites, the formation of chelated and carbamate-type bonding motifs, while observations are also consistent with Li intercalation. The lithium content can be tuned even further through additional mechanochemical intercalation with metallic lithium without compromising chemical or thermal stability.

These results establish a viable pathway toward simplifying solid-state battery architectures by integrating ion conduction and mechanical separation within a single, low-toxicity material derived from abundant precursors. The solvent- and byproduct-free mechanochemical strategy provides a sustainable and industrially scalable framework for the development of solid-state lithium-ion conductors and high-tensile-strength polymer composites, with particular relevance for dendrite-suppressive separators and protective anode coatings in lithium-metal batteries.

Beyond energy storage, the presence of paramagnetic domains associated with frustrated, edge-embedded $\bullet$CO$_2^-$ radical anions broadens the functional scope of LBNCO, enabling its application as a spin-label material and as an initiator for radical polymerization, where it may simultaneously act as initiator and nanofiller. Extension of this approach to alternative charge carriers (Na, K, Mg, Ca) represents a promising direction for expanding the applicability of LBNCO-type materials toward advanced energy and thermally robust material systems.

\section*{Materials and Methods} \label{MatMet}
Hexagonal boron nitride (hBN) was purchased from Sigma-Aldrich (\#255475, $98\%$), lithium oxalate (Li$_{2}$C$_{2}$O$_{4}$) from Fischer Scientific (\#AA1342618, $99{+}\%$). The Al$_2$O$_3$ used as a dilutant for the concentration-dependent ESR measurement was purchased from Sigma-Aldrich (\#229423, $99.99\%$) $^{13}$C enriched oxalic acid ($99\%$ $^{13}$C, CLM-2002-PK, $98\%$) and $^{6}$Li ($95\%$ $^{6}$Li, LLM-827-PK, $98\%$) was purchased from Cambridge Isotope Laboratories. $^{6}$Li$_{2}$$^{13}$C$_{2}$O$_{4}$ was synthesized from these isotopes by first reacting $^{6}$Li with dropwise added cold DI water then reacting the resulting $^{6}$LiOH solution with the $^{13}$C enriched oxalic acid in a $2{:}1$ molar ratio. After evaporation of the water, XRD confirmed the isotope-enriched Li$_{2}$C$_{2}$O$_{4}$ product. 

A SPEX 8000M high-energy ball mill, equipped with a $1/3$~hp ($245$~W) motor and operated at $115$ V, $60$ Hz, and $1725$ rpm resulting in a clamp speed of $1060$ back-and-forth cycles, was used to carry out the mechanochemical reaction of hBN and Li$_{2}$C$_{2}$O$_{4}$ \cite{Siva23}. In a typical reaction, $2.0$ g of a $6{:}1$ molar mixture of hBN and Li$_{2}$C$_{2}$O$_{4}$ was filled in a stainless steel milling jar (Cole-Parmer SamplePrep 8007; 440C steel; $65$~mL) containing $45$ g of stainless steel balls of $1/8"$ ($3.175$~mm) diameter each. The jar was sealed under Ar gas atmosphere in a glovebox (MBraun Unilab Pro, operated under $0.1$ ppm O$_2$, $0.1$ ppm H$_2$O, source gas: UHP Ar ($5.0$), Airgas). The milling was carried out until the XRD pattern of Li$_{2}$C$_{2}$O$_{4}$ completely disappeared which required $5$ hours. After every hour of milling, a $30$ minute pause was held in order to cool the jar and the machine. Five batches of materials have been made for the detailed characterization listed below, after reaction parameter initialization. The number of samples and/or sampling methods are described at the end of each method.

The mechanical energy input was estimated with the standard collision model of shaker mills, in which each ball-wall impact transfers the ball kinetic energy $E=\tfrac{1}{2}m_{\mathrm{b}}v^{2}$ \cite{Basset94, Delogu00}. The clamp oscillates at $f \approx 17.67$~Hz ($1060$~cycles$/$min) with a $5.9$~cm throw, giving peak vial velocities of $v\approx3.3$~m$/$s, consistent with ball impact velocities of $1.8{-}3.3$~m$/$s measured for this type of high-energy mill \cite{Basset94}. As each ball undergoes two wall impacts per cycle \cite{Delogu00}, the ball charge $M_{\mathrm{b}}=45$~g (${\approx}340$ balls; ball-to-powder mass ratio $22.5{:}1$) delivers $0.2{-}0.7$~mJ per impact and a mechanical power of
    \begin{equation}
    P = \tfrac{1}{2} M_{\mathrm{b}} v^{2} \, (2f) \approx 3{-}9~\mathrm{W}.
    \end{equation}
Over the $t=5$~h of net milling, the cumulative specific energy dose \cite{Takacs04, Ward05} delivered to the $m_{\mathrm{p}}=2.0$~g charge is
    \begin{equation}
    D = \frac{Pt}{m_{\mathrm{p}}} \approx 20{-}80~\mathrm{kJ\,g^{-1}},
    \end{equation}
corresponding to $3{-}10$~eV per atom of the reactant mixture.

The milling of the same mixture of reactants ($2.0$ g powder) was also carried out in a planetary ball mill (Changsha DECO-PBM-V-0.4L; $110$~V/$60$~Hz variant, $750$ W) using a zirconia jar with zirconia balls ($100$ g of balls, each of $5$ mm diameter) in a $100$ mL airtight jar sealed under UHP Ar atmosphere, at close to maximum rotation speed. It took $92$ h until the XRD pattern of Li$_{2}$C$_{2}$O$_{4}$ fully disappeared from the product.

For the planetary mill, the same per-impact energy, $E=\tfrac{1}{2}m_{\mathrm{b}}v^{2}$, applies, but the ball velocity is set by the combined evolution of the disk and rotation of the jar, following the kinematic model of Burgio \textit{et al.} \cite{Burgio91}. At the maximum speeds of $400/800$~rpm (disk/jar, fixed $1{:}2$ ratio), with a revolution radius of $r\approx6$~cm and $m_{\mathrm{b}}=0.39$~g ZrO$_{2}$ balls (${\approx}255$ balls; ball-to-powder mass ratio $50{:}1$), ballistic impacts would deliver $1{-}3.5$~mJ per hit and an upper-bound power of ${\sim}10$~W, which would drive the reaction to completion within hours. The observed $92$~h instead corresponds to an effective power of only ${\approx}0.1{-}0.5$~W, i.e., the same cumulative dose delivered $92/5 \approx 18$ times more slowly than in the shaker mill \cite{Takacs04}. The difference has a simple origin: at the fixed $1{:}2$ speed ratio, the centrifugal field of the jar rotation exceeds that of the disk revolution, so the balls remain pressed against the jar wall and energy is transferred by rolling friction rather than by ballistic impacts -- a roughly hundredfold less efficient mode of mechanical activation \cite{Takacs06}.

Calcination of LBNCO was carried out in an alumina crucible within a tube furnace at temperatures between $350$ and $1000~^{\circ}$C under the flow of N$_{2}$ gas for $8$ hours.

High-resolution transmission electron microscopy (HR-TEM) was performed at the Notre Dame Integrated Imaging Facility using a Spectra 300 (S)TEM (Thermo Fisher Scientific), operated at an accelerating voltage of $300$ kV. HR-TEM images were captured using a Ceta 16M CMOS camera. Two batches were characterized in detail. EELS maps were recorded with a Gatan K3 detector. The EELS energy resolution was $1.5$ eV, as determined by the FWHM of the zero-loss peak (ZLP), while the spectra were collected with an energy dispersion of $0.18$ eV/channel.

Powder X-ray diffraction (XRD) patterns were measured using a Bruker D2 Phaser device at Cu K$\alpha$ X-ray radiation ($1.5406$ {\AA} wavelength) using the $\Theta-2\Theta$ geometry. The PXRD pattern of all batches was recorded; no distinguishable differences were found between batches. Phase composition was evaluated through Rietveld refinement.

X-ray photoelectron spectroscopy (XPS) measurements were performed at the Materials Characterization Facility (MCF) of the University of Notre Dame, using a PHI 5000 Versa Probe II equipped with monochromatic Al K$\alpha$ X-rays ($15$ kV, $25$ W). Three batches were scanned.

Thermogravimetric analysis (TGA) and differential scanning calorimetry (DSC) measurements were performed on a commercial Mettler Toledo TGA/DSC 3+ HT/1600 instrument at the Notre Dame Materials Characterization Facility research core facility (MCF). Approximately $12$ mg of the investigated material was placed in a $40~\mu$L aluminum crucible with a pierced lid, which was heated between $30-640~^\circ$C with a sweep rate of $1.5~^\circ$C/min. The carrier gas with a $100$ mL/min flow was chosen to be Ar. An empty crucible run was performed prior to the sample measurement and used as a baseline. The instrument is coupled to a Pfeiffer Thermo Star mass spectrometer that continuously samples the gas composition during the heating. Three batches were scanned.

Infrared absorption spectra were recorded at the University of Pannonia, using a Bruker Vertex 70 type Fourier-transform infrared (FTIR) spectrometer equipped with a single-reflection, diamond ATR accessory with atmospheric compensation. The final spectra of the powdered samples were acquired without any further sample preparation by averaging $512$ scans, using a resolution of $2~\text{cm}^{-1}$ and a room-temperature DTGS detector. All batches were scanned. No distinguishable differences were found between batches.

Raman measurements were performed using an NRS-5100, Jasco confocal Raman microscope equipped with a $\times100$ long working distance objective and a $785$ nm laser source at the Notre Dame Materials Characterization Facility Research core facility (MCF). Additional measurements at $457$, $532$, and $633$ nm were carried out using a WITec alpha300R confocal Raman microscope and a Zeiss LD EC Epiplan-Neofluar $50\times$/$0.55$ long working distance objective and with a $600$ lines/mm grating. All batches were scanned. No distinguishable differences were found between batches.

Absorption measurements were performed using a JASCO V-770 UV-Vis spectrometer equipped with an integrating sphere at the Notre Dame Materials Characterization Facility Research core facility (MCF). Two batches were scanned.

Initial $^{11}$B and $^{7}$Li solid-state NMR spectra of LBNCO and its $350~^{\circ}$C calcined derivative were measured and analyzed at Argonne National Laboratory, using a $500$ MHz Bruker Advance III spectrometer ($11.7$ T) with a $2.5$ mm MAS probe operating at $30$ kHz spinning speed. A rotor-synchronized echo pulse sequence ($90^\circ \tau - 180^\circ \tau_{\text{acq}}$), where $\tau = 1/\nu_{\text{r}}$ (spinning frequency), was used for all the acquisitions. $^{7}$Li experiment was performed with a $\pi/2$ pulse width of $2.5$~$\mu$s and pulse delay of $5$ s and was referenced with 1 M LiCl at $0$ ppm. For $^{11}$B MAS NMR, a $\pi/12$ pulse width of $2.6$~$\mu$s and a pulse delay of $2$ s were used. $0.1$ M boric acid was used as a secondary reference at $18.8$ ppm, similarly to Ref. \cite{zhang2016experimental}. Comprehensive $^{7}$Li, $^{11}$B (CT selective), $^{14}$N and $^{13}$C magic angle spinning nuclear magnetic resonance (MAS NMR) spectra were recorded at room temperature on a Bruker Avance II $400$ MHz spectrometer with a magnetic field of $9.4$ T at the University of Pannonia. $^{13}$C and $^{14}$N MAS spectra were acquired with a $4$ mm Bruker double channel ($^{1}$H,X) probe while $^{7}$Li, $^{11}$B MAS spectra were recorded with a $2.5$ mm Bruker double resonance MAS probe. $^{13}$C with high-efficiency $^{1}$H decoupling experiments (SPINAL) were carried with $\pi/2$ pulse width of $7~\mu$s for $^{13}$C and relaxation delay of $10$ s at spinning rate of $6$ kHz for Li$_{2}$C$_{2}$O$_{4}$ and $12$ kHz for LBNCO. As a chemical shift reference, external alpha glycine, having a chemical shift of the carbonyl signal at $176.5$ ppm, was applied. $^{14}$N spectra were measured with a $\pi/2$ pulse length of $3.3~\mu$s and recycle delay of $300$ s, the number of scans was $24$ and internal ammonium chloride with chemical shift of $0$ ppm was used as reference. $^{7}$Li (selective and non-selective) MAS spectra were acquired with a $1.1$ $\mu$s length of $\pi/2$ pulse and an $11$ kHz spinning rate; LiCl ($0$ ppm) was used as the reference. One batch was characterized.

For $^{11}$B quantitative measurement, the pulse length ($0.25~\mu$s) was less than the $\pi/2$ pulse of the given RF field strength, and a recycle delay of $5$ s was chosen at a spinning rate of $16$ kHz. As a chemical shift reference, cubic boron nitride ($\delta=1.6$ ppm, \cite{jeschke1998comprehensive}) was applied. To quantify the amount of different chemical environments for quadrupolar $^{11}$B, the line-shape fitting application of Topspin 4.1.4 (Bruker) was used, and the central transition of broadband spectra was fitted with the QUAD central model.

Initial ESR measurement and analysis of LBNCO (without heat treatment or isotope substitution) was carried out at Argonne National Laboratory using a continuous wave (cw) Bruker Elexsys II E500 X-band ($9.5$ GHz) spectrometer with $100$ kHz magnetic field modulation and phase-sensitive detection. A standard rectangular resonator (TE$_{102}$, Bruker ER4102ST) was used. Measurements were performed at ambient temperature ($T=295$ K). Comprehensive continuous wave ESR was performed at the University of Notre Dame, Stavropoulos Center for Complex Quantum Matter, using a commercial Bruker Elexsys II E500 spectrometer in the X-band ($0.33$ T, $9.5$ GHz). Approximately $10-12$ mg of material was placed in a high-quality quartz sample tube (ATS Life Sciences Wilmad 707-SQ-250M), evacuated to a vacuum of $10^{-6}$ mbar, and then sealed with $\sim 40$ mbar of research-grade ($6.0$) He gas. All measurements were performed at room temperature in the ER 4122SHQE (Super High Q) resonator using $0.2$ mW microwave power ($30$ dB attenuation) to avoid saturation effects. The samples showed minimal effect on the $Q$-factor of the cavity. The amplitude of the modulation was chosen to be $1$ G to avoid line distortion, with a standard modulation frequency of $100$ kHz. The obtained spectra were evaluated using fitting derivative Lorentzian and Gaussian curves. The $g$-factor was calibrated using Mn:MgO with $1.5$ ppm Mn$^{2+}$ standard with $g=2.0014$, and the susceptibility was calculated using a CuSO$_4 \cdot 5$ H$_2$O standard. All batches were scanned. No distinguishable differences were found between batches.

For electrical transport measurements, the sample powder was pressed into a $1/8"$ diameter pellet under a uniaxial pressure of $600$~MPa. Electrical contacts were made using conductive silver paint and $50~\mu$m, $99.995\%$ pure gold wires (Fisher Scientific) attached to the pellet sides. Two-point electrical measurements were performed with the Electrical Transport Option (ETO) of a Quantum Design 14 T Dynacool PPMS. In this high-resistivity configuration, a $10$~V peak-to-peak sinusoidal excitation at $1$~Hz was applied to the sample, and the current response was recorded with a current amplifier. The resistance was determined from the slope of a linear fit to the voltage-current characteristics. The electronic transport measurements were conducted at the University of Notre Dame, Stavropoulos Center for Complex Quantum Matter. Three batches were examined. No distinguishable differences were found between batches.

For impedance spectroscopy measurements, the samples were pressed into a $6$ mm diameter pellet with a uniaxial pressure of $600$ MPa. The measurements were performed with an Omicron Lab Bode 100 Vector Network Analyzer integrated with the Multi-Function Probe (MFP) option of a Quantum Design 14 T Dynacool PPMS. The measurement circuit contained an Omicron Lab B-AMP 12 amplifier, in a Series-thru arrangement. The source voltage was $0$ dBm, between $10$ Hz and $50$ MHz, at a detection bandwidth of $300$ Hz. Data for the LBNCO material were processed with FFT filtering, as the high impedance values approach the detection limit of the device, introducing significant low-frequency noise. For fitting data, an equivalent Randles circuit was used \cite{Lazanas2025_Randles}.

DFT calculations were performed using Quantum ESPRESSO \cite{QEgiannozzi2009quantum, QEgiannozzi2017advanced, QEgiannozzi2020quantum} with GIPAW for NMR shielding \cite{GIPAWvarini2013enhancement}. The GIPAW extension of the Perdew--Burke--Ernzerhof (PBE) functional was used with ultrasoft pseudopotentials (Martin--Troullier for Li). For NMR calculations, plane-wave cutoffs were $51$ Ry (wavefunction) and $612$ Ry (density). The $k$-point grids were $8\times1\times3$ (zigzag model) and $5\times2\times2$ (armchair model), corresponding to a $k$-space resolution of $\Delta k\approx0.3-0.4$ $1/$\AA. Structures were relaxed to $<0.0001$ Ry/$a_0$ in force and $<0.01$ kbar in stress. For comparison, NMR spectrum parameters were also calculated using norm-conserving pseudopotentials and a much larger, $12\times4\times8$ $k$-points grid ($\Delta k\approx0.12$ $1/$\AA) with wavefunction and density cutoffs of $100$ and $400$ Ry, respectively. Tables S2 and S3 of the Supporting Information show that reasonably good agreement was achieved for the chemical shifts as obtained by the two computational frameworks, further indicating sufficient numerical stability of the results. Calculated chemical shifts of hBN, Li$_2$C$_2$O$_4$, and LiBOB were in reasonable agreement with experimental values, validating the computational setup.


For the calculation of the ESR $g$-factors, relaxed structures and wavefunctions obtained with norm-conserving pseudopotentials were used. The ESR calculations used fully relaxed geometries obtained by constraining the wavefunctions to two or four unpaired electrons per simulation cell. These constraints reflect the fact that the simulation cells contained two formula units of Li$_2$(BN)$_6$C$_2$O$_4$ and two interlayer B(sp$^3$)$-$C(sp$^3$) or C(sp$^3$)$-$C(sp$^3$) bonds. Isotropic $g$-factors were computed from $g$-tensors as $g_{\text{iso}}=(g_{\text{xx}}+g_{\text{yy}}+g_{\text{zz}})/3$.Structures and electron densities are visualized using the VESTA software \cite{vesta}.

\section{Acknowledgements}
We kindly thank Prof. J{\'a}nos Rohonczy (E{\"o}tv{\"o}s Lor\'and University, Budapest, Hungary) for NMR-related discussions; Dr. Davide Ceresoli (University of Milan, Italy) for NMR and ESR calculations-related advice; and Abrasive Tools (Poltava, Ukraine) for a generous donation of cBN samples. This research used resources of the National Energy Research Scientific Computing Center, a DOE Office of Science User Facility supported by the Office of Science of the U.S. Department of Energy under Contract No. DE-AC02-05CH11231 using NERSC award SBIR-ERCAP0033652. The EPR work at Argonne National Laboratory was supported by the U.S. Department of Energy, Office of Science, Office of Basic Energy Sciences, Division of Chemical Sciences, Geosciences, and Biosciences, through Argonne National Laboratory under Contract No. DE-AC02-06CH11357. The materials discovery phase of this research was funded by the U.S. National Science Foundation under an STTR Phase I grant to Boron Nitride Power LLC and IIT (Award Number 2109286, with K.N. as PI). Additional private funding was provided by Boron Nitride Power LLC. F.S. acknowledges the National Research, Development and Innovation Office of Hungary (NKFIH) Grants Nr. 149457 and 2026-4.1.1-MISSZI\'OK-2026-00030. Zs.B acknowledges the support provided by the Ministry of Culture and Innovation of Hungary from the National Research, Development and Innovation Fund, financed under the 2024-1.1.1-KKV-FÓKUSZ-2024-00082 project. We gratefully acknowledge Maksym Zhukovskyi and Alex Sheardy for their assistance with the HRTEM and STEM-EELS measurements performed at NDIIF.

\section*{Author contributions}
Conceptualization and synthesis design were carried out by KN. Exploratory synthesis of FBNs, synthesis of LBNCO, XRD, and initial FTIR measurements were carried out by VRT and SR. HRTEM analysis was performed by BGM and DB. STEM-EELS measurements were analyzed by LV under the supervision of DB and BGM. RF-impedance measurements were performed and analyzed by LV under the supervision of BGM, FS, and LF. Electrochemical cell preparation and testing were carried out by AS under the supervision of LS and KN. DSC-TGA-MS measurements and analysis were performed by AN and BGM under the supervision of DB. Comprehensive IR spectra were measured by BZs and analyzed by BZs, DB, and KN. Preliminary NMR spectra were measured and analyzed by FDK. Comprehensive NMR spectra were recorded and analyzed by MK and GSz. BD and BGM obtained the Raman spectra. BD and BJM performed the XPS measurements. Preliminary ESR measurements and analysis were carried out by JN, MYT, and OGP. Comprehensive ESR measurements and analysis were carried out by BGM under the supervision of FS and LF. AN conducted the transport experiments and analysis under the supervision of LF. CC performed the DLS measurements and analysis, as well as general laboratory assistance to project members at IIT. DB performed the UV-Vis measurements and analysis. Theoretical models were developed, and DFT calculations were carried out by KN. KN oversaw the project. The development of the first draft was led by KN with contributions from all authors, especially from BGM, DB, AN, BZs, MK, and GSz. BGM, AN, DB, KN, and LF designed and perfected the figures. A substantial reorganization of the first draft and the finalized overall design of the paper was done by LF in collaboration with all authors. All authors contributed to the writing of the manuscript.

\section*{Competing interests}
The authors declare no competing interests.

\section*{Additional information}
Supporting information is available for this paper.

\section*{Data Availability Statement}
The data that support the findings of this study are available from the corresponding authors upon request.

\bibliography{references}

\clearpage

\onecolumngrid
\appendix
\renewcommand{\appendixname}{}
\renewcommand{\thesection}{S\arabic{section}}
\renewcommand{\thesubsection}{S\arabic{subsection}}
\renewcommand\thefigure{S\arabic{figure}}
\setcounter{figure}{0}
\renewcommand\thetable{S\arabic{table}}
\setcounter{table}{0}
\renewcommand*{\thepage}{S\arabic{page}}
\setcounter{page}{1}

\section*{Supplementary Material}

The Supporting Information contains XRD reproducibility and reaction-progress analyses, including quantitative phase analysis; complete TEM line profiles; XPS spectra of the C~1s and O~1s regions; Raman spectra; ESR linewidth analysis and spin-labeling benchmarks; additional HR-TEM images and STEM-EELS elemental maps; TEM- and dynamic light scattering (DLS)-derived particle-size distributions; electrochemical cycling and symmetric-cell measurements of LBNCO/PVDF membranes; FTIR, UV-Vis, and optical-microscopy studies of the reactivity of the LBNCO radical centers; additional thermal analyses (TGA, DSC, and MS); UV-Vis absorption and photoluminescence spectra; comparative FTIR spectra; further details of the theoretical calculations of the NMR and ESR parameters; and comparisons with representative Li-ion conductors and previously reported hBN functionalization strategies.

\section{Reproducibility and scale-up}

\begin{figure}[!htp]
    \includegraphics[width=0.6\columnwidth]{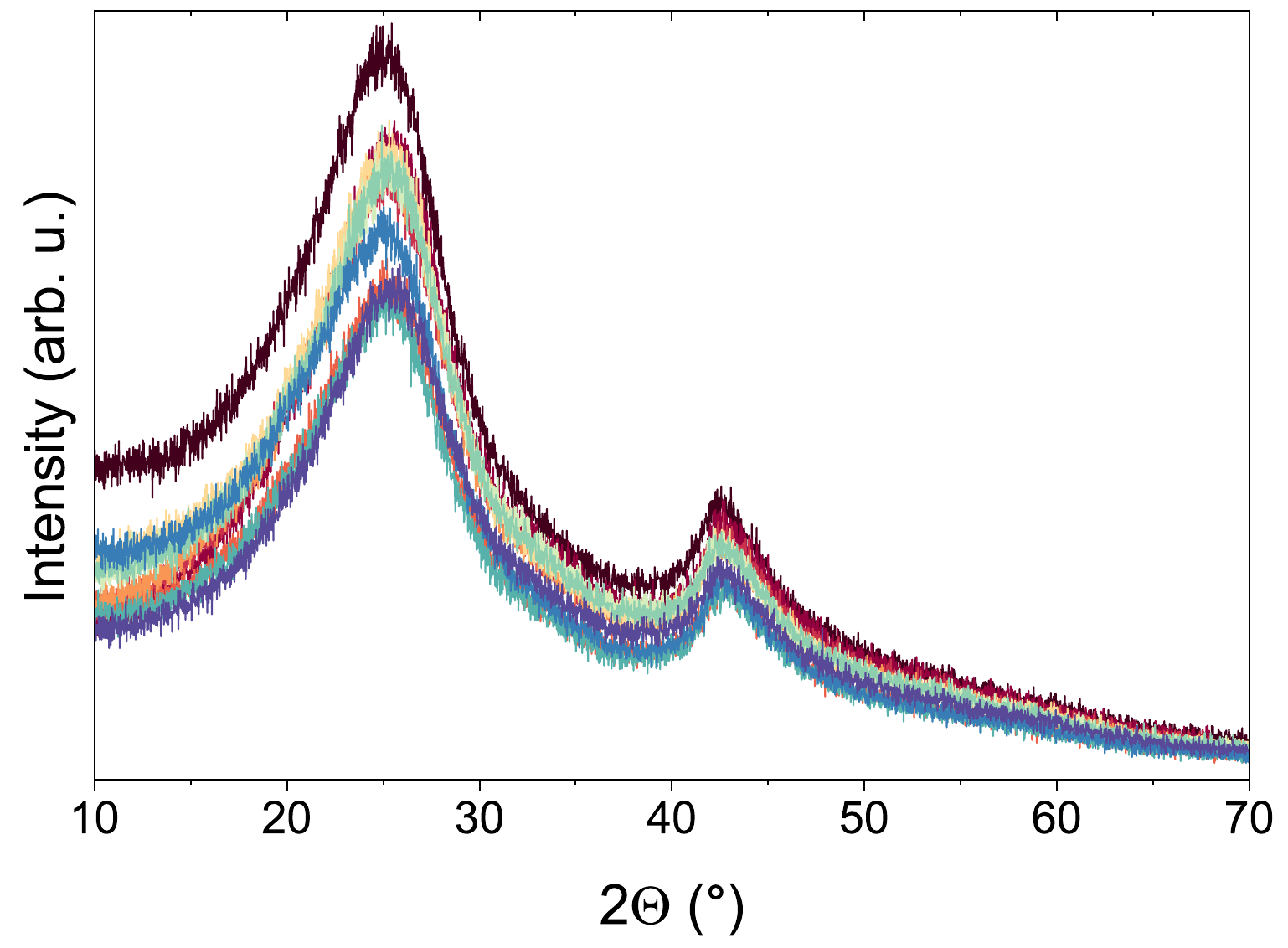}
    \caption{Reproducibility of the functionalization by XRD.}
    \label{FigSI:XRD-reprod}
\end{figure}

To assess the reproducibility of the mechanochemical functionalization, XRD patterns from $11$ independently prepared batches synthesized under identical conditions are shown in Fig. \ref{FigSI:XRD-reprod}. The diffraction patterns exhibit no discernible variation in peak positions or relative intensities across the different batches, demonstrating the high reproducibility of the synthesis. All samples were prepared under an Ar atmosphere and measured in air immediately after synthesis.

While scale-up of the LBNCO synthesis has not yet been demonstrated in the present study, high-energy ball milling (HEBM) is widely regarded as a scalable approach for materials synthesis \cite{Ali2025BallMilling}. Successful scale-up of HEBM-based synthesis has been reported for a range of material classes, including titanium diboride nanosheets \cite{Gaykwad2023LargeScale}, transition-metal vanadates \cite{Li2023Vanadates}, magnesium-based hydrogen-storage alloys \cite{Shao2012Nanotechnology}, and powder-metallurgical processing of titanium- and magnesium-based alloys for aerospace and automotive applications \cite{NanomaterialPowder2025HighEnergyBallMill}.

On this basis, we consider attrition milling to be a promising route for future scale-up of the present LBNCO synthesis. Its continuous or high-throughput operation, larger processing volumes, and efficient energy transfer make it a natural next step toward demonstrating the practical scalability of this mechanochemical functionalization strategy.

\section{Reaction progress followed by XRD}
\begin{figure}[!htp]
    \includegraphics[width=0.9\columnwidth]{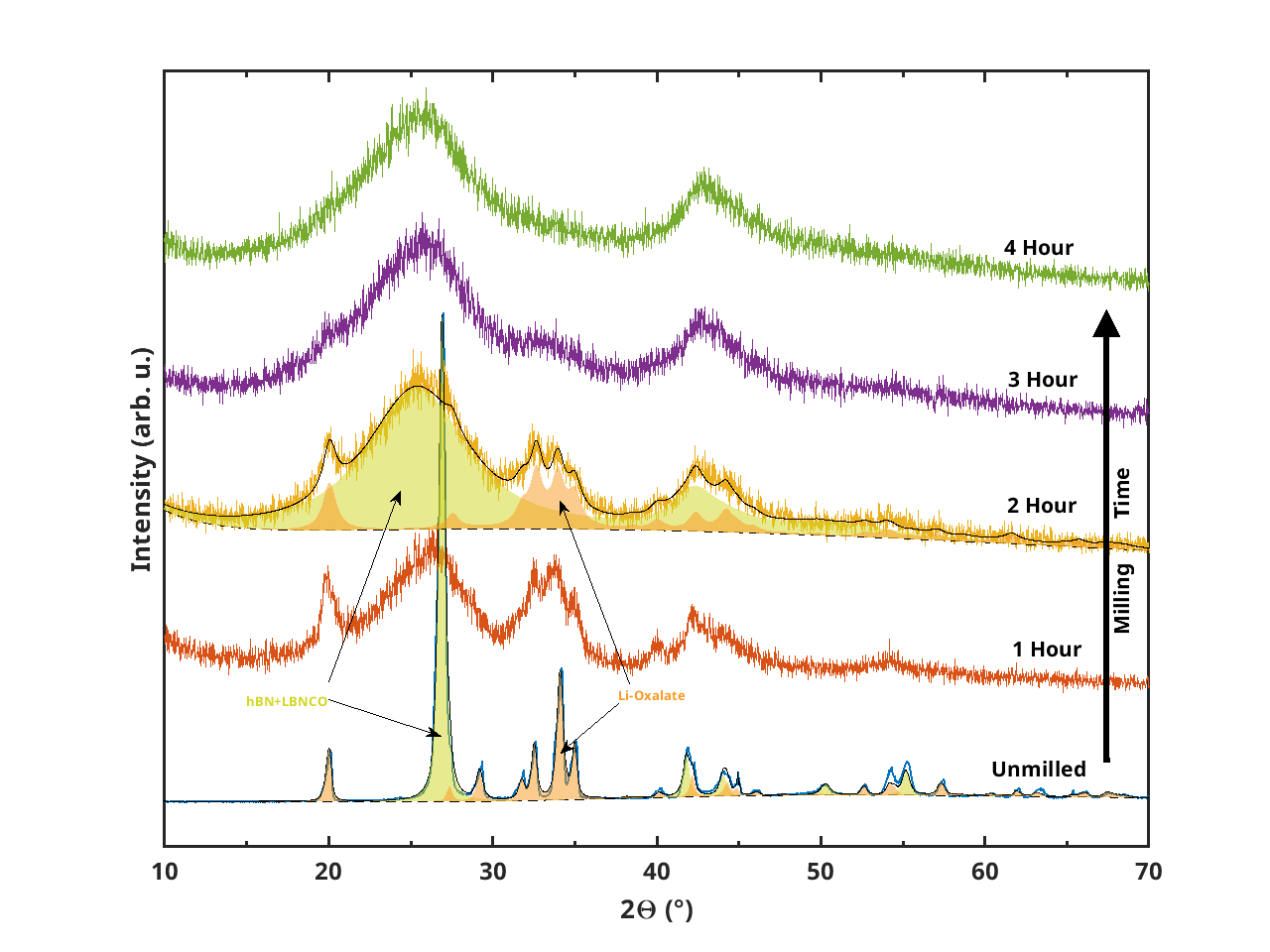}
    \caption{Comparison of milling-induced changes in XRD reflection intensities of pristine hBN-Li$_2$C$_2$O$_4$ mixture and LBNCO as a function of the milling time.}
    \label{FigSI:XRD-mill}
\end{figure}

\begin{figure}[!htp]
    \includegraphics[width=0.7\columnwidth]{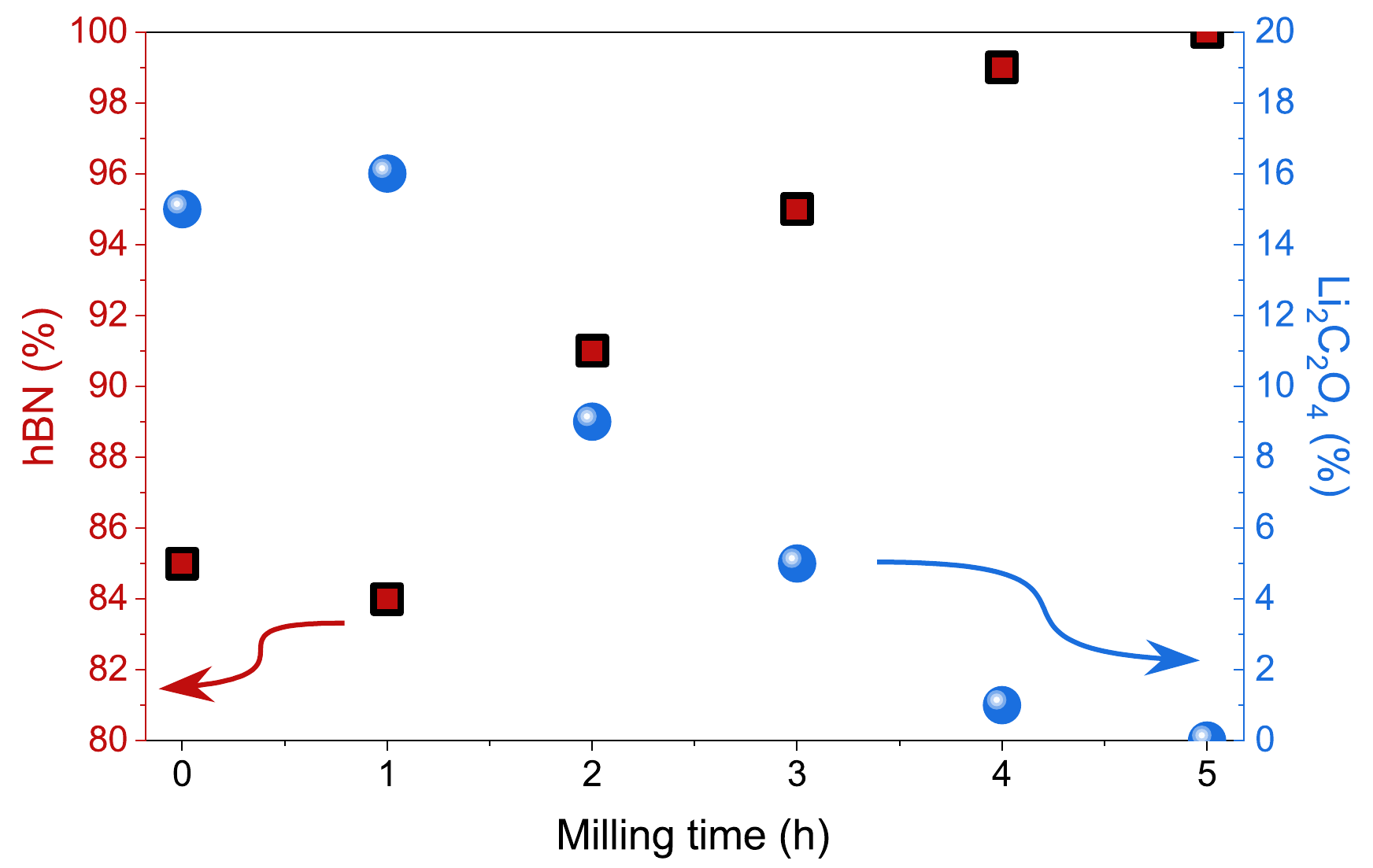}
    \caption{Fraction of hBN+LBNCO and Li-oxalate as a function of the milling time.}
    \label{figSI:XRD-milling_time}
\end{figure}

The progress of the mechanochemical reaction was followed by XRD (Fig.~\ref{FigSI:XRD-mill}). Diffractograms were analyzed with Rietveld full profile refinement. The corresponding quantitative phase analysis is given in Table~\ref{TabSI:composition} and plotted in Fig.~\ref{figSI:XRD-milling_time}. The reaction shows an induction period of about one hour, during which the composition remains unchanged within uncertainty, followed by continuous conversion to completion at $5$~hours. Such an induction stage is characteristic of mechanochemical solid-state reactions, where initial milling is consumed in particle refinement and the creation of fresh reactive interfaces before the reaction proper begins.

\begin{table}[!htp]
    \caption{Relative phase composition of hBN+LBNCO and Li-oxalate as a function of milling time. The error of the composition is around $\pm1\%$.}
    \centering
    \begin{tabular}{l|c|c}
             Milling time & hBN+LBNCO & Li-oxalate \\ 
              \hline
       Unmilled & 85$\%$ &  15$\%$\\       
       1 hour & 84$\%$ &  16$\%$\\
       2 hours & 91$\%$ &  9$\%$\\
       3 hours & 95$\%$ &  5$\%$\\
       4 hours & 99$\%$ &  1$\%$\\
       5 hours&100$\%$ &  0$\%$\\
    \end{tabular}
    \label{TabSI:composition}
\end{table}

\begin{figure}[!htp]
    \includegraphics[width=0.8\columnwidth]{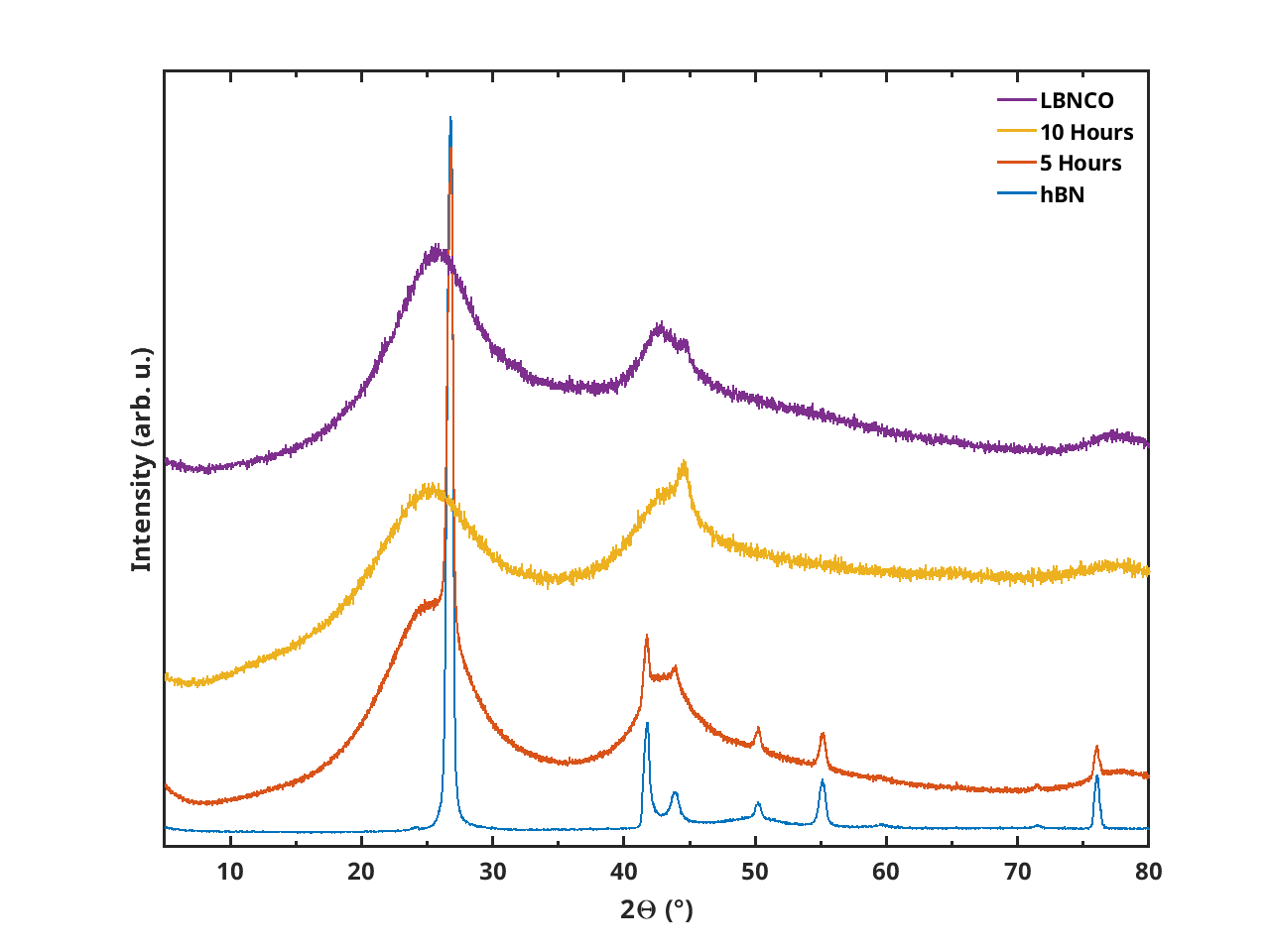}
    \caption{Comparison of pristine hBN, ball-milled hBN, and lithium-oxalate-functionalized hBN (LBNCO).}
    \label{FigSI:XRD-oxalate-milling}
\end{figure}

For comparison, pure hBN was milled separately under identical conditions (Fig.~\ref{FigSI:XRD-oxalate-milling}). Sharp reflections characteristic of well-ordered, non-turbostratic hBN remain evident after $5$~h of milling and disappear only after $10$~h. These results demonstrate that pristine hBN undergoes structural disordering more slowly than the hBN-Li$_2$C$_2$O$_4$ mixture under the same milling conditions.

\section{Complete TEM lineprofiles} \label{sec:fullTEM}

\begin{figure}[!htp]
    \includegraphics[width=0.45\columnwidth]{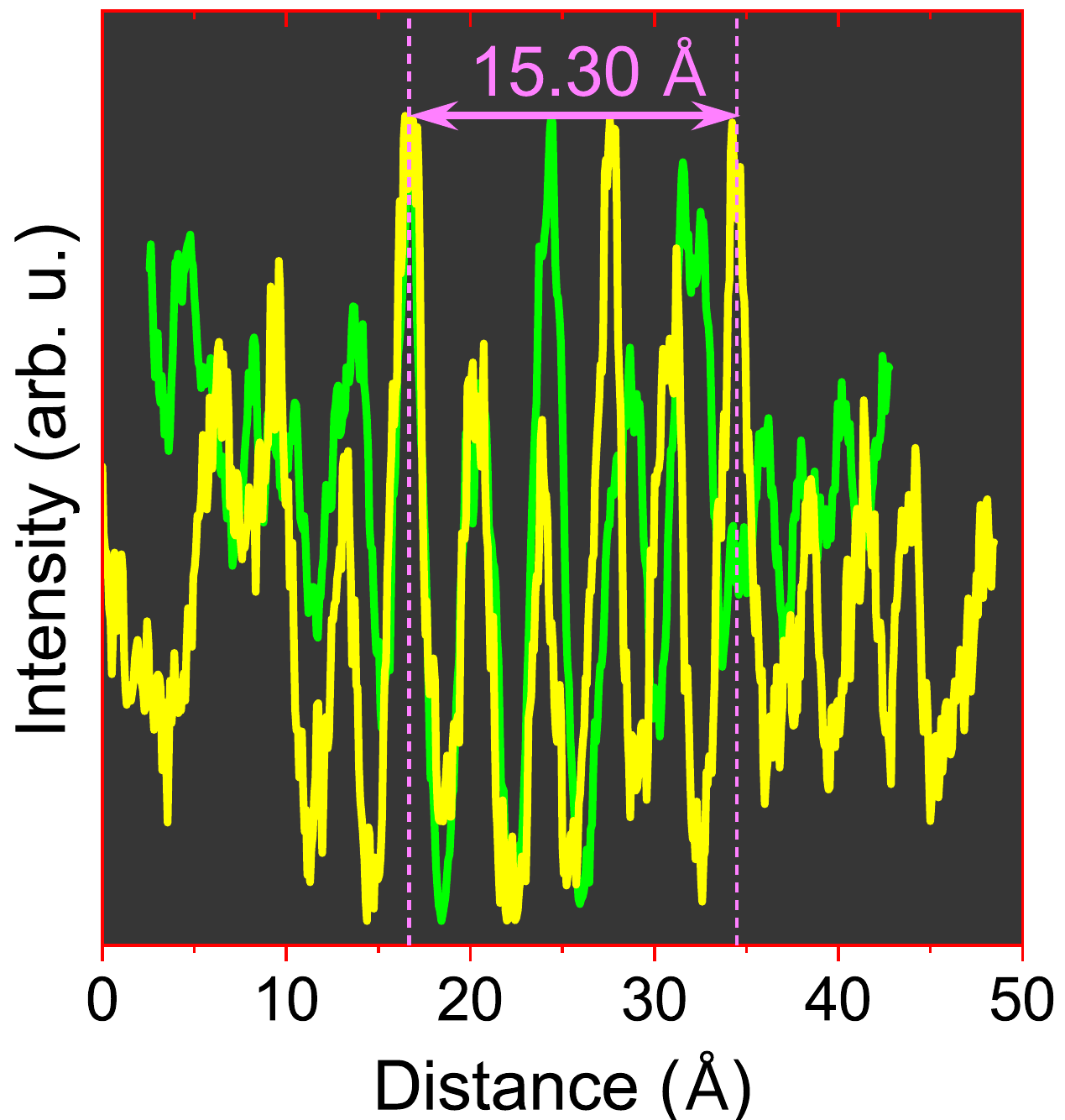}
    \caption{The full TEM line profiles corresponding to the image shown in Fig. 1 of the manuscript.}
    \label{FigSI:TEM-fullprofile}
\end{figure}

The full TEM line profiles for the yellow and green curves of Fig. 1 are shown in Fig. \ref{FigSI:TEM-fullprofile}.

\section{HRTEM-EELS analysis} \label{sec:EELS}

\begin{figure}[!htp]
    \centering
    \begin{minipage}[t]{0.48\columnwidth}
    \vspace{0pt}
        \centering
        \includegraphics[width=\linewidth]{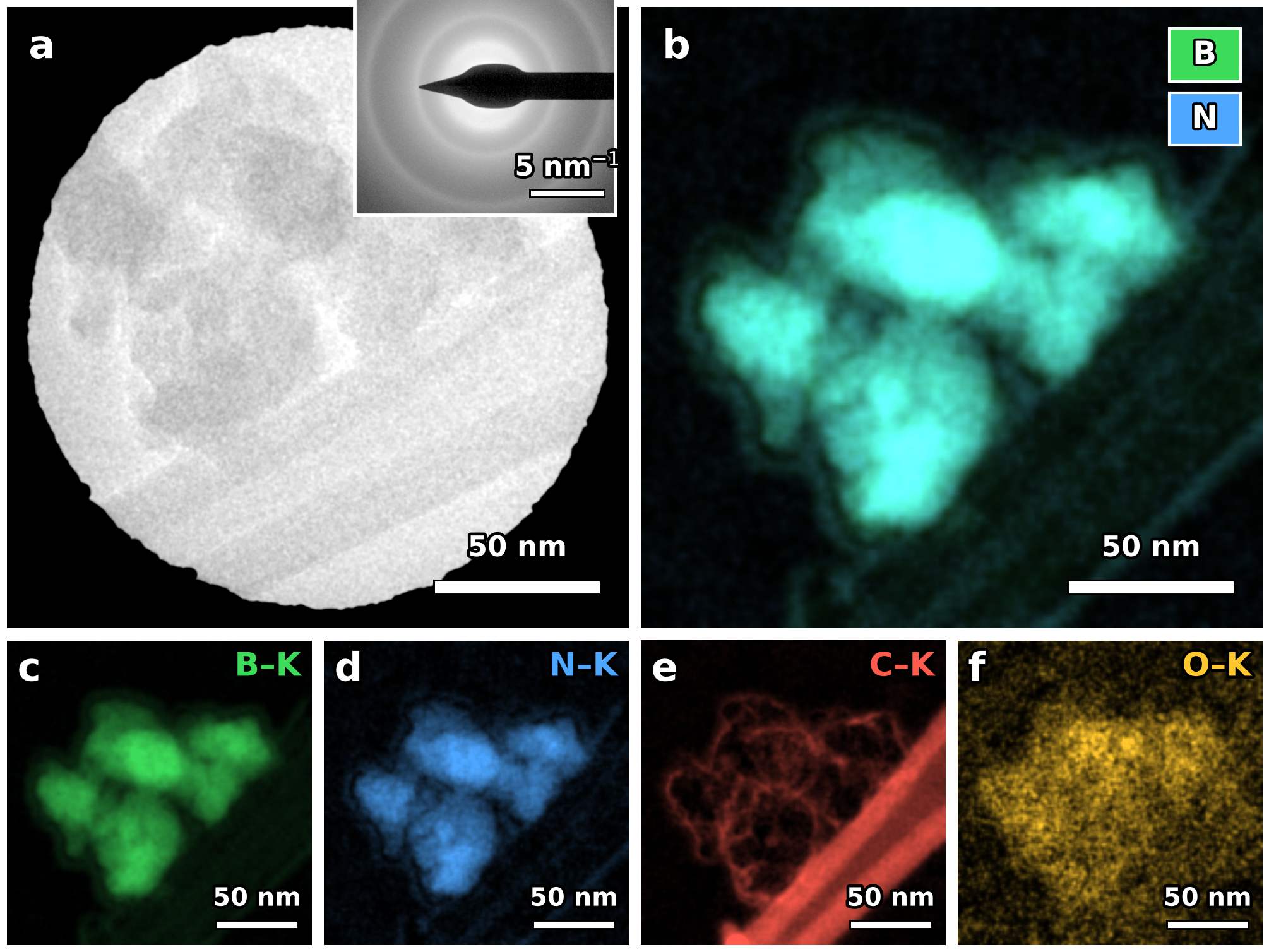}
        \caption{Pure hBN ball-milled for $5$~h under conditions identical to the LBNCO synthesis, without lithium oxalate or metallic Li. a)~Bright-field TEM image of the mapped region, recorded through the selected-area aperture that defines the diffraction area, with the corresponding SAED pattern inset. The (100) and (110) broad rings appear at ($d = 2.13$ and $1.24$~\AA), and an additional diffuse halo near $d \approx 3.3$~\AA{} is present, indicating randomly oriented, turbostratically disordered fragments produced by milling. b)~Overlay of B (green) and N (blue). c-f)~B-K, N-K, C-K and O-K signals in arbitrary units. B and N are uniformly co-localized over the particles, C is confined to the lacey carbon support film, and O follows the particle outlines, consistent with a thin surface oxide. No Li-K signal is observed in this sample.}
        \label{fig:hBN_EELS}
    \end{minipage}\hspace{0.02\columnwidth}
    \begin{minipage}[t]{0.48\columnwidth}
    \vspace{0pt}
        \centering
        \includegraphics[width=\linewidth]{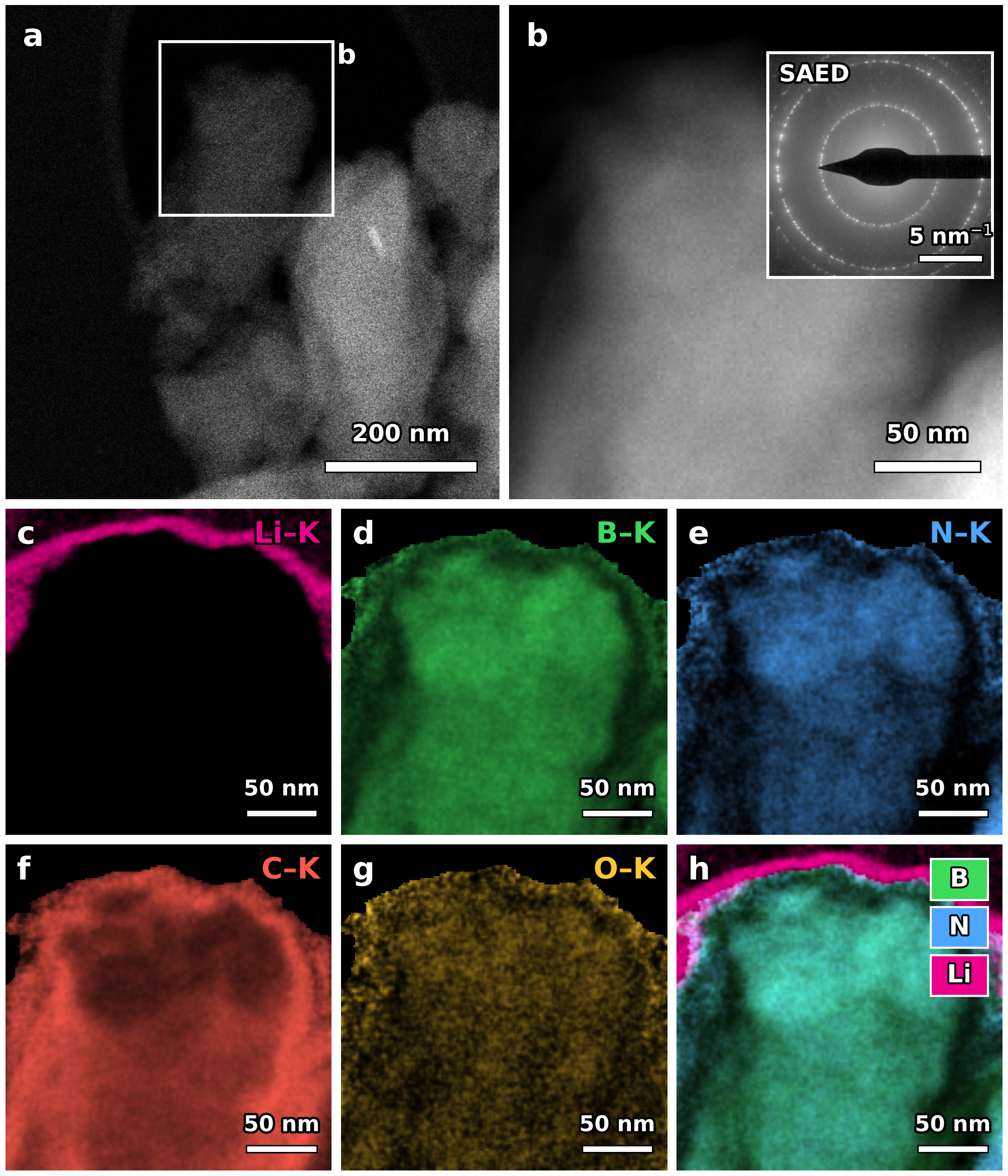}
        \caption{STEM-EELS elemental mapping and selected-area electron diffraction of LBNCO. a)~ADF-STEM survey; the white square marks the area shown in~b). b)~ADF-STEM image of the mapped region with the corresponding SAED pattern inset. The rings index to hBN~(100), $d = 2.13$~\AA, and (110), $d = 1.24$~\AA; (002) is essentially absent. We chose areas where flakes lie flat with their basal planes normal to the beam. c-g)~Li-K, B-K, N-K, C-K and O-K maps, respectively. h)~Overlay of B (green), N (blue) and Li (magenta). The imaged objects are agglomerates of several stacked and overlapping flakes rather than isolated single sheets. Nevertheless, the edge localization of Li is visible at the edge of the agglomerate, where crystal edges concentrate.}
        \label{fig:LBNCO_EELS}
    \end{minipage}
\end{figure}

HRTEM-EELS analysis was performed on ball-milled hBN powder and LBNCO (Figures \ref{fig:hBN_EELS} and \ref{fig:LBNCO_EELS}). Fig.~\ref{fig:hBN_EELS}a shows the aggregated hBN flakes of the milled hBN powder, and the corresponding diffraction rings coming from the random flake orientations (powder average). The EELS elemental maps (Fig.~\ref{fig:hBN_EELS}b-f confirm that boron and nitrogen are uniformly co-localized throughout the particles. Carbon is predominantly associated with the lacey carbon support film, whereas oxygen follows the surfaces and outlines of the hBN particles, consistent with a thin surface oxide formed during or after milling. No detectable lithium signal is observed in the milled hBN reference.

LBNCO exhibits a similar morphology, consisting of agglomerates of stacked and overlapping hBN flakes (Fig.~\ref{fig:LBNCO_EELS}a and b. In the SAED pattern, the same ring spacings appear as in the milled hBN sample, with more well-defined polycrystalline components. Fig.~\ref{fig:LBNCO_EELS}c-h show the EELS elemental maps, with boron and nitrogen tracing the entire area of the flakes, and carbon and oxygen appearing as contaminants over the flake areas, and in higher concentrations around the edges. In marked contrast to the pristine milled hBN reference, however, LBNCO exhibits a prominent lithium signal that is preferentially localized at the perimeter of the agglomerates, where the edges of the constituent hBN flakes are concentrated. Although the overlapping morphology does not allow individual flake edges to be resolved throughout the agglomerate, the preferential localization of lithium, together with the increased carbon and oxygen signals near the edges, is consistent with the molecular structures proposed in Fig.~\ref{FigSI:other-ESR-structures}. The HRTEM-EELS measurements therefore support a picture in which lithium-oxalate functionalization occurs preferentially at or near the edges of the ball-milled hBN flakes rather than uniformly across their basal planes.

\clearpage

\section{C and O XPS} \label{sec:CO-XPS}

\begin{figure}[!htp]
    \includegraphics[width=0.7\columnwidth]{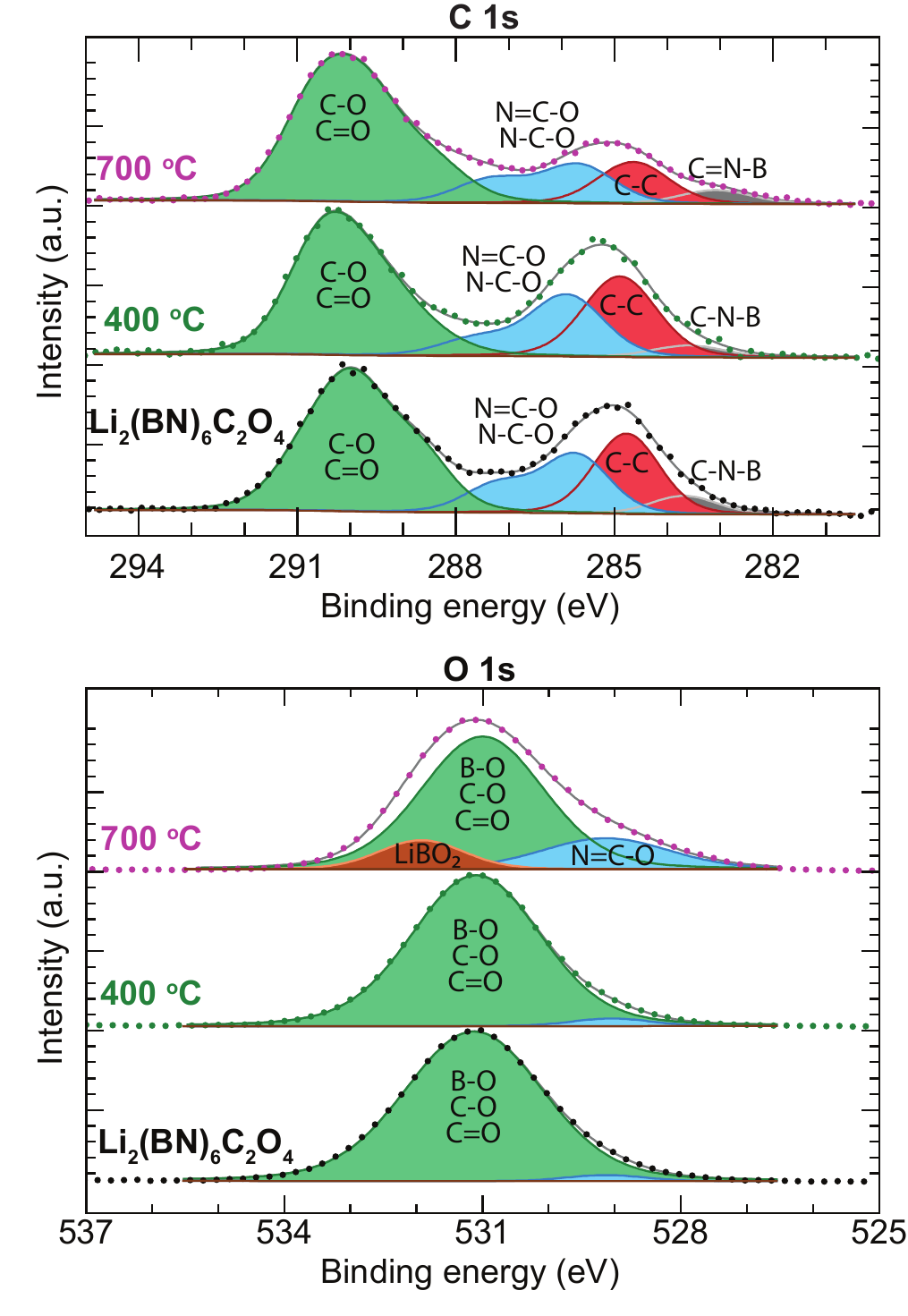}
    \caption{
        C 1s and O 1s XPS spectra and their deconvolution of LBNCO and its $400~^{\circ}$C and $700~^{\circ}$C calcined derivatives.}
    \label{FigSI:XPS-C-O}
\end{figure}

X-ray Photoelectron Spectroscopy (XPS) spectra of LBNCO and its calcined derivatives at $400~^{\circ}$C and $700~^{\circ}$C are presented in Fig.~\ref{FigSI:XPS-C-O} for C 1s and O 1s core levels. The detailed discussion of the observed spectra can be found in Section III of the main text.

\section{Raman spectroscopy} \label{Ramansection}

\begin{figure}[!htp]
    \includegraphics[width=0.6\columnwidth]{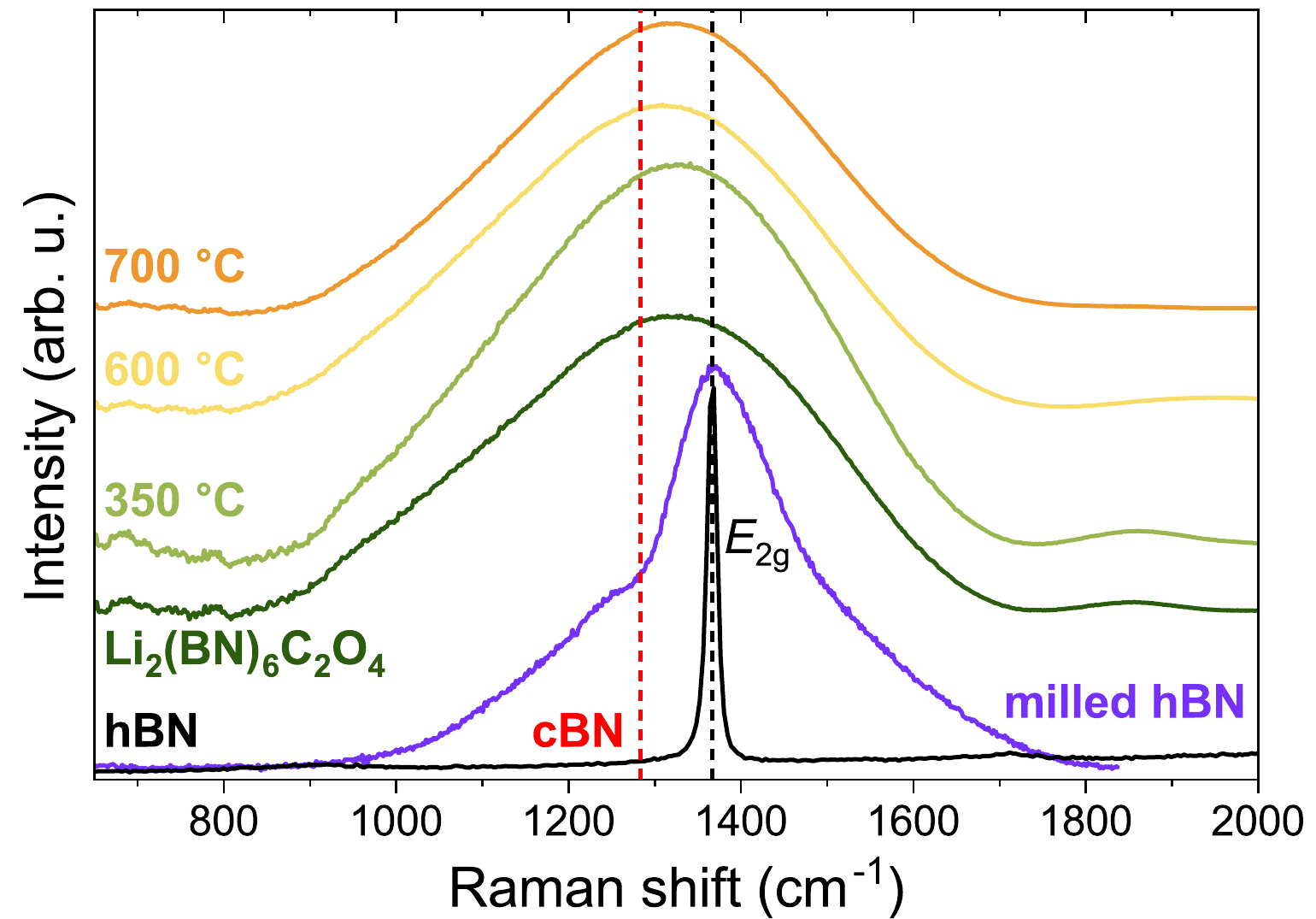}
    \caption{
        Raman spectra of hBN, LBNCO, and its $350$, $600$, and $700~^{\circ}$C calcined derivatives acquired with a $785$ nm laser excitation. Black and {\color{red}red} vertical dashed lines denote the $E_{2\text{g}}$ mode of hBN at $1367$ cm$^{-1}$ and the LO mode of cBN at $1283$ cm$^{-1}$, respectively. The luminescence backgrounds were subtracted from all spectra for visibility.}
    \label{FigSI:Raman}
\end{figure}

Figure \ref{FigSI:Raman} presents the Raman spectra of hBN, milled hBN, and BNCO samples at various calcination stages, recorded under $785$ nm excitation. High-energy ball milling resulted in spectral broadening due to the reduction in particle size. Following the mechanochemical reaction, the Raman spectra became even broader and nearly featureless, resembling the corresponding powder X-ray diffractograms.

\section{ESR linewidths} \label{sec:ESRlw}

\begin{figure}[!htp]
    \includegraphics[width=0.6\columnwidth]{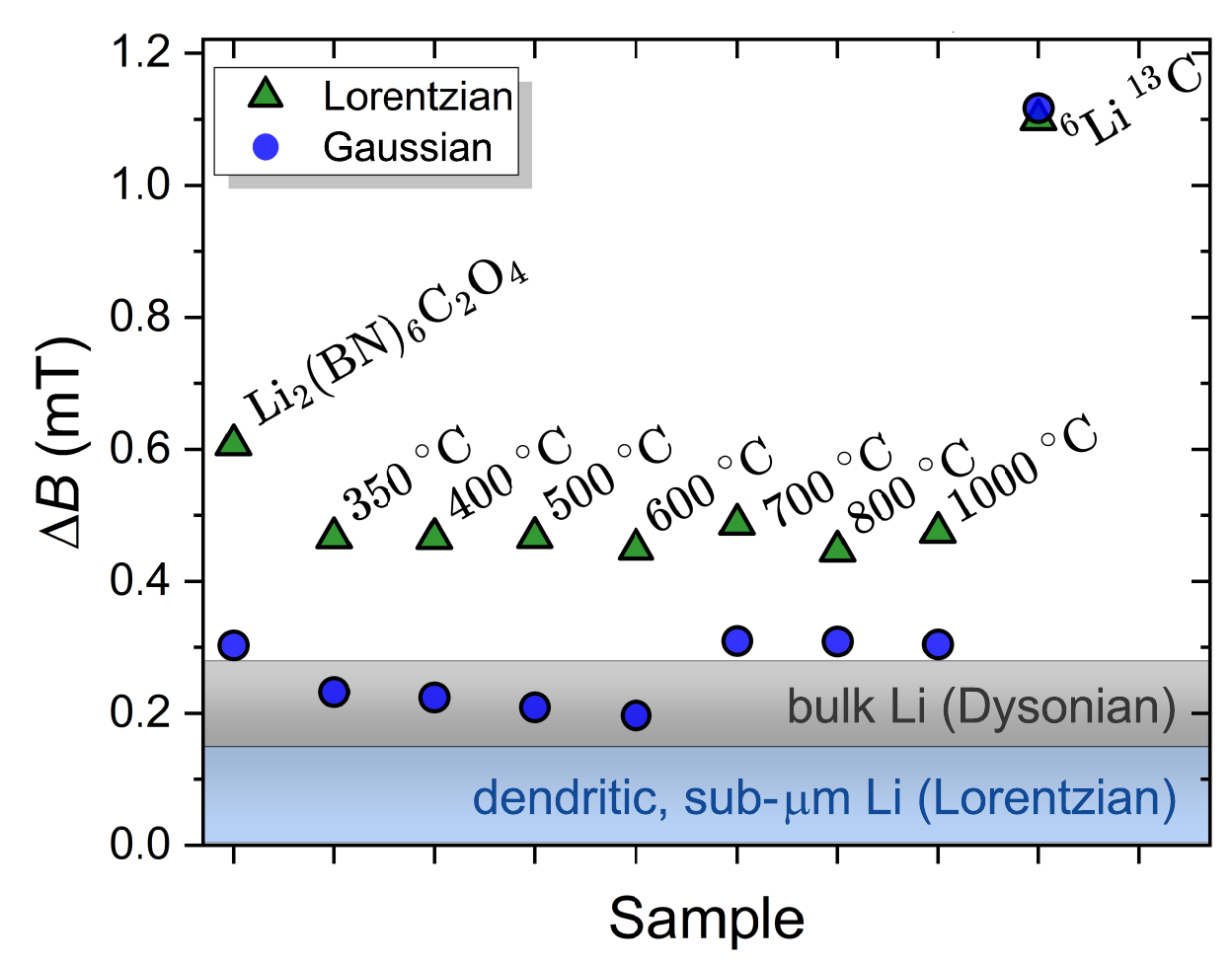}
    \caption{
        ESR linewidth (half-width at half-maximum) for LBNCO and its calcined derivatives. While the linewidths only change slightly by the thermal treatment, isotope substitution has a pronounced effect in broadening both components of the observed ESR line. This could be understood through the hyperfine coupling to $I=1/2$ $^{13}$C nuclei compared to the $I=0$ $^{12}$C atoms, and also a clear signature that the electron spins are close to the carbon atoms. The grey and blue colored areas represent the expected linewidth for metallic Li species with the corresponding lineshape in braces.}
    \label{FigSI:ESR-LW}
\end{figure}

The ESR linewidths, $\Delta B=\text{HWHM}$ (half-width at half-maximum) for LBNCO and its calcined derivatives are presented in Fig. \ref{FigSI:ESR-LW}. The expected ESR linewidths of metallic Li species are also indicated in the plot for comparison. Although the linewidth reported for bulk metallic Li is comparable in magnitude to the Gaussian linewidth observed in LBNCO, bulk Li typically exhibits a characteristic Dysonian lineshape arising from the skin effect in conducting metals; no such asymmetry is observed in the present spectra. Moreover, metallic Li (Li$^0$) can only be regarded as bulk-like from an ESR perspective when the characteristic particle dimension exceeds the microwave skin depth, which is approximately ${\sim}1.1-1.3~\mu$m at X-band frequencies. In contrast, both XRD and TEM analyses indicate that the largest characteristic particle dimensions in the present material are only on the order of several tens of nanometers. Dendritic and sub-$\mu$m metallic Li, on the other hand, typically produce a symmetric Lorentzian signal with a substantially narrower linewidth than that observed here. Thus, neither the lineshape nor the linewidth is consistent with an assignment to metallic Li \cite{Feher1955PR, Damay1976PRB, Niemoller2018SciRep, Dutoit2021NatCom, Kang2024ChemMater, Dutoit2025ChemMater}. Furthermore, alternative highly ionized Li species, such as Li$^{2+}$ or Li$^{3+}$, as well as molecular ions such as Li$_2^+$ and Li$_3^+$, are not expected to be thermodynamically stable under the present solid-state conditions. Their contribution to the observed ESR signal can therefore be reasonably excluded. Anionic Li$^-$ can likewise be excluded as the origin of the observed ESR signal, since its $1s^2 2s^2$ electron configuration is closed-shell and therefore ESR-silent \cite{evans1966introduction}. Given that the detected electron is residing on the carbon atom of the carboxyl group, Li$^+$ remains the only chemically plausible oxidation state of lithium in LBNCO under the present conditions. While calcination only results in subtle changes, isotope enrichment with $^{13}$C and $^{6}$Li significantly broadens both components, suggesting that both types of unpaired electrons are associated with carbon atoms, as $^{13}$C has a finite nuclear spin of $I=1/2$ and therefore, a finite hyperfine coupling arises, while $^{12}$C is a zero-spin nucleus. The resolution of individual hyperfine lines is prohibited by inhomogeneous broadening. This broadening is consistent with a mechanism involving fragmentation of the interlayer B(sp$^3$)$-$C(sp$^3$) and C(sp$^3$)$-$C(sp$^3$) bonds of the armchair edge (whose presence is indicated by XPS and $^{13}$C NMR measurements).

\section{Spin-label benchmark}

\begin{figure}[!hp]
    \includegraphics[width=0.5\columnwidth]{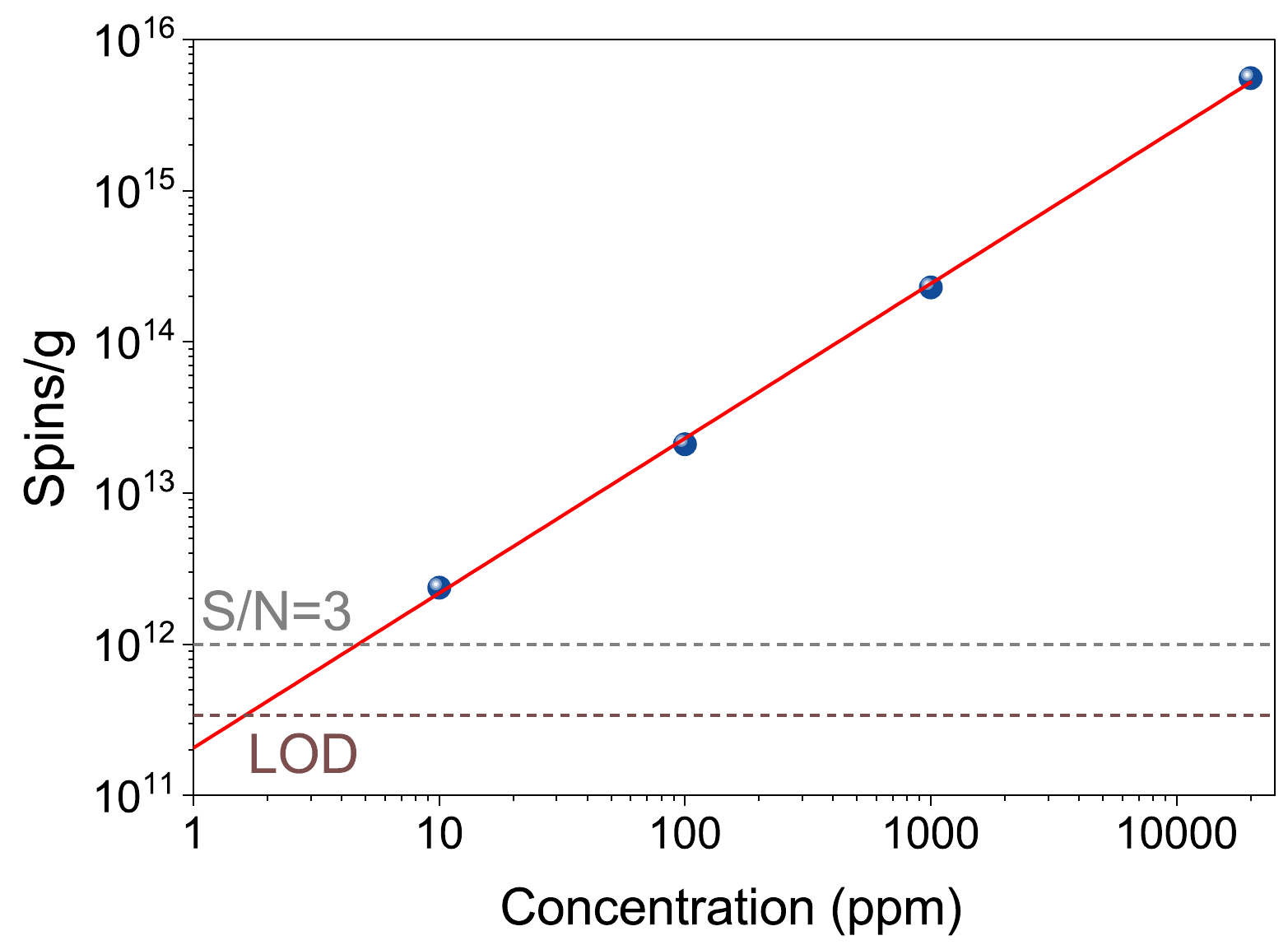}
    \caption{Spins/g calculated from fitted ESR intensity at various concentrations of LBNCO and Al$_2$O$_3$. A typical signal-to-noise (S/N) ratio of $3$ can be achieved at around $5$ ppm concentration, while the absolute limit of detection calculated for the Bruker spectrometer and the LBNCO linewidth is around $2$ ppm.}
    \label{FigSI:ESR-LOD}
\end{figure}

To benchmark the spin-labeling capabilities of {LBNCO}, we prepared diluted samples with concentrations of $20{,}000$, $1{,}000$, $100$, and $10$ ppm using ESR-silent Al$_2$O$_3$ as the diluent. The corresponding calculated spin mass densities are shown in Fig. \ref{FigSI:ESR-LOD}. Even at the lowest concentration of $10$ ppm, a signal-to-noise (S/N) ratio of approximately $4{:}1$ was obtained from an average of $128$ spectra. Linear extrapolation of the concentration-dependent S/N ratio yields an analytical detection threshold of $\textrm{S/N}=3$ at $5(1)$ ppm and an absolute limit of detection of $2(1)$ ppm. These detection limits were determined using the manufacturer-specified sensitivity of $1\times10^9$ spins/G for the Bruker Elexsys spectrometer equipped with an SHQE cavity.

These results demonstrate that LBNCO remains readily detectable at concentrations as low as $5$ ppm when dispersed in otherwise ESR-silent, low-microwave-loss host materials, for example polymers or inorganic oxides. Such low loading fractions enable robust ESR-based tracking while introducing only a negligible perturbation to the intrinsic physical properties of the host material.

\section{Electrochemical performance of the LBNCO/PVDF membrane}
\label{SecSI:electrochem}

\begin{figure}[!hp]
    \includegraphics[width=0.48\columnwidth]{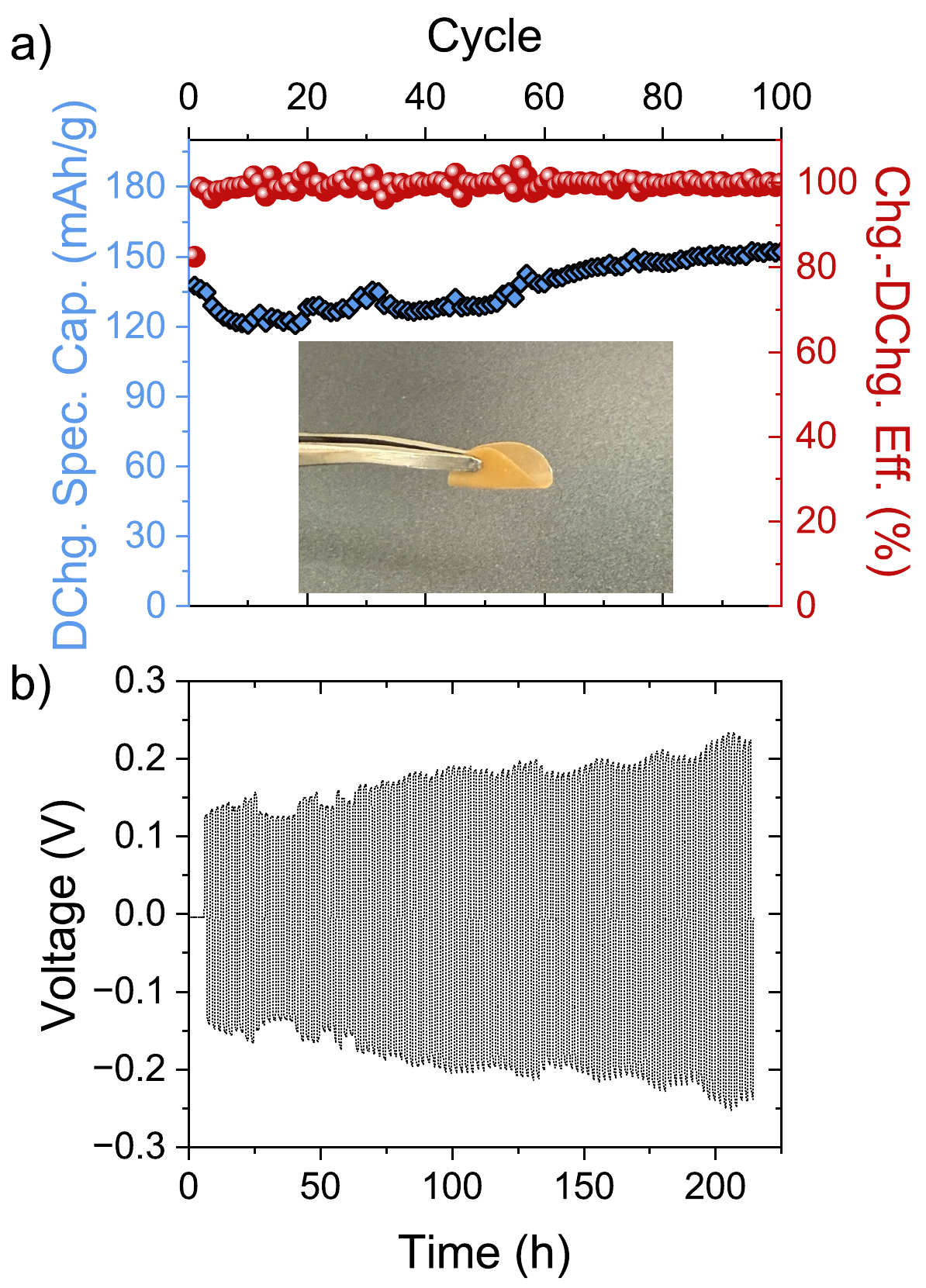}
    \caption{a) Cycling performance -- Discharge Specific Capacity (left) and Charge-Discharge Specific Efficiency (right) -- of the membrane in a LiFePO$_4$ (LFP)-Li cell over $100$ cycles. The increase in reversible capacity is attributed to the formation and stabilization of the solid electrolyte interphase (SEI) at the lithium-metal anode. b) Symmetric-cell measurements showing reduced interfacial resistance and polarization, consistent with improved Li-ion transport and more stable Li plating/stripping behavior.}
    \label{FigSI:cell_cycling}
\end{figure}

To assess the influence of LBNCO on the electrochemical performance of lithium-metal batteries under operating conditions, LBNCO was incorporated into polyvinylidene difluoride-based (PVDF) polymer solid electrolyte (PSE) membranes. The LBNCO/PVDF membranes were fabricated using a conventional slurry-casting method, with the membrane thickness controlled to approximately $48~\mu$m. Figure \ref{FigSI:cell_cycling}a shows the cycling performance of cells employing the membrane, an LiFePO$_4$ (LFP) cathode, and a lithium-metal anode over $100$ cycles. The LFP cathode had an active-material loading of $2.25$~mg/cm$^2$. Electrochemical cycling was performed in 2032-type coin cells at a current rate of $0.5$~C within a voltage window of $2.8-4.0$~V. The cells show a gradual increase in reversible capacity upon cycling, consistent with progressive stabilization of the electrode-electrolyte interface and the formation of a stable solid electrolyte interphase (SEI) at the lithium-metal anode.

The symmetric-cell measurements (the membrane sandwiched between Li-disks; current density of $0.5$ mA/cm$^{2}$) shown in Fig. \ref{FigSI:cell_cycling}b further indicate that incorporation of LBNCO reduces interfacial resistance and polarization, consistent with improved Li-ion transport and more stable lithium plating/stripping behavior. These observations provide preliminary evidence that functionalized hBN species, such as LBNCO, may help suppress nonuniform lithium deposition and dendrite formation. A comprehensive investigation of the electrochemical performance, interfacial processes, and underlying transport mechanisms is beyond the scope of the present work and will be reported separately in a forthcoming study.


\section{Reactivity of LBNCO radicals}

\begin{figure}[!htp]
    \includegraphics[width=0.95\columnwidth]{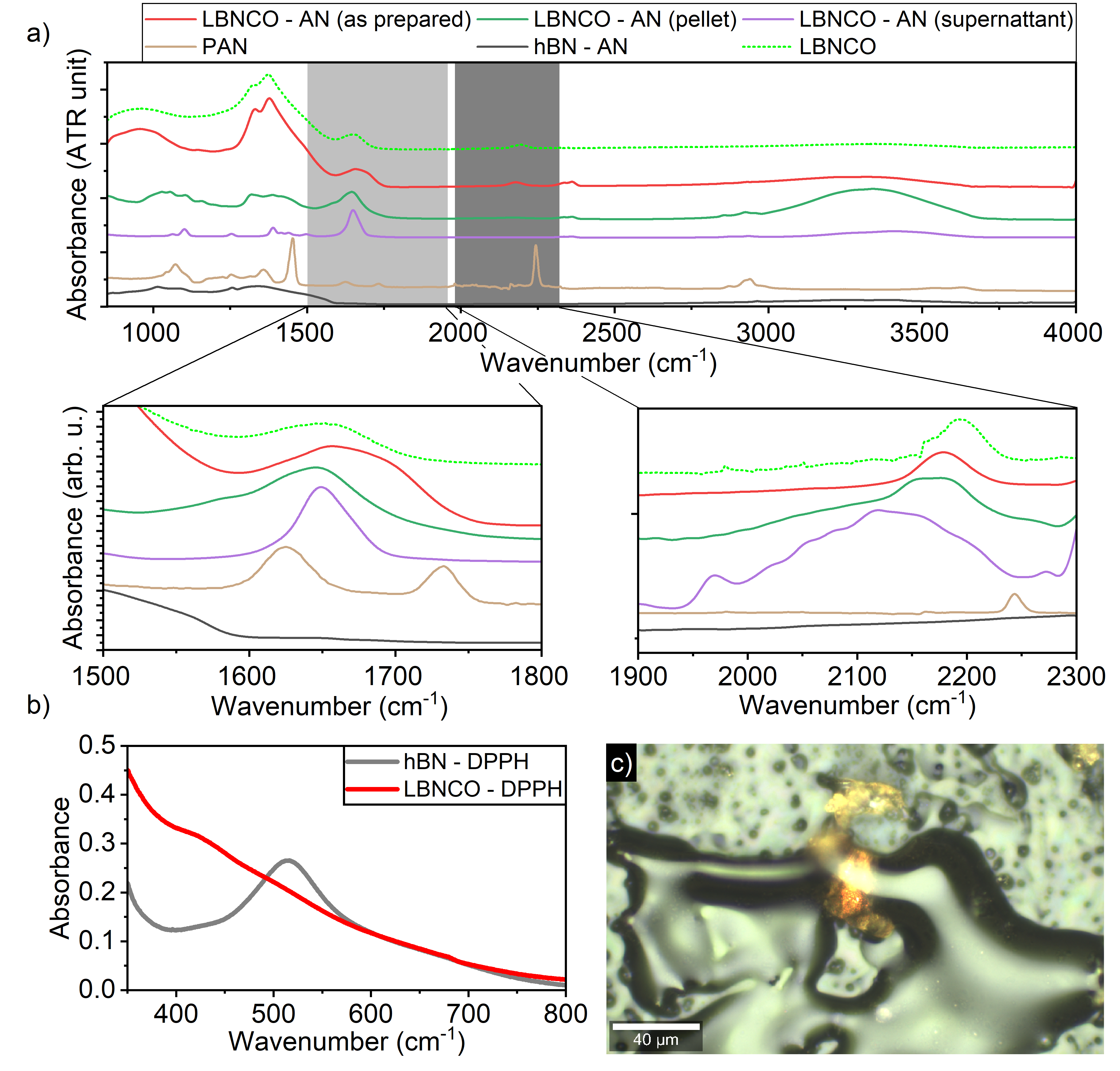}
    \caption{a) FTIR spectra of PAN, LBNCO, the products of the LBNCO-AN reaction (as prepared, pellet and supernatant) and the hBN-AN control; the spectra are scaled for comparison. The two panels on the right magnify the regions in which the samples differ. In the nitrile stretching region, the LBNCO band is downshifted and broadened, and a new band appears at $2270$~cm$^{-1}$ after the reaction. In the $1500-1800$~cm$^{-1}$ region, the LBNCO-AN spectrum is broadened towards higher wavenumbers, the pellet is essentially unchanged, the supernatant shows a sharp band attributed to residual solvent, PAN shows two bands, and the hBN-AN control shows no feature. b) UV-Vis absorption spectra of DPPH in absolute ethanol after contact with LBNCO and with milled hBN. c) Optical micrograph of the LBNCO-AN product: yellow LBNCO particles surrounded by a transparent phase.}
    \label{FigSI:PAN}
\end{figure}

The reactivity of LBNCO radicals was probed in two preliminary experiments (Fig.~\ref{FigSI:PAN}). Refluxing LBNCO with acrylonitrile (AN) in isopropanol produced a mass gain of about $13$ mg. In contrast, hBN milled for $10$~h under otherwise identical conditions showed no measurable mass gain, indicating that the uptake requires the functionalized material. EPR spectra recorded before and after the reaction showed a change in the relative weights of the overlapping Gaussian and Lorentzian components, indicating that the paramagnetic centers themselves are modified upon contact with the monomer.

The FTIR data are consistent with AN attachment at the surface rather than bulk polymer formation. The characteristic nitrile band of AN and PAN is not observed at its usual position near $2240$~cm$^{-1}$; instead, the nitrile band of LBNCO is downshifted and broadened, and a new band appears at $2270$~cm$^{-1}$ after the reaction (Fig.~\ref{FigSI:PAN}a). Coordination of nitrile groups to Li$^+$ shifts $\nu$(C$\equiv$N) to higher wavenumbers by $15$--$30$~cm$^{-1}$ \cite{wang1996position, sim2017development}, and the observed position matches this range; therefore, this new band is assigned to nitrile groups bound to lithium at the LBNCO surface. Together with the low mass gain, this suggests that AN reacts with the surface radical sites, but that chain propagation remains limited. Optical microscopy of the product shows the yellow LBNCO particles surrounded by a transparent phase (Fig.~\ref{FigSI:PAN}c). We note that identification is hampered by the overlap of the product bands with those of LBNCO and of residual solvent, and that the sample was dried at elevated temperature, which may have altered any polymer formed; these results are therefore preliminary, and a detailed study of the polymerization behavior is left to future work.

LBNCO was also suspended in an ethanolic solution of the stable radical DPPH (2,2-diphenyl-1-picrylhydrazyl). The suspension turned amber within $30$~min and the DPPH absorption band near $510$~nm disappeared, while a weak feature appeared near $420$~nm (Fig.~\ref{FigSI:PAN}b), as expected for reduction of DPPH$\bullet$ to the corresponding hydrazine. When milled hBN was used, no change in the absorption spectrum of DPPH was detected.

Taken together, these experiments demonstrate that the radical centers of LBNCO are stable but chemically accessible and reactive towards both a vinyl monomer and a stable radical. Limited propagation might be improved with reaction condition optimization. However, a filler that reacts with monomer at its surface without polymerizing independently can be incorporated during polymerization and become covalently integrated into the matrix, which is the behavior required both for polymer membranes in batteries and for spin labeling. A quantitative assessment of the reaction efficiency and the fate of the radical centers is beyond the scope of the present work.

\clearpage

\section{Thermal decomposition} \label{sec:xps-tgams}

\begin{figure*}[htp]
    \includegraphics[width=0.8\textwidth]{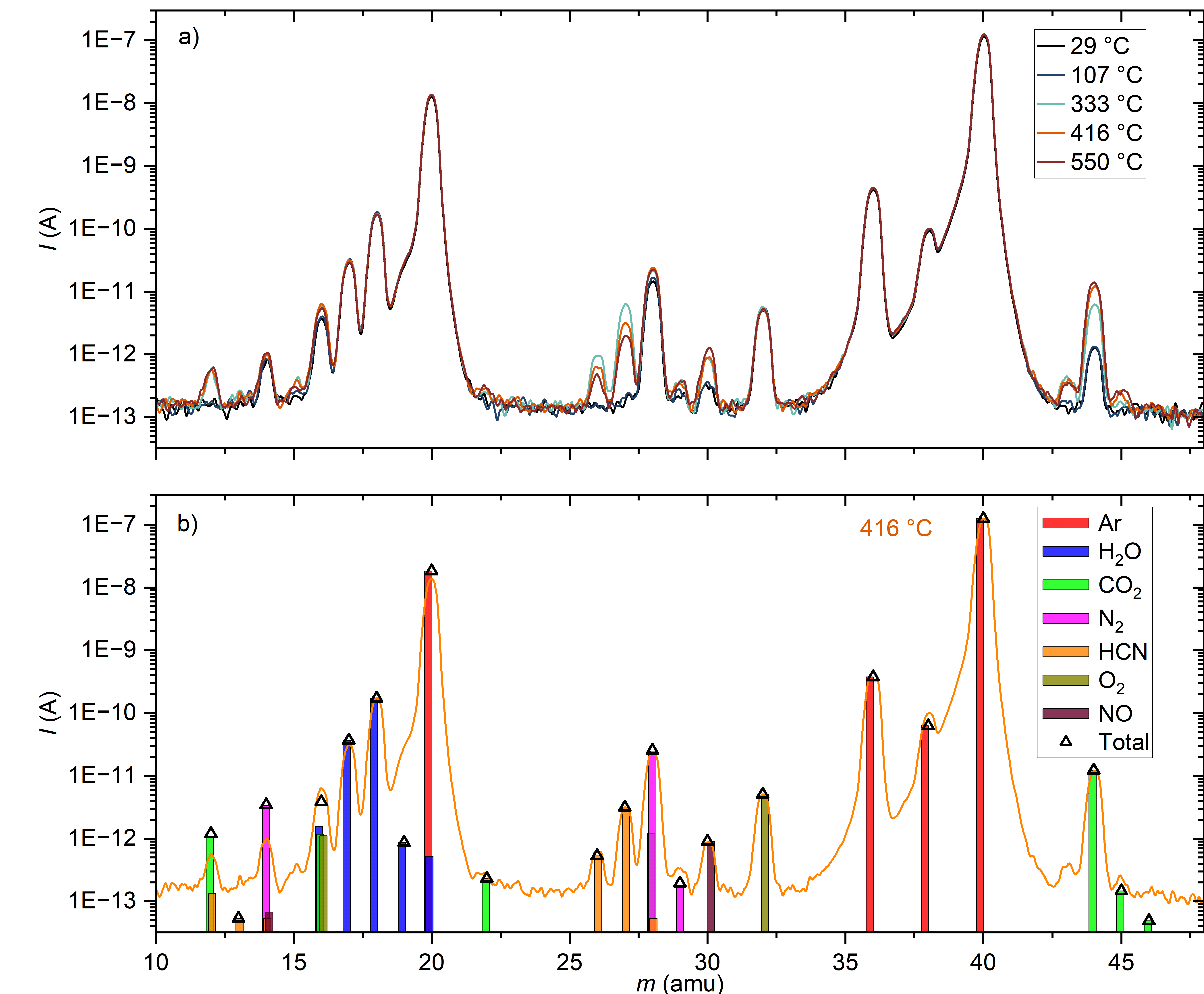}
    \caption{Detailed analysis of the mass spectrometry results. a) Mass spectra obtained during the TGA measurement. Spectra measured at the initial and the mass-loss peak temperatures indicated in Fig.~2a. b) Mass spectrum recorded at the mass-loss peak at $416~^\circ$C. Colored peaks are identified as indicated in the legend.}
    \label{FigSI:Fig_MS_peak_analysis}
\end{figure*}

Fig.~\ref{FigSI:Fig_MS_peak_analysis} depicts the in-depth analysis of the mass spectroscopy measurement presented in Fig.~2a and discussed in the main text. Here, the Ar carrier gas produces several ion current peaks that remain constant throughout the entire temperature ramp. In contrast, other peaks exhibit temperature-dependent behavior, resulting in the mass loss of the material. 

Fig.~\ref{FigSI:Fig_MS_peak_analysis}b depicts the mass spectrum (orange line) measured at $416~^\circ$C. The colored peaks are assigned to elements as indicated in the legend based on mass spectra from the National Institute of Standards and Technology's Mass Spectrometry Data Center (NIST MSDC).

\clearpage

\begin{figure}[!htp]
    \includegraphics[width=0.7\linewidth]{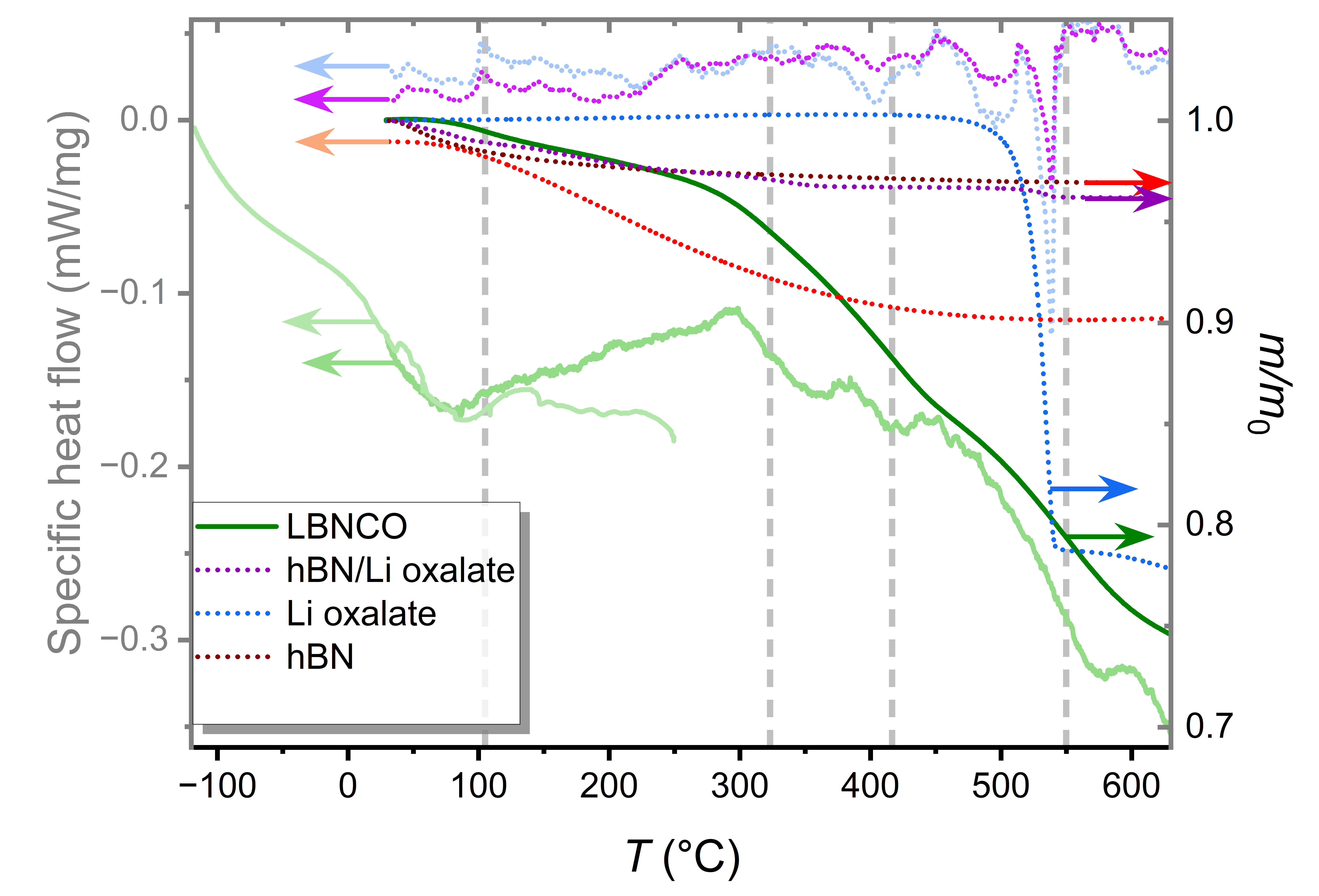}
    \caption{Comparison of TGA mass-change from the upper panel of Fig. 2a (olive line), the TGA data of the reactants: hBN milled under identical conditions (red line), Li oxalate (blue line), and the reaction mixture of the components (purple line), together with the corresponding DSC data (plotted with lighter colors). DSC measurement of LBNCO was also recorded with a higher-precision device. DSC data are vertically shifted for clarity. Vertical dashed lines are transitions identified in TGA and MS measurements. The DSC measurements reveal no sharp phase transitions in the synthesized material. Milled hBN showed marginal changes. Li oxalate showed decomposition at around $540~^{\circ}$C.
    \label{FigSI:DSC} 
    }
\end{figure}

Fig.~\ref{FigSI:DSC} presents the DSC measurement (green line) corresponding to the TGA and MS measurements shown in Fig.~2a of the main article. An additional DSC curve is displayed as a comparison, which was obtained with a Mettler Toledo DSC 1 instrument that has a higher sensitivity to small changes in heat flow. The high-precision measurement was taken while heating approximately $9$~mg of LBNCO from $-120$ to $250~^{\circ}$C with a sweep rate of $5~^\circ$C/min. The carrier gas of N$_2$ flow was set to $200$ mL/min. The two curves are matched at room temperature for clarity. The mass-loss events identified from the mass change peaks in Fig.~2a are depicted by vertical dashed lines. The specific heat flow exhibits small changes over the investigated temperature ranges. Meanwhile, the temperature of the main mass-loss events shows negligible temperatures of the main mass-loss events show negligible correlation, indicating no sharp phase transitions in the material.

\section{TEM analysis}

\begin{figure}[htp]
    \includegraphics[width=\linewidth]{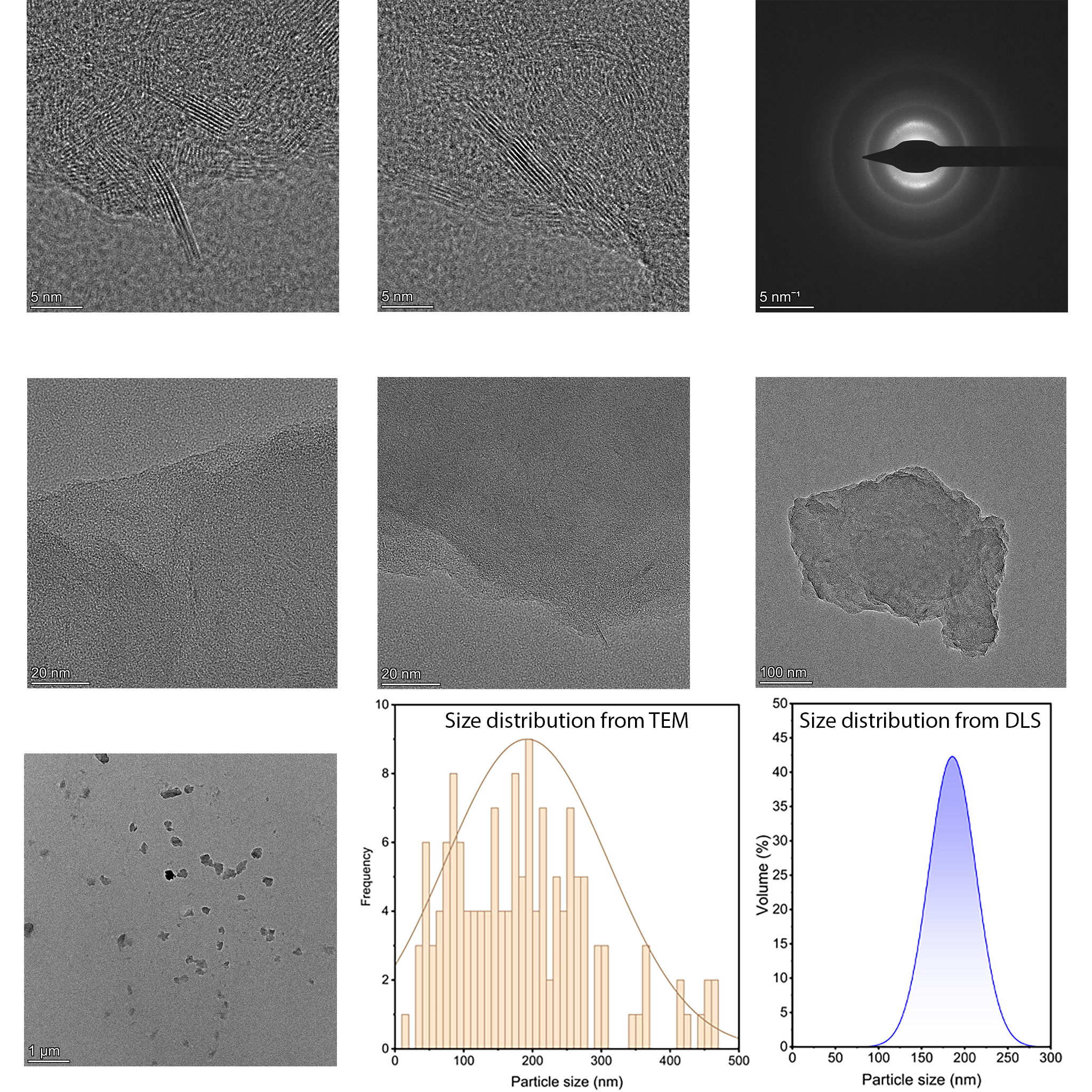}
    \caption{HR-TEM images, diffraction, and size distribution of LBNCO.
    \label{fig:TEM} 
    }
\end{figure}

Additional HR-TEM images acquired from the LBNCO material are shown in Fig. \ref{fig:TEM}. Size distributions obtained from electron microscopy image analysis and DLS experiments show a relatively good agreement and yield a mean particle size of ${\sim}200$~nm.

\clearpage

\section{Optical properties}

\begin{figure}[!htp]
    \includegraphics[width=0.6\linewidth]{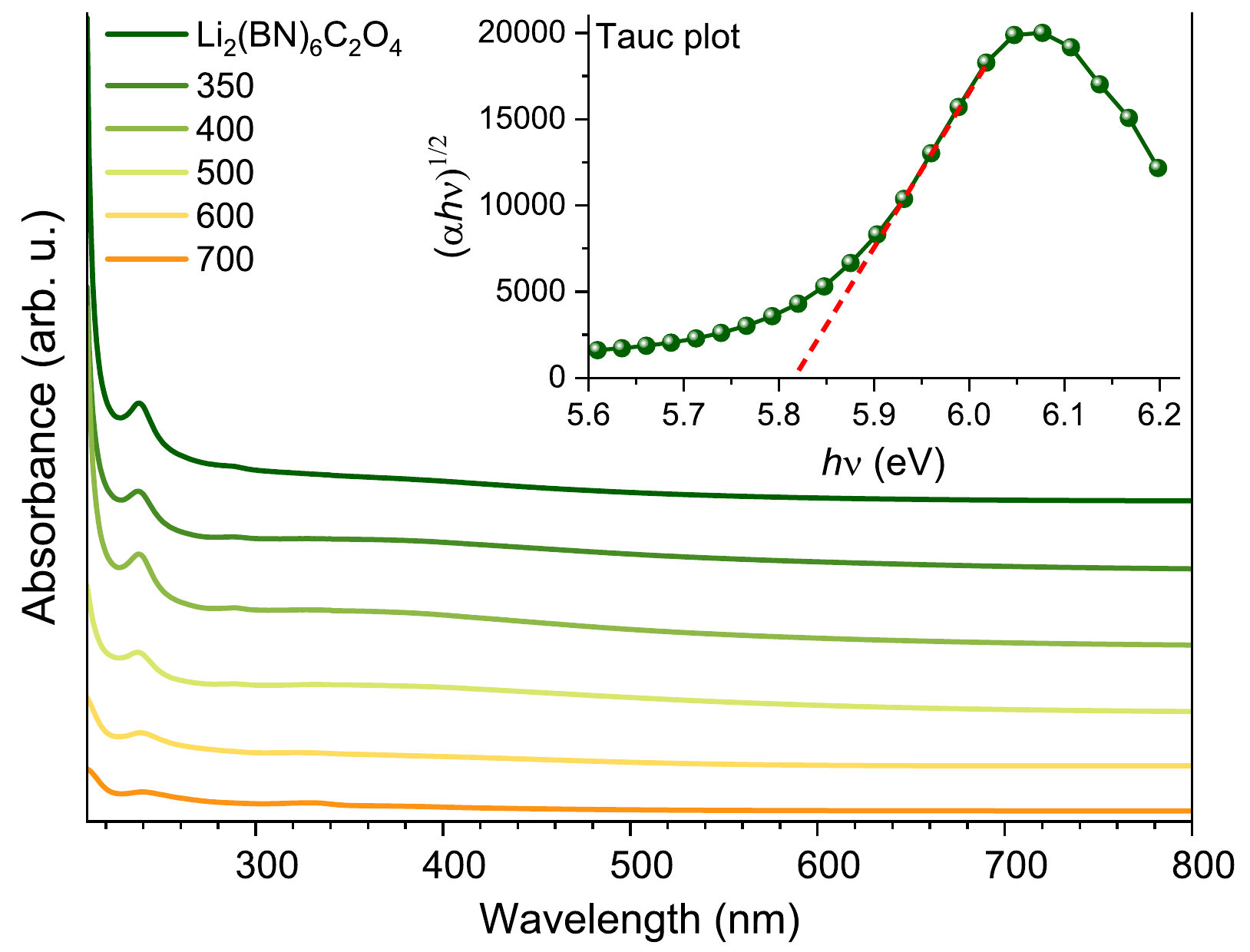}
    \caption{UV-Visible absorption spectra of LBNCO and its calcined derivatives. Spectra are vertically offset for clarity. The inset shows the calculated Tauc plot for LBNCO.
    \label{fig:abs} 
    }
\end{figure}

UV-Visible absorption spectra of LBNCO and its calcined derivatives are presented in Fig. \ref{fig:abs}. Using the Tauc plot, the bulk bandgap is estimated to be in the order of $5.82-6.06$ eV.

\begin{figure}[!htp]
    \includegraphics[width=0.6\linewidth]{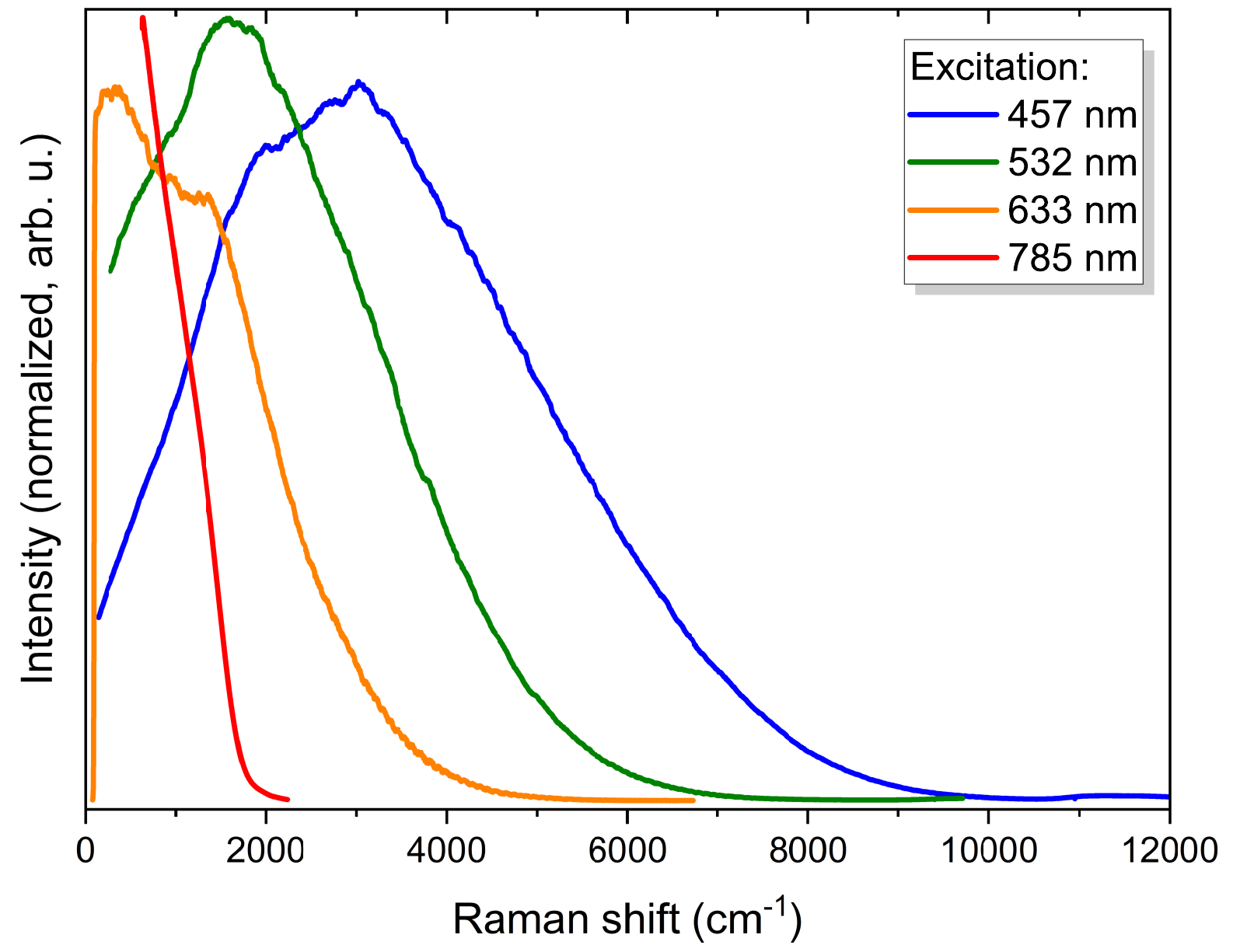}
    \caption{Photoluminescence spectra of LBNCO under different excitation wavelengths recorded with confocal Raman microscopes. The Raman shift is measured relative to the excitation wavelengths.
    \label{fig:PL} 
    }
\end{figure}

Figure \ref{fig:PL} shows the photoluminescence spectrum of LBNCO under $457$, $532$, $633$, and $785$ nm laser excitation wavelengths obtained with Raman spectrometers.

\begin{figure}[!htp]
    \includegraphics[width=0.6\linewidth]{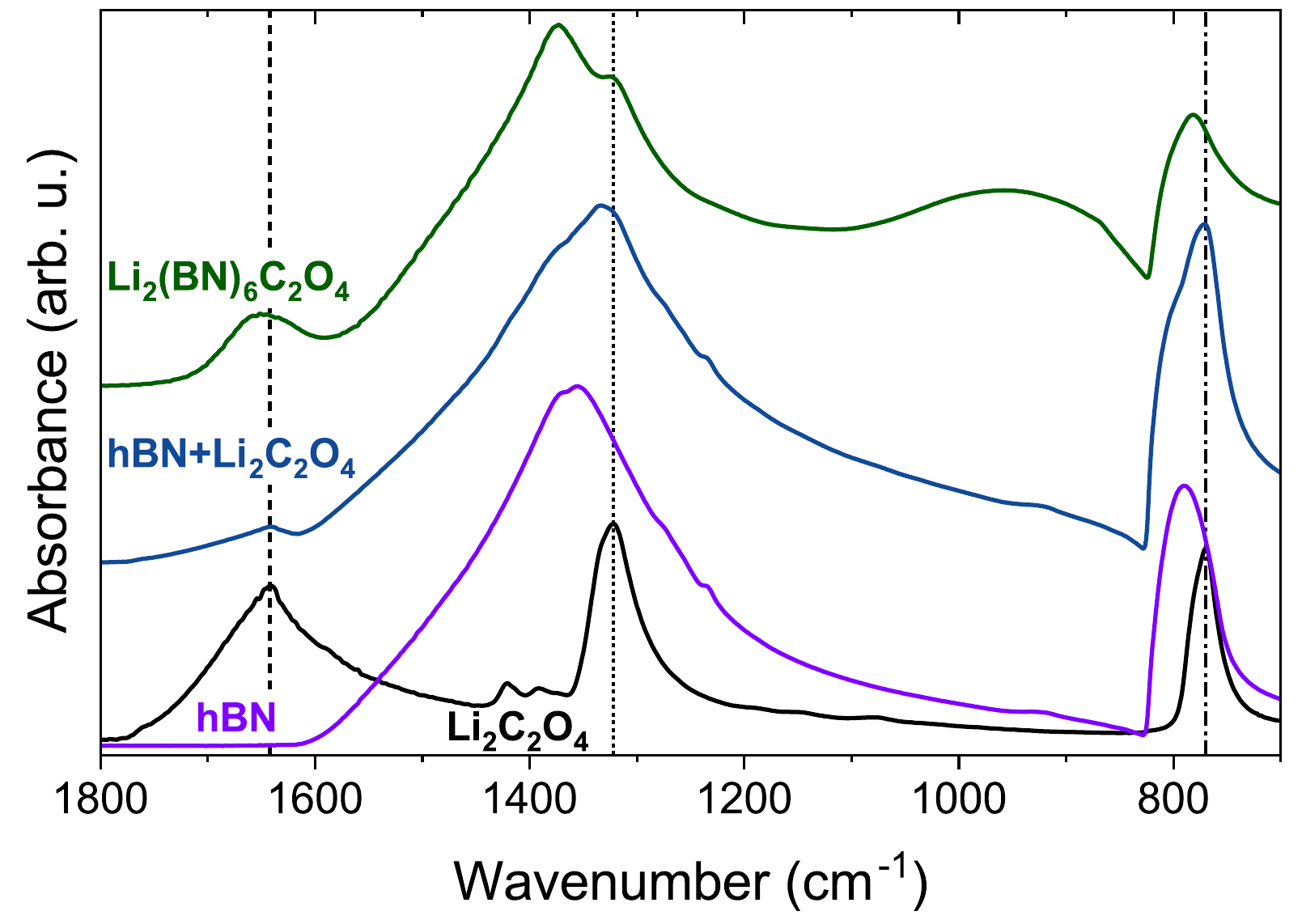}
    \caption{Comparative FTIR absorption spectra of pure Li$_2$C$_2$O$_4$, hBN, the mixture of these two (hBN+Li$_2$C$_2$O$_4$) and the reaction product LBNCO. Vertical dashed lines are vibrational modes of the lithium oxalate. The differences in peak positions and shapes in the LBNCO and the as-mixed samples suggest that nearly all of the Li$_2$C$_2$O$_4$ was consumed during the reaction.
    \label{figSI:IRCompare} 
    }
\end{figure}

A comparison between a physical mixture of hBN and Li$_2$C$_2$O$_4$ and the reacted LBNCO is displayed in Fig. \ref{figSI:IRCompare}. For the mix, the materials were gently ground together in a mortar. Please note the difference between the obtained new material and the starting materials in terms of peak positions and lineshapes.

\section{NMR parameters calculated using DFT} \label{NMRSupMat}

The NMR parameters calculated using DFT with norm-conserving and ultrasoft pseudopotentials are summarized in Tables \ref{NMRcalcNC} and \ref{NMRcalcUS}, respectively. Experimental data of hBN and cBN are from Ref. \cite{jeschke1998comprehensive}. $\delta_{\rm iso}$ and $\delta_{\rm CSA}$ are the isotropic chemical shift and the chemical shift anisotropy, respectively. $C_{\rm Q}$ is the quadrupole coupling and $\eta_{\rm Q}$ the anisotropy parameter.

The $^{11}$B and $^{14}$N isotropic chemical shifts were referenced to that of cBN. $^{13}$C and $^{7}$Li chemical shifts were referenced to $\gamma$-glycine and Li$_2$C$_2$O$_4$, respectively. The referencing was carried out according to $\delta_{\rm iso, cal}(i)=\delta_{\rm iso,ref}(t_{i})+[\sigma_{\rm iso,ref}(t_{i})-\sigma_{\rm iso,cal}(i)]$ where $\delta_{\rm iso, cal}(i)$, $\delta_{\rm iso,ref}(t_{i})$, $\sigma_{\rm iso,ref}(t_{i})$ and $\sigma_{\rm iso,cal}(i)$ are the calculated isotropic chemical shift of atom $i$, the reference experimental chemical shift of the type of atom $i$ ($t_i$), and the calculated isotropic shielding for the type of atom $i$ in the reference and in the model structure, respectively.

On top of using cBN as a reference for the $^{11}$B chemical shifts, an additional uniform shift of $-8.8$ and $-14.3$ ppm was applied to the $^{11}$B spectra of the zigzag model as calculated using norm-conserving and ultrasoft pseudopotentials, respectively. This additional uniform correction had to be applied to achieve the best agreement between the two computational models and with experimental observations. As the experimental spectrum indicates only one peak for B(sp$^{3}$) atoms, the additional uniform correction for the $^{11}$B chemical shifts of the zigzag model was as large as what was needed to achieve identical chemical shifts for the B(sp$^{3}$) atoms of the zigzag and armchair(B$-$C) models. After this additional correction, the $^{11}$B chemical shifts agreed within $2$ ppm between the two computational methods for the zigzag model. In general, a uniform shift of the calculated spectrum is allowed on the basis of the empirical observation that it accounts for the errors in the model. This correction may be different in the case of the armchair and zigzag models.

The calculated $^{14}$N chemical shift of hBN was greater by about $10$ ppm than the experimental value (Tables \ref{NMRcalcNC} and \ref{NMRcalcUS}). The calculated chemical shifts from both computational models agreed within $2$ ppm for the armchair models, while they differed more for the zigzag model. As no N(sp$^{3}$) was experimentally observed (and none present in the models), a similar correction of the calculated $^{14}$N chemical shifts as in the case of $^{11}$B was not possible. Instead, a uniform, $-7.0$ ppm shift was applied to the ultrasoft spectrum, leading to a good (within $2$ ppm and for one atom in $3$ ppm) agreement with the norm-conserving one. Four N atoms of the zigzag model thus have similar calculated chemical shifts as in the armchair models, and the other two zigzag nitrogens appear to be extremely outlying. These are the two different kinds of outermost N-s, one in the B-terminated and one in the N-terminated edge of the zigzag model. These latter two N-s were not detected experimentally, though, for a yet unknown reason.

It is possible that the calculated $^{14}$N spectra are more sensitive to the accuracy of the type of density functional than the $^{11}$B ones. The density functional applied here is unable to reproduce the correct layer distance of hBN, and more accurate functionals (such as SCAN \cite{sun2015strongly}) are not available for NMR calculations yet. 

The calculated $^{13}$C chemical shift of Li$_{2}$C$_{2}$O$_{4}$ is in good agreement (within $1.2$ ppm) with the measured one, while that of LiBOB shows a greater difference ($7-17$ ppm), when using the same (C(sp$^{2}$) $\gamma$-glycine) reference. This latter difference also indicates the need for different references for the two types of oxalates (free and chelate), similarly to the above-described $^{11}$B spectra. Note that LiBOB showed three slightly different shifts due to small asymmetries in the relaxed structure. 

\begin{table}[t] 
{\scriptsize
     \caption{Calculated and experimental NMR spectrum parameters. Calculations used norm-conserving pseudopotentials.}
     \label{NMRcalcNC}
     \begin{tabular}{|c|c|c|c|c|c|c|c|c|c|}
     \hline   
           Nucleus & Sample & \multicolumn{2}{|c|}{$\delta_{\rm iso}$ (ppm)} & \multicolumn{2}{|c|}{$\delta_{\rm CSA}$ (ppm)} & \multicolumn{2}{|c|}{$C_{\rm Q}$ (kHz)} & \multicolumn{2}{|c|}{$\eta_{\rm Q}$}   \\
           \cline{3-10}
                   &     & calc & expt & calc & expt & calc & expt & calc & expt  \\
          \hline \hline
                   & hBN &  29.8 &30.4 & 29.9 & - & 3510 &  2936 & 0.00 & - \\
                   & cBN &   1.6 & 1.6 & 0    & 0 &   5  & $<$50 & 1.00 & - \\
                   &LiBOB& 10.3  & 7.3 &-1.7  & - & -623 & -     & 0.86 & - \\
                   \cline{2-10}
                   & armchair(B$-$C) 
                   & 26.7 & - & 26.5 & - & 3229 & - & 0.29 & - \\ 
                 & & 29.0 & - & 27.8 & - & 3439 & - & 0.06 & - \\ 
                 & & 27.7 & - & 29.0 & - & 3473 & - & 0.01 & -  \\ 
                 & & 27.7 & - & 28.1 & - & 3358 & - & 0.06 & - \\ 
                 & & 27.8 & - &-26.5 & - & 3263 & - & 0.31 & - \\  
                 & &  3.2 & - & -4.7 & - & 1551 & - & 0.94 & - \\ 
                   \cline{2-10}
         $^{11}$B  & armchair(C$-$C) 
                   & 25.4 & - & 17.7 & - & 3144 & - & 0.33 & - \\ 
                 & & 27.7 & - & 28.0 & - & 3365 & - & 0.04 & -  \\ 
                 & & 27.4 & - & 29.1 & - & 3464 & - & 0.01 & - \\ 
                 & & 26.2 & - & 17.4 & - & 3115 & - & 0.31 & - \\  
                 & & 29.0 & - & 27.9 & - & 3334 & - & 0.03 & - \\ 
                 & & 28.0 & - & 29.1 & - & 3411 & - & 0.01 & - \\ 
                   \cline{2-10}
                   & zigzag 
                   & 24.0 & - & 39.2 & - & 3264 & - & 0.13 & - \\ 
                 & & 18.5 & - & 37.0 & - & 3400 & - & 0.00 & - \\ 
                 & & 14.7 & - & 42.1 & - & 3402 & - & 0.06 & - \\ 
                 & & 14.0 & - & 40.8 & - & 3396 & - & 0.09 & - \\ 
                 & & 16.2 & - & 34.2 & - & 3306 & - & 0.04 & - \\ 
                 & &  3.2 & - &-21.0 & - &-2796 & - & 0.53 & - \\ 
          \hline      \hline   
                   & hBN &  71.2 &  61.2 & 152 & 160 &  -66  &  140  & 0.00 & - \\ 
                   & cBN & -17.6 & -17.6 & 0   & 0   &   -5  &    0  & 0.03 & - \\
                   \cline{2-10}
                   & armchair(B$-$C) 
                   & 86.0 & - & 76.7 & - &2252  & - &0.73 & - \\ 
                 & & 69.7 & - &146.2 & - &-225  & - &0.89 & - \\ 
                 & & 71.9 & - &149.7 & - &-135  & - &0.68 & - \\ 
                 & & 63.2 & - &120.2 & - &-803  & - &0.39 & - \\ 
                 & & 74.5 & - &149.6 & - & 238  & - &0.87 & - \\ 
                 & & 91.9 & - & 96.7 & - &2381  & - &0.82 & - \\ 
                   \cline{2-10}
        $^{14}$N   & armchair(C$-$C) 
                   &84.4  & - & 78.2  & - &2482  & - &0.76 & - \\ 
                 & &66.0  & - &144.4  & - & 228  & - &0.91 & - \\ 
                 & &70.8  & - &149.7  & - & 135  & - &0.66 & - \\ 
                 & &70.2  & - &143.4  & - &-342  & - &0.06 & - \\ 
                 & &74.1  & - &147.7  & - & 255  & - &0.62 & - \\ 
                 & &85.6  & - & 77.4  & - &2581  & - &0.77 & - \\ 
                   \cline{2-10}
                   & zigzag 
                   &261.5 & - &343.9  & - &3024  & - &0.94  & - \\ 
                 & & 86.3 & - &170.2  & - &-396  & - &0.62  & - \\ 
                 & & 74.7 & - &165.3  & - &-255  & - &0.25  & - \\ 
                 & & 70.5 & - &159.5  & - & 209  & - &0.64  & - \\ 
                 & & 66.2 & - &160.8  & - & 308  & - &0.55  & - \\ 
                 & & 18.3 & - & 32.2  & - &-1491 & - &0.28  & - \\ 
          \hline   \hline  
& Li$_{2}$C$_{2}$O$_{4}$ &171.9 &170.7 & -75.7 & -81.7 & 0 & - & 0.17  & - \\ 
& LiBOB          &167.7 & 158.4 &-81.7 & -     & 0 & - & 0.98& - \\
&                &173.5 & 158.4 &-83.5 & -     & 0 & - & 0.73& - \\
&                &175.1 & 158.4 &-76.4 & -     & 0 & - & 0.70& - \\
                   \cline{2-10}    
                   &  armchair(B$-$C)
                   & 166.3 & 168.4 & 56.2 & - & 0 & - & 0.85 & - \\  
                 & & 129.0 & 127.0 & -27.8 & - & 0 & - & 0.71 & - \\ 
                   \cline{2-10}
        $^{13}$C   &  armchair(C$-$C) 
                   & 124.5 & 127.0 & 20.3 & - & 0 & - & 0.87 & - \\  
                 & & 122.0 & 127.0 & 22.6 & - & 0 & - & 0.74 & - \\  
                   \cline{2-10}
                   &  zigzag 
                   & 165.8 & 168.4 & 61.2 & - & 0 & - & 0.56 & - \\  
                 & & 165.1 & 168.4 & 60.6 & - & 0 & - & 0.56 & - \\  
          \hline \hline
& Li$_{2}$C$_{2}$O$_{4}$ & 0.6 & 1.1 &-3.1 & - & -185 & 96 & 0.66 & 0.56 \\ 
& LiBOB                  & 0.2 & 0.5 & 6.1 & - & 206  & - & 0.63 & - \\
                   \cline{2-10}
                   & armchair(B$-$C) 
                   & 0.7 & - & 5.0 & - & 335 & - & 0.90 & - \\ 
                 & &-0.2 & - &-3.5 & - & 238 & - & 0.84 & - \\ 
                   \cline{2-10}
      $^{7}$Li     & armchair(C$-$C) 
                   & 2.5 & - & 4.7 & - &-173 & - & 0.88 & - \\ 
                 & & 0.2 & - & 2.6 & - & 158 & - & 0.63 & - \\ 
                   \cline{2-10}
                   & zigzag 
                   & 0.5 & - & -2.5 & - &-203 & - & 0.04 & - \\
                 & &-1.0 & - &-22.1 & - &-459 & - & 0.52 & - \\
          \hline
     \end{tabular}
 }
 \end{table}

\begin{table}[t]
{\scriptsize
     \caption{Calculated and experimental NMR spectrum parameters. Calculations used ultrasoft pseudopotentials.}
     \label{NMRcalcUS}
     \begin{tabular}{|c|c|c|c|c|c|c|c|c|c|}
     \hline   
           Nucleus & Sample & \multicolumn{2}{|c|}{$\delta_{\rm iso}$ (ppm)} & \multicolumn{2}{|c|}{$\delta_{\rm CSA}$ (ppm)} & \multicolumn{2}{|c|}{$C_{\rm Q}$ (kHz)} & \multicolumn{2}{|c|}{$\eta_{\rm Q}$}   \\
           \cline{3-10}
                   &     & calc & expt & calc & expt & calc & expt & calc & expt  \\
          \hline \hline
                   & hBN & 29.9   &30.4 & 29.8 & - & 3235 &  2936 & 0.0  & - \\
                   & cBN &  1.6   & 1.6 &  0.0 & 0 &    5 & $<$50 & 0.95 & - \\ 
                   &LiBOB& 10.3   & 7.3 &-1.2  & - &-596  & -     & 0.72 & - \\
                   \cline{2-10}
                   & armchair(B$-$C)    
                   &26.9 & - &17.9  & - &2893  & - &0.39  & - \\ 
                 & &29.4 & - &27.5  & - &3157  & - &0.06  & - \\ 
                 & &27.9 & - &28.6  & - &3188  & - &0.02  & -  \\ 
                 & &28.0 & - &27.7  & - &3081  & - &0.05  & - \\ 
                 & &28.1 & - &-26.2 & - &2997  & - &0.32  & - \\  
                 & & 2.7 & - &-4.8  & - &1664  & - &0.93  & - \\ 
                   \cline{2-10}
     $^{11}$B      & armchair(C$-$C)   
                   &25.8  & - &16.6  & - &2810  & - &0.43  & - \\ 
                 & &28.7  & - &26.9  & - &3065  & - &0.02  & -  \\ 
                 & &27.9  & - &28.0  & - &3171  & - &0.01  & - \\ 
                 & &26.4  & - &16.9  & - &2821  & - &0.38  & - \\  
                 & &29.4  & - &27.1  & - &3060  & - &0.02  & - \\ 
                 & &28.4  & - &28.1  & - &3140  & - &0.03  & - \\ 
                   \cline{2-10}
                   & zigzag
                   &24.5  & - &23.4  & - &3006  & - &0.15  & - \\ 
                 & &17.3  & - &24.7  & - &3099  & - &0.01  & - \\ 
                 & &16.4  & - &24.3  & - &3086  & - &0.04  & - \\ 
                 & &15.9  & - &23.7  & - &3069  & - &0.06  & - \\ 
                 & &14.7  & - &22.6  & - &2965  & - &0.02  & - \\ 
                 & & 2.7  & - &27.8  & - &-2648 & - &0.53  & - \\ 
          \hline \hline        
                   & hBN & 72.6  &  61.2 &153.4  & 160 &66  &  140  &0.01  & - \\ 
                   & cBN &-17.6  & -17.6 & 0     & 0   & 5  &    0  &0.03  & - \\ \cline{2-10}
                   & armchair(B$-$C)
                   &86.0  & - & 77.7   & - &2147  & - &0.68 & - \\ 
                 & &70.6  & - &147.0   & - & 127  & - &0.45 & - \\ 
                 & &72.7  & - &150.5   & - & 158  & - &0.49 & - \\ 
                 & &64.2  & - &121.1   & - &-694  & - &0.57 & - \\ 
                 & &75.2  & - &150.3   & - & 199  & - &0.77 & - \\ 
                 & &89.8  & - & 94.9   & - &2241  & - &0.84 & - \\ 
                   \cline{2-10}
    $^{14}$N       & armchair(C$-$C)  
                   &85.5  & - & 78.0  & - &2348  & - &0.72 & - \\ 
                 & &65.6  & - &142.9  & - & 142  & - &0.54 & - \\ 
                 & &69.1  & - &146.5  & - & 114  & - &0.96 & - \\ 
                 & &68.6  & - &142.1  & - &-161  & - &0.12 & - \\ 
                 & &71.4  & - &145.3  & - & 178  & - &0.61 & - \\ 
                 & &87.6  & - &77.3   & - &2363  & - &0.71 & - \\ 
                   \cline{2-10}
                   & zigzag
                   &262.2 & - &345.3  & - & 3194  & - &0.86  & - \\ 
                 & & 86.5 & - &153.6  & - & -443  & - &0.99  & - \\ 
                 & & 72.7 & - &147.1  & - & -264  & - &0.06  & - \\ 
                 & & 70.5 & - &144.9  & - & -219  & - &0.40  & - \\ 
                 & & 67.4 & - &141.0  & - &  220  & - &0.25  & - \\ 
                 & & 15.3 & - & 88.2  & - &-1286  & - &0.32  & - \\ 
          \hline \hline
& Li$_{2}$C$_{2}$O$_{4}$ & 171.9 & 170.7 & -77.1 & -81.7 & 0 & - & 0.28 & - \\ 
& LiBOB     & 165.0      & 158.4 &-82.5  & -     & 0 & - & 0.92& -\\
&           & 170.6      & 158.4 &-84.5  & -     & 0 & - & 0.79& -\\
&           & 172.4      & 158.4 &-76.6  & -     & 0 & - & 0.77& -\\
                   \cline{2-10}  
                   &  armchair(B$-$C)
                   &168.0  & 168.4 & 54.5  & - &0  & - &0.92  & - \\  
                 & &128.8  & 127.0 &-25.6  & - &0  & - &0.70  & - \\ 
                   \cline{2-10}
    $^{13}$C       &  armchair(C$-$C)
                   &123.3  & 127.0 &19.5  & - &0  & - &0.94  & - \\  
                 & &122.5  & 127.0 &-24.3 & - &0  & - &0.78  & - \\  
                   \cline{2-10}
                   &  zigzag
                   &169.0  & 168.4 &61.3  & - &0  & - &0.70  & - \\  
                 & &169.3  & 168.4 &61.7  & - &0  & - &0.69  & - \\  
          \hline \hline& 
Li$_{2}$C$_{2}$O$_{4}$ &1.1  &1.1  &-3.3 & - &-130 & 96 & 0.62 & 0.56 \\ 
& LiBOB                &0.2  &0.5  & 6.1 & - & 121 & - & 0.80 & - \\
                   \cline{2-10}
                   & armchair(B$-$C) 
                   &1.2 & - & 5.3 & - & 275 & - &0.74  & - \\ 
                 & &0.1 & - &-3.7 & - &-166 & - &0.97  & - \\ 
                   \cline{2-10}
    $^{7}$Li       & armchair(C$-$C) 
                   &2.5 & - &-6.7 & - & 210 & - &0.11  & - \\ 
                 & &0.5 & - & 8.3 & - &-293 & - &0.84  & - \\ 
                   \cline{2-10}
                   & zigzag 
                   &3.9 & - &-7.5 & - &-190 & - &0.43  & - \\
                 & &0.9 & - &-8.9 & - &-191 & - &0.60  & - \\
          \hline
     \end{tabular}
 }
 \end{table}

\clearpage

\section{Calculated ESR parameters}

\begin{figure}[!htp]
    \centering
    \includegraphics[width=\linewidth]{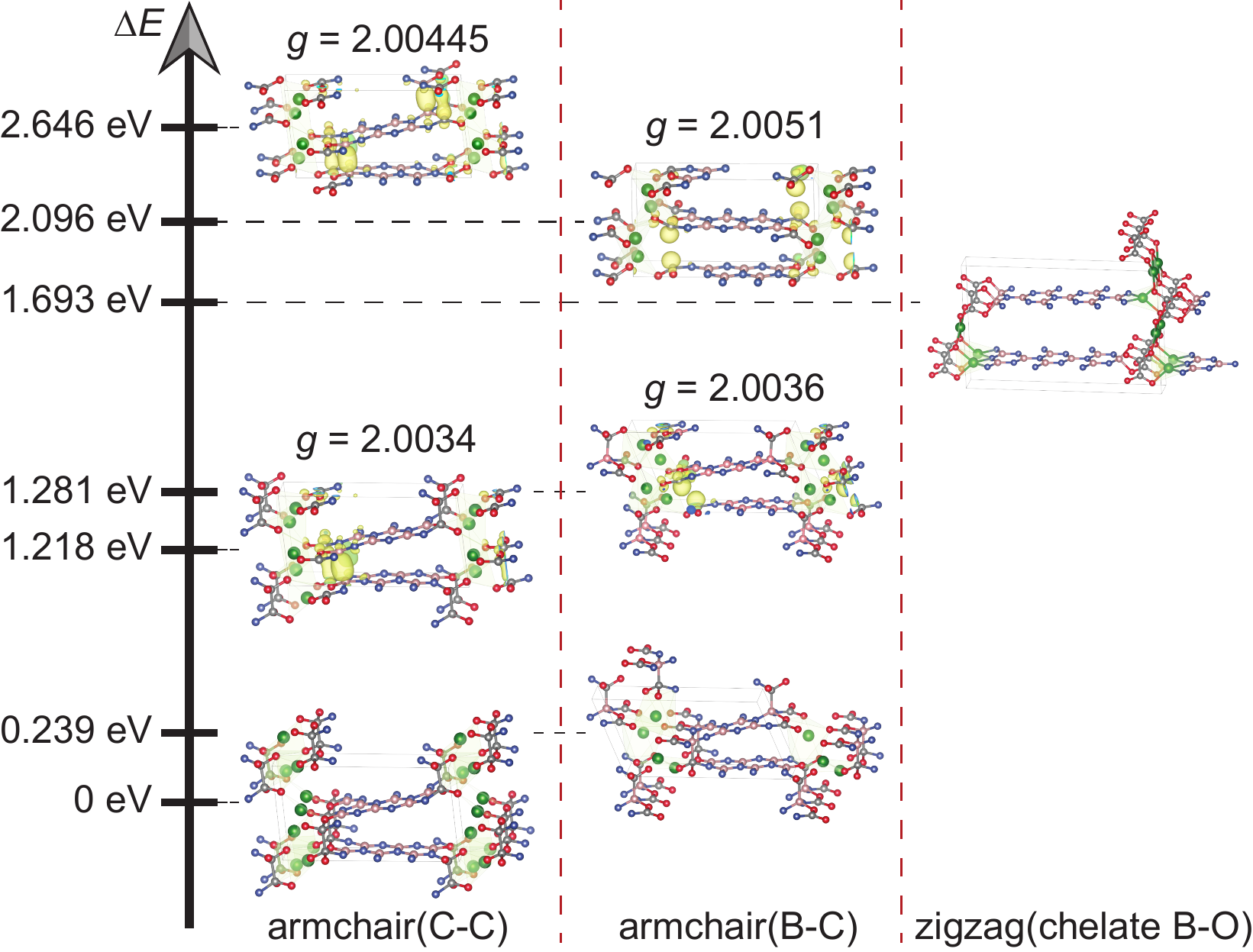}
    \caption{
        Calculated structures and excited radical states with $\Delta E$ energy differences and $g$-factors given.}
    \label{FigSI:other-ESR-structures}
\end{figure}

The calculated ESR parameters are summarized in Table \ref{DFTstructures}. Structures, radical excitations, and electron densities are shown in Fig. \ref{FigSI:other-ESR-structures}.

\begin{table}[htp]
{
     \caption{Calculated (DFT) relative total energies per formula unit ($\Delta E$), nearest interlayer bond lengths and the $g_{\rm iso}$ values of fully relaxed armchair(B/C$-$C) and zigzag(chelate B$-$O) Li$_{2}$(BN)$_{6}$C$_{2}$O$_{4}$ structures at various number of unpaired electrons ($N_{\rm s}$) per unit cell (two formula units).}
     \label{DFTstructures}
     \begin{tabular}{|c|c|c|c|c|c|c|c|c|c|}
     \hline   
           Structure & $N_{\rm s}$ &  $\Delta E$ (eV) & \multicolumn{2}{|c|}{C-C (\AA)} & \multicolumn{2}{|c|}{B-B (\AA)} & \multicolumn{2}{|c|}{B-C (\AA)} & $g_{\rm iso}$   \\
           \hline
           armchair(C$-$C) 
           & 0 & 0.000 & 1.68 & 1.68 & 2.74 & 2.74 & 3.30 & 3.30 &  0.000 \\    
           & 2 & 1.281 & 1.67 & 2.60 & 2.75 & 2.20 & 3.29 & 3.34 &  2.0034 \\
           & 4 & 2.646 & 2.52 & 2.52 & 2.17 & 2.17 & 3.33 & 3.33 &  2.00445 \\
           \hline
           armchair(B$-$C) 
           & 0 & 0.239  & 1.70 & 1.70 &   2.86 & 2.86  &  1.70 & 1.70  & 0.000 \\    
           & 2 & 1.218  & 3.41 & 3.31 &   2.91 & 3.55  &  1.70 & 3.46  & 2.0036 \\
           & 4 & 2.096  & 3.84 & 3.84 &   3.89 & 3.87  &  3.80 & 3.81  & 2.0051 \\
          \hline
           zigzag(chelate B-O)
           & 0 & 1.693  & -    & -    &   -    & -     &  -    & -     & 0.000 \\
          \hline
     \end{tabular}
}     
\end{table}

\section{Comparison with other Li-ion conductors and different functionalization strategies}

Table~\ref{tab:SSE_comparison} places the ion-transport properties of LBNCO in the context of representative solid-state Li-ion conductors and BN-based composite electrolytes. Conventional sulfide and oxide ceramics generally exhibit substantially higher room-temperature ionic conductivities and lower activation energies, whereas polymer-based electrolytes can reach comparable conductivities only with mobile counter-anions and therefore typically show Li$^+$ transference numbers well below unity. In contrast, LBNCO is intrinsically lithiated and is designed as a single-ion conductor, distinguishing it from both salt-containing polymer composites and conventional BN fillers. Its comparatively high activation energy indicates that Li$^+$ transport remains strongly thermally activated, while the measured conductivity demonstrates that covalent functionalization of hBN can create an ionically conducting phase without introducing a conventional lithium salt.

\begin{table*}[htbp]
\centering
\caption{Room-temperature ionic conductivity ($\sigma_\mathrm{RT}$), activation energy ($E_\mathrm{a}$), and Li$^+$ transference number ($t_{\mathrm{Li}^+}$) of LBNCO compared with representative solid-state Li-ion conductors and BN-based composite electrolytes. Note that polymer- and gel-based composites contain mobile anions (dual-ion conductors), whereas ceramic electrolytes and LBNCO are single-ion conductors by design.}
\label{tab:SSE_comparison}
\begin{tabular}{lllccc}
\hline
Material & Class & $\sigma_\mathrm{RT}$ (S\,cm$^{-1}$) & $E_\mathrm{a}$ (meV) & $t_{\mathrm{Li}^+}$ & Ref. \\
\hline
Li$_{10}$GeP$_2$S$_{12}$ (LGPS) & sulfide ceramic & $1.2\times10^{-2}$ & 250 & ${\sim}1$ & \cite{Kamaya2011} \\
Li$_6$PS$_5$Cl (argyrodite) & sulfide ceramic & $1.3\times10^{-3}$ & ${\sim}300-400$ & ${\sim}1$ & \cite{Boulineau2012} \\
Li$_7$La$_3$Zr$_2$O$_{12}$ (LLZO) & oxide garnet & ${\sim}3\times10^{-4}$ & $300-340$ & ${\sim}1$ & \cite{Murugan2007} \\
Li$_{1.3}$Al$_{0.3}$Ti$_{1.7}$(PO$_4$)$_3$ (LATP) & NASICON oxide & $7\times10^{-4}$ & ${\sim}300$ & ${\sim}1$ & \cite{Aono1990} \\
LiPON (thin film) & amorphous oxynitride & $2.3\times10^{-6}$ & 550 & ${\sim}1$ & \cite{Yu1997} \\
PEO-LiTFSI & polymer (dual-ion) & $10^{-7}$--$10^{-5}$ & VTF$^{a}$ & $0.1-0.3$ & \cite{Xue2015} \\
Single-ion triblock copolymer & polymer (single-ion) & $1.3\times10^{-5}$ ($60~^\circ$C) & --$^{a}$ & $>0.85$ & \cite{Bouchet2013} \\
\hline
PVDF-HFP/LiTFSI + 1\,wt\% h-BN & BN-polymer composite & $1.8\times10^{-3}$ & -- & -- & \cite{Zhang2019} \\
PEO/LiTFSI + FBN nanosheets & BN-polymer composite & $1.1\times10^{-4}$ ($30~^\circ$C) & -- & 0.56 & \cite{Wang2023} \\
\hline
LBNCO & FBN (single-ion) & $5\times10^{-11}$ & 816 & TBD & this work \\
Li-doped LBNCO & lithiated FBN (single-ion) & $1\times10^{-8}$ & 850 & TBD & this work \\
\hline
\multicolumn{6}{l}{\footnotesize $^{a}$PEO-based electrolytes above their melting point do not follow Arrhenius behavior; an apparent Arrhenius $E_\mathrm{a}$ is not directly comparable.} \\
\hline
\end{tabular}
\label{TableSI:prepertycompare}
\end{table*}

This distinction is also evident from the comparison of hBN functionalization strategies summarized in Table~\ref{tab:FBN_comparison}. Previous covalent modification approaches primarily introduced O-, N-, C-, or F-containing groups to improve dispersibility, interfacial compatibility, or electronic properties, and ionic transport was generally not a design objective. Importantly, these methods do not incorporate lithium as an intrinsic component of the functionalized hBN structure. The lithium-oxalate mechanochemical route developed here therefore differs conceptually by combining covalent edge functionalization with Li-containing species in a single material. This should also be distinguished from previously reported Li intercalation into hBN, which relies on interstitial lithium introduced under comparatively severe processing conditions and does not produce a covalently functionalized Li-bearing phase. Together, these comparisons position LBNCO as a distinct class of functionalized hBN in which chemical modification and Li-ion transport are directly coupled.

\begin{sidewaystable}[htbp]
\centering
{\scriptsize
\caption{Comparison of covalent hBN functionalization strategies. Prior approaches introduce H-, O-, N-, or C-based groups but no lithium; ionic conductivity is generally not reported (n.r.) since these materials were developed for dispersibility, composites, or band-gap engineering rather than ion transport.}
\label{tab:FBN_comparison}
\begin{tabular}{llllllc}%
\hline
Method & Functional group & Funct. degree & Li content & Ionic conductivity & Thermal stability & Ref. \\
\hline
Oxygen-radical grafting (solution) & --O-$t$Bu $\rightarrow$ --OH & ${\sim}4$\,at\% B (TGA) & none & n.r. (insulator) & -- & \cite{Sainsbury2012} \\
Sonication-assisted hydrolysis & --OH & low (edge/defect) & none & n.r. & -- & \cite{Lin2010} \\
Urea-assisted ball milling & --NH$_2$ (edge) & n.q.; yield ${\sim}85$\% & none & n.r. & -- & \cite{Lei2015} \\
Hydroxide-assisted ball milling & --OH (edge) & n.q. & none & n.r. & -- & \cite{Lee2015} \\
Carbene addition (solution) & $>$CBr$_2$ & low & none & n.r. & -- & \cite{Sainsbury2014} \\
Fluorination/oxidation for electrodes & --F, --O & varies & electrochem.\ only$^{a}$ & n.r. & -- & \cite{Zhang2016} \\
\hline
Lithium-oxalate mechanochemistry & oxalate (edge) + Li & \emph{- \%} & \emph{- wt\% or Li:B} & \emph{- $\sigma_\mathrm{RT}$} & \emph{- $T_\mathrm{dec}$} & this work \\
Li-doped LBNCO & oxalate + excess Li & \emph{-} & \emph{-} & \emph{-} & \emph{-} & this work \\
\hline
\hline
Li intercalation (1523\,K anneal or milling+heat) & interstitial Li (no covalent group) & n/a & Li(BN)$_9$--Li(BN)$_3$; $d_{002}=3.76$~\AA & n.r.; reverts to insulating hBN upon Li removal & requires $>$1500\,K synthesis & \cite{Sumiyoshi2010,Kim2016b} \\
Li at hBN/hBN interfaces (DFT) & none & n/a & thermodynamically unstable ($E_I>0$) & -- & -- & \cite{Shirodkar2016,altintas2011intercalation} \\
\multicolumn{7}{l}{\footnotesize n.q. = not quantified; n.r. = not reported. $^{a}$Li introduced only during electrochemical cycling as electrode material, not intrinsic to the functionalized phase.} \\
\hline
\end{tabular} }
\label{TableSi:reactioincompare}
\end{sidewaystable}



\end{document}